\documentclass[a4paper,11pt]{article}
\usepackage{graphicx}

\usepackage{jheppub} 

\usepackage[T1]{fontenc} 
\usepackage{atlasphysics}
\usepackage{subfigure}
\usepackage{verbatim}
\usepackage{rotating}
\usepackage{float}
\usepackage{bm}            
\usepackage{hyperref}     
\usepackage{multirow}

 \usepackage{preprintcover}

 \PreprintCoverPaperTitle{Search for squarks and gluinos in events with isolated leptons, jets and missing transverse momentum at $\sqrt{s}=8$ \TeV~with the ATLAS detector}

 \PreprintIdNumber{CERN-PH-EP-2014-278}

 \PreprintCoverAbstract{
 The results of a search for supersymmetry
   in final states containing at least one isolated lepton (electron or muon),
     jets and large missing transverse momentum with the ATLAS detector at the Large Hadron Collider are reported.
       The search is based on proton--proton collision data at a centre-of-mass
         energy $\sqrt{s} = 8$~TeV collected in 2012,
	   corresponding to an integrated luminosity of 20~\ifb.
	     No significant excess above the Standard Model expectation is observed.
	       Limits are set on supersymmetric particle masses for various supersymmetric models. Depending on the model,
	           the search excludes gluino masses up to 1.32 \TeV~and squark masses up to 840 \GeV.
		     Limits are also set on the parameters of a minimal universal extra dimension model,
		       excluding a compactification radius of $1/R_{\mathrm{c}}=950$ \GeV~for a cut-off scale times radius ($\Lambda R_{\mathrm{c}}$) of approximately 30.
 }

 \PreprintJournalName{JHEP}

\usepackage{datetime}
\date{\currenttime}

\title{Search for squarks and gluinos in events with isolated leptons, jets and missing transverse momentum at $\sqrt{s}=8$ \TeV~with the ATLAS detector}

\abstract{
  The results of a search for supersymmetry 
  in final states containing at least one isolated lepton (electron or muon),
  jets and large missing transverse momentum with the ATLAS detector at the Large Hadron Collider are reported.
  The search is based on proton--proton collision data at a centre-of-mass
  energy $\sqrt{s} = 8$~TeV collected in 2012,
  corresponding to an integrated luminosity of 20~\ifb.
  No significant excess above the Standard Model expectation is observed.
  Limits are set on supersymmetric particle masses for various supersymmetric models. Depending on the model,
    the search excludes gluino masses up to 1.32 \TeV~and squark masses up to 840 \GeV.
  Limits are also set on the parameters of a minimal universal extra dimension model, 
  excluding a compactification radius of $1/R_{\mathrm{c}}=950$ \GeV~for a cut-off scale times radius ($\Lambda R_{\mathrm{c}}$) of approximately 30.
}

\begin{document}

\newcommand{\figref}[1]{Figure~\ref{#1}}
\newcommand{\figsref}[1]{Figures.~\ref{#1}}
\newcommand{\Secref}[1]{Section~\ref{#1}}
\newcommand{\tabref}[1]{Table~\ref{#1}}

\newcommand{\meff}{\ensuremath{m_{\mathrm{eff}}}} 
\newcommand{\mt}{\ensuremath{m_\mathrm{T}}}
\newcommand{\mttwo}{\ensuremath{m_\mathrm{T2}}}
\newcommand{\mc}{\ensuremath{m_\mathrm{C}}}
\newcommand{\mct}{\ensuremath{m_\mathrm{CT}}}
\def\ptl{\ensuremath{p_{\mathrm{T}}^\ell}}
\def\ptlone{\ensuremath{p_{\mathrm{T}}^{\ell_1}}}
\def\ptltwo{\ensuremath{p_{\mathrm{T}}^{\ell_2}}}
\def\ptj{\ensuremath{p_{\mathrm{T}}^\mathrm{jet}}}
\def\dPhimin{\ensuremath{\Delta \phi_{\rm min}}}
\newcommand{\meffincl}{\ensuremath{m_{\mathrm{eff}}^{\mathrm{incl}}}}

\def\Ptmiss{\ensuremath{\vec{E}_\mathrm{T}^{\mathrm{miss}}}}

\def\lsim{\mathrel{\rlap{\lower4pt\hbox{\hskip1pt$\sim$}}
    \raise1pt\hbox{$<$}}}                
\def\gsim{\mathrel{\rlap{\lower4pt\hbox{\hskip1pt$\sim$}}
    \raise1pt\hbox{$>$}}}                

\makeatletter
\renewcommand{\p@subfigure}{\arabic{figure}}
\renewcommand{\thesubfigure}{\alph{subfigure}}
\renewcommand{\@thesubfigure}{(\thesubfigure)}
\makeatother

\maketitle
\flushbottom

\section{Introduction}
Supersymmetry (SUSY) \cite{Miyazawa:1966,Ramond:1971gb,Golfand:1971iw,Neveu:1971rx,Neveu:1971iv,Gervais:1971ji,Volkov:1973ix,Wess:1973kz,Wess:1974tw} postulates the existence of particles (sparticles) which differ by half a unit of spin from their Standard Model (SM) partners.
The squarks ($\tilde{q}_{\rm L}$ and $\tilde{q}_{\rm R}$) and sleptons ($\tilde{\ell}_{\rm L}$ and $\tilde{\ell}_{\rm R}$) are the scalar partners of the left-handed and right-handed quarks and leptons, 
the gluinos ($\tilde{g}$) are the fermionic partners of the gluons, and the charginos (${\tilde{\chi}}_{i}^{\pm}$ with $i=1,2$) and neutralinos
 (${\tilde{\chi}}_{i}^{0}$ with $i=1,2,3,4$) are the mass eigenstates (ordered from the lightest to the heaviest) formed from the linear superpositions of the SUSY partners of the
Higgs and electroweak gauge bosons. An attractive feature of SUSY is that it can solve the SM hierarchy problem  
\cite{Witten:1981nf,Dine:1981za,Dimopoulos:1981au,Sakai:1981gr,Kaul:1981hi,Dimopoulos:1981zb}  if the gluino, higgsino and top squark masses 
are not much higher than the \TeV~scale. 
 
If strongly interacting sparticles exist at the \TeV~scale, they should be accessible 
at the Large Hadron Collider (LHC). In the minimal supersymmetric extension of the SM   
such particles decay into jets, possibly leptons, and
the lightest sparticle (LSP). If the LSP is stable owing to R-parity conservation \cite{Fayet:1976et,Fayet:1977yc,Farrar:1978xj,Fayet:1979sa,Dimopoulos:1981zb} and only weakly 
interacting, it escapes detection, leading to missing transverse momentum
(${\boldsymbol p}_{\mathrm{T}}^\mathrm{miss}$ and its magnitude \met) in the final state. In this scenario, the LSP can be a dark-matter candidate.
Significant \met~can also arise in R-parity-violating scenarios in which the LSP decays to final states containing neutrinos or 
in scenarios where neutrinos are present in the cascade decay chains of the produced sparticles.

This paper presents a search with the ATLAS detector \cite{Aad:2008zzm,CSCbook} for SUSY in final states containing jets, 
at least one isolated lepton (electron or muon) and large \met. Different search channels are used in order to
cover a broad parameter space: the events are selected by different requirements on the transverse momentum (\pT) of the leptons,  
either using low-\pT~leptons (referred to as the ``soft'' lepton selection), or high-\pT~leptons (referred to as the ``hard'' lepton selection). 
Each of these categories is further subdivided into a single-lepton and a dilepton search channel.  
The soft-lepton and hard-lepton channels are complementary, being more sensitive to supersymmetric spectra with small or large mass splittings, respectively, while the
different lepton multiplicities cover different production and decay modes. To enhance the sensitivity to gluino or squark production, 
high and low jet multiplicity signal regions are defined. 

Previous searches in these final states have been conducted by the ATLAS~\cite{paper7tev,razor_7tev} and CMS~\cite{CMS_1lep7TeV} collaborations using their full
2011 dataset at a centre-of-mass energy of 7 \TeV.
In this paper, the analysis is performed on the full 2012 ATLAS dataset at a centre-of-mass energy of 8 \TeV, corresponding to an integrated luminosity of up to 20.3~\ifb.
All signal regions defined in this search are optimised for this dataset.

The paper is organised as follows. After a brief description of the ATLAS detector in section \ref{sec:detector}, the simulation of the background and signal processes used in the analysis is detailed in
section \ref{simulsection}. Section \ref{sec:trigger} discusses the trigger strategy and the dataset used, while the object reconstruction and 
the event selection are addressed in sections \ref{objectreco} and \ref{sec:EventSelection}.
The background estimation and the systematic uncertainties are discussed in sections \ref{bkgestimate} and \ref{sec:systuncert}. The fitting procedure used is 
described in section \ref{sec:bkgfit} and the results are presented in section \ref{sec:results}. Finally, section \ref{conclusion} presents the conclusions.

\section{The ATLAS detector}\label{sec:detector}

ATLAS is a multi-purpose detector which provides a nearly full solid angle coverage around the interaction point.\footnote{The nominal $pp$ 
interaction point at the centre of the detector is defined as the origin of a right-handed coordinate system.  
The positive $x$-axis is defined by the direction from the interaction point to the centre of the LHC ring, with the
positive $y$-axis pointing upwards, while the beam direction defines the $z$-axis. The azimuthal angle $\phi$ is measured around the beam axis and the polar angle $\theta$ is the angle from
the $z$-axis. The pseudorapidity is defined as $\eta = -\ln \tan(\theta/2)$.}
It consists of a tracking system (inner detector or ID) surrounded by a thin superconducting solenoid providing a 2 T magnetic field,
 electromagnetic and hadronic calorimeters and a muon spectrometer (MS). The ID consists of pixel and silicon microstrip detectors covering the pseudorapidity region $|\eta|<2.5$, 
surrounded by the transition radiation tracker (TRT) which provides electron identification in the region $|\eta|<2.0$. The calorimeters cover $|\eta|<4.9$, the 
forward region ($3.2<|\eta|<4.9$) being instrumented with a liquid-argon (LAr) calorimeter for both the electromagnetic and hadronic measurements. In the central region, a high-granularity lead/LAr  
electromagnetic calorimeter covers $|\eta|<3.2$, while the hadronic calorimeter uses two different detector technologies, 
with scintillator tiles ($|\eta|<1.7$) or LAr ($1.5<|\eta|<3.2$) as active medium.
The MS is based on three large superconducting toroids arranged with an eight-fold azimuthal coil symmetry around the calorimeters, and a system of three
layers of precision tracking chambers providing coverage over $|\eta|<2.7$, while dedicated fast chambers allow triggering over $|\eta|<2.4$.  
The ATLAS trigger system \cite{atlastrigger} consists of three levels; the first level (L1) is a hardware-based system, while the second and
third levels are software-based systems and are collectively referred to as the High Level Trigger (HLT).

\section{SUSY signal modelling and simulated event samples}
\label{simulsection}

\subsection{Signal event samples}
\label{simulsignal}
The signal models considered cover simplified \cite{Alwall:2008ag,Alves:2011wf} and phenomenological SUSY models, as well as a minimal Universal Extra 
Dimension (mUED) scenario \cite{Cheng:2000,Cheng:2002}. Some of these models were also probed by other ATLAS searches based on the 8 \TeV~$pp$ 
dataset, using different final-state selections \cite{2lepss,0lepmultijet,0lepton,3bj,paperStrongTau}.
The simplified models studied here include the pair production of gluinos or first- and second-generation squarks with different hypotheses for  
their decay chains, as well as gluino-mediated top squark pair production. In these models, the LSP is always the lightest neutralino.
The phenomenological models include scenarios for minimal super-gravity-mediated SUSY breaking (mSUGRA/CMSSM) \cite{msugra1,msugra2,msugra3,msugra4,msugra5,Kane:1993td}, bilinear R-parity violation (bRPV) \cite{brpv}, natural gauge mediation (nGM) \cite{ngm} 
and a non-universal Higgs-boson mass with gaugino mediation (NUHMG) \cite{nuhmg}. 

\subsubsection{Simplified models}

The topologies of the simplified models considered in this paper are illustrated in figure \ref{fig:feynman}.

In these simplified models, all the sparticles which do not directly enter the production and decay chain are effectively
decoupled.

The first category of simplified models focuses on the pair production of left-handed squarks
or of gluinos, the latter assuming degenerate first- and second-generation squarks. 
This category of models is subdivided into three different decay chains: ``one-step'' models,  
``two-step'' models with sleptons, and ``two-step'' models without sleptons. 

\begin{figure}[htb]
\centering
\includegraphics[width=0.25\textwidth]{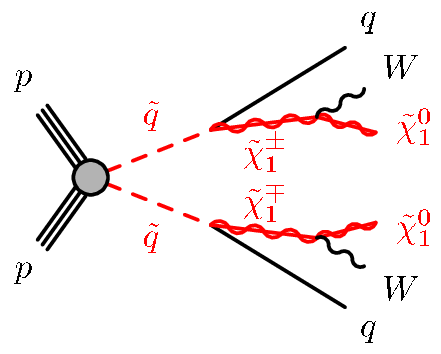}
\qquad
\includegraphics[width=0.25\textwidth]{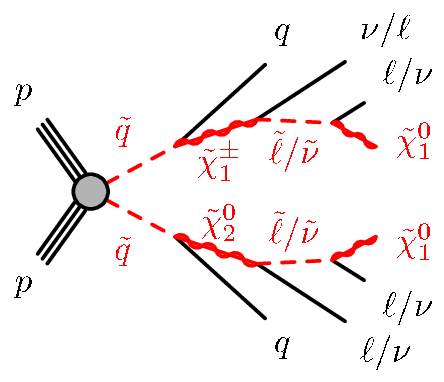}
\qquad
\includegraphics[width=0.25\textwidth]{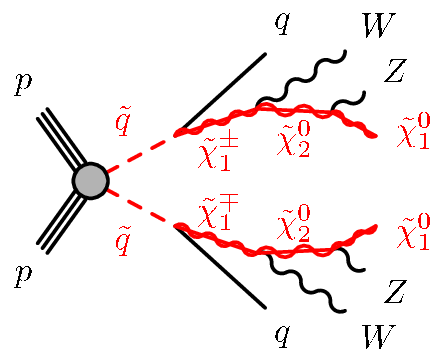}
\includegraphics[width=0.25\textwidth]{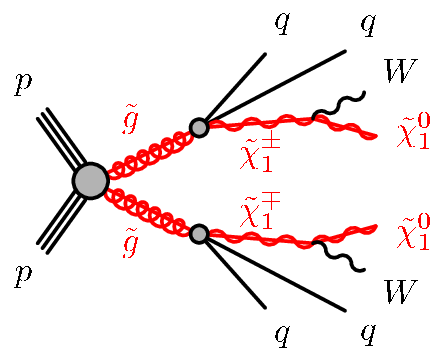}
\qquad
\includegraphics[width=0.25\textwidth]{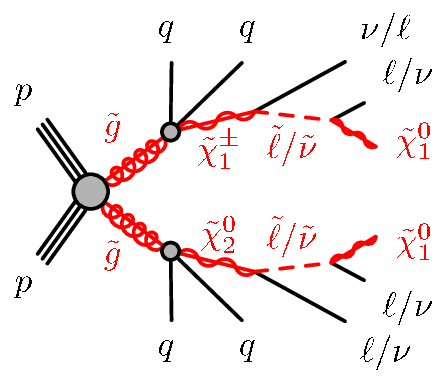}
\qquad
\includegraphics[width=0.25\textwidth]{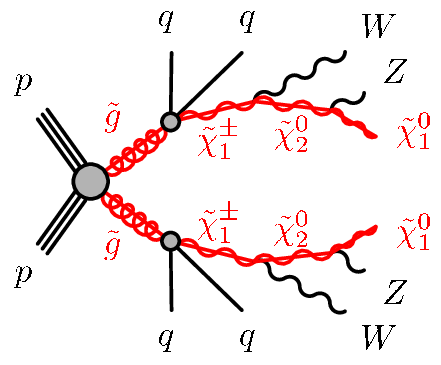}
\qquad
\includegraphics[width=0.25\textwidth]{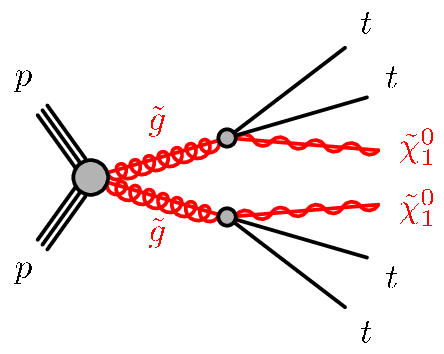}
\qquad
\includegraphics[width=0.25\textwidth]{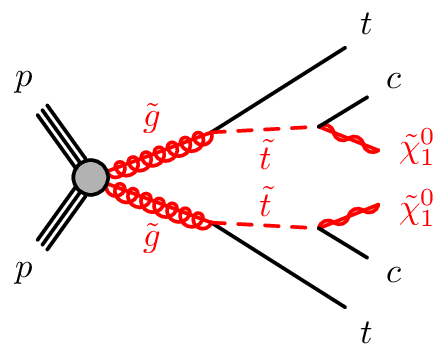}
\caption{Examples of the decay topologies of the $\tilde{q}_{\rm L}$ (top) or $\tilde{g}$ (middle) pair production, in the simplified model with ``one step'' (left) and ``two steps'' with (centre) or without (right) sleptons.
The bottom diagrams show examples of the topologies considered for gluino-mediated production of top squarks.
} \label{fig:feynman}
\end{figure}

In the ``one-step'' models, the pair-produced strongly interacting sparticles 
decay via the lighter chargino into a $W$ boson and the lightest neutralino.  
The free parameters in these models are chosen to be the mass of the squark/gluino
 and either the mass of the chargino, with a fixed $\ninoone$ mass 
set to 60~\GeV, or the mass of the $\ninoone$, with the chargino mass set to $m_{\chinoonepm}=(m_{\gluino/\squark}+m_{\ninoone})/2$.

In the ``two-step'' models with sleptons, the strongly interacting sparticles decay with equal probability via
either the lightest chargino or the next-to-lightest neutralino. These subsequently decay via left-handed sleptons (or sneutrinos)
which decay into a lepton (or neutrino) and the lightest neutralino. In these models, the free parameters
are chosen to be the initial sparticle mass and the $\ninoone$ mass. The masses of the intermediate charginos/neutralinos are set to be equal, $m_{\chinoonepm,\ninotwo}=(m_{\gluino/\squark}+m_{\ninoone})/2$, 
while the slepton and sneutrino masses (all three lepton flavours are mass degenerate in this model) are set to $m_{\sleptonL,\snu}=(m_{\chinoonepm/\ninotwo}+m_{\ninoone})/2$.

Finally, in the ``two-step'' models without sleptons, the initial sparticle
decays via the lighter chargino, which itself decays into a $W$ boson and the next-to-lightest neutralino. The latter finally
decays into a $Z$ boson and the $\ninoone$. The lighter chargino mass is fixed at $m_{\chinoonepm}=(m_{\gluino/\squark}+m_{\ninoone})/2$ 
and the next-to-lightest neutralino mass is set to be $m_{\ninotwo}=(m_{\chinoonepm}+m_{\ninoone})/2$.
This signature could be realised in the Minimal Supersymmetric Standard Model (MSSM) in a region of parameter space where additional decay modes, 
not contained in the simplified model, may lead to a significant reduction in the cross section times branching fraction of the $WZ$ signature.

The second category of simplified models considers the gluino-mediated production of top squarks.\footnote{In these models, 
the $\stop$ mixing angle is taken to be $56^{\circ}$, but the value of this mixing angle has no impact on the results of the analyses presented in this paper.} In these models, the 
lightest squark is the lightest top squark mass
eigenstate $\stopone$ formed from the mixing of $\stop_L$ and $\stop_R$, and
the squarks of all other flavours are effectively decoupled.  
Two models are considered in this specific search for
gluino-mediated top squark production.
In the first model, $\stopone$ is effectively decoupled and its mass is set to 2.5 \TeV, a mass for which there is no current sensitivity to direct production. 
Each gluino decays with 100\% branching fraction to a top quark and a virtual top squark, the latter exclusively decaying to a top quark and the $\ninoone$, 
leading to a final state with a pair of top quarks and a neutralino, $\gluino \to t\bar t\ninoone$. The mass of the gluino is a free parameter and is varied up to 1.4 \TeV, 
a value representative of the expected reach of the analysis.
This final state is therefore characterised by the presence of four top quarks (decaying to four $b$-jets and four $W$ bosons) and two $\ninoone$. 
In the second model, the gluino is heavier than the $\stopone$, and the mass gap between the $\stopone$ and the $\ninoone$ is smaller than the $W$ boson mass and fixed to 20~\GeV. Gluinos decay to a top quark and a top squark, 
$\gluino \to \bar t \stopone$, and the $\stopone$ is set to exclusively decay to a charm quark and the $\ninoone$, $\stopone \to c \ninoone$. Using gluino-mediated production 
to probe this decay is particularly interesting as it is complementary to the direct pair production of $\stopone$, which is more 
difficult to extract from the background for this specific decay mode of $\stopone$ (see ref.~\cite{charmstop}).
This final state is therefore characterised by the presence of two top quarks (decaying to two $b$-jets and two $W$ bosons), two $c$-quarks
and two $\ninoone$. 

\subsubsection{Phenomenological models}
Phenomenological models are also considered in this paper. The mSUGRA/CMSSM model 
is specified by five parameters: the universal scalar mass $m_0$, the universal gaugino mass $m_{1/2}$, the universal trilinear scalar coupling $A_0$, 
the ratio $\tan\beta$ of the vacuum expectation values of the two Higgs fields, and the sign of the higgsino mass parameter $\mu$.
In the mSUGRA/CMSSM model studied here, the values $\tan\beta = 30$, $A_{0} = -2m_0$ and $\mu > 0$ were chosen, such that 
the lightest scalar Higgs boson mass is approximately 125 \GeV~in most of the $(m_0$,$m_{1/2})$ parameter space studied.

A bRPV scenario is also studied; it uses the same parameters as the mSUGRA/CMSSM model, but with non-zero bilinear R-parity-violating couplings, which
are determined by a fit to atmospheric and solar neutrino data \cite{brpv2} under the tree-level dominance scenario \cite{neutrinobrpv}. In this scenario, the 
$\ninoone$ LSP decays promptly to $W\mu$, $W\tau$, $Z\nu$ or $h\nu$ (where
the $W/Z/h$ boson can either be on shell or off shell) with branching fractions which are weakly dependent on $m_0$ and $m_{1/2}$ but which 
are typically of the order of 20--40\%, 20--40\%, 20--30\% and 0--20\%, respectively. 

The nGM scenario differs from the general gauge mediation models \cite{ggm,ggm2} in that all 
sparticles that are not relevant to the tuning of the Higgs sector are decoupled. The relevant sparticles are thus the higgsinos, one 
or two light top squarks, a light gluino and a very light gravitino ($\gravino$) LSP. This configuration results in minimal fine tuning while obeying 
all current collider constraints. The sparticles that play no role in fine tuning can subsequently be reintroduced while retaining the naturalness of the model.
In the model considered here, and described in detail in ref. \cite{paperStrongTau}, the stau ($\stau$) is assumed to be the next-to-lightest SUSY particle (NLSP), and the gluino is assumed to be the only light coloured sparticle. 
Therefore, the only relevant production process in this model is gluino
pair production followed by two possible decay chains: $ \gluino \rightarrow g \ninoonetwo \rightarrow g \stau \tau \rightarrow g \tau \tau \gravino$ and 
$ \gluino \rightarrow q q' \chinoonepm \rightarrow q q' \nu_{\tau} \stau \rightarrow  q q' \nu_{\tau} \tau \gravino$, 
where $q$ and $q'$~are almost exclusively top or bottom quarks.
The exact proportion of the two processes depends on the mass of the decoupled squarks, with the first process only happening for low gluino masses. 
The higgsino mass parameter $\mu$ is set to 400~\GeV, which fixes the mass of the chargino and the neutralinos, such that strong production is the dominant process at the LHC. 
A range of signals with varying gluino and stau masses are studied. 
The lightest Higgs-boson mass is specifically set to 125~\GeV. 

NUHMG is an R-parity-conserving model with the tau-sneutrino as the NLSP. There are six parameters which can be varied to obtain different phenomenologies: 
$\tan\beta$, $m_{1/2}$, $A_0$ and the sign of $\mu$, defined above, as well as the squared mass terms of the two Higgs doublets: $m^2_{H_1}$ and $m^2_{H_2}$. 
These parameters are set as follows: $\tan\beta=10$, $\mu>0$, $m^2_{H_2}=0$; $m_{1/2}$ and $m^2_{H_1}$ are chosen such that the NLSP
is a tau-sneutrino with properties satisfying Big Bang Nucleosynthesis constraints (see ref.~\cite{nuhmg}); $A_0$ is chosen to maximise the mass of the lightest Higgs boson (in NUHMG models, the 
Higgs boson mass obtained is usually lower than the measured value: varying $A_0$ allows the models considered here to minimise this difference to the level of a few \GeV).
In this model, there is a significant  production of gluinos and squarks throughout the parameter space studied. 
The gluino decays mainly to a first- or second-generation quark/squark pair $q\squark$ ($\approx50$\%), but also to $t\stop$~($\approx30$\%) or $b\sbottom$~($\approx20$\%), while the squark cascade decay typically involves charginos, neutralinos and/or sleptons. 

This analysis also considers the mUED model, which is the minimal extension of the SM with one additional universal spatial dimension.
The properties of the model depend on only three parameters: the compactification radius $R_{\mathrm{c}}$, the cut-off scale $\Lambda$ and the Higgs 
boson mass $m_h$. In this model, the mass spectrum is naturally degenerate and the decay chain of the Kaluza--Klein (KK) quark 
to the lightest KK particle, the KK photon, gives a signature very similar to the supersymmetric 
decay chain of a squark to the lightest neutralino. 
Signal events for this model are generated with a Higgs-boson mass of 125~\GeV.

\subsubsection{Event generation}
\textsc{SUSY-HIT} and \textsc{SDECAY} 1.3b \cite{ref:susyhit,ref:sdecay}, interfaced to \textsc{SOFTSUSY} 3.1.6 \cite{ref:softsusy},
are used to calculate the sparticle mass spectra and decay tables, and to ensure consistent
electroweak symmetry breaking in the mSUGRA/CMSSM models.
All the simplified models except the gluino-mediated top squark production are generated with up to one extra parton in the matrix element using
\madgraph5 1.3.33~\cite{Alwall:2011uj} interfaced to \pythia6.426 \cite{PYTHIA}; MLM matching \cite{Alwall:2007fs} is applied with a scale parameter
that is set to a quarter of the mass of the lightest sparticle in the hard-scattering matrix element.
\textsc{Herwig++} 2.5.2 \cite{Bahr:2008pv} is used to generate the mUED, mSUGRA and nGM samples, as well as the samples for the simplified model with gluino-mediated top squark production.
Finally, the NUHMG and bRPV samples are generated with \pythia6.426.
The ATLAS underlying-event tune AUET2 is used \cite{AUET2} for \madgraph5 and \pythia6 samples while the CTEQ6L1-UE-EE-3 tune \cite{UEEE3} is used for \textsc{Herwig++} samples.
The parton distribution functions (PDFs) from CTEQ6L1 \cite{Pumplin:2002vw} are used for all signal samples.

For all except the mUED sample, the signal cross sections are calculated at next-to-leading order (NLO) in the strong coupling constant, 
adding the resummation of soft gluon emission at next-to-leading-logarithmic accuracy 
(NLO+NLL)~\cite{Beenakker:1996ch,Kulesza:2008jb,Kulesza:2009kq,Beenakker:2009ha,Beenakker:2011fu}.
The nominal cross section 
is taken from an envelope of cross-section predictions 
using different PDF sets and factorisation and renormalisation scales, as described in ref.~\cite{Kramer:2012bx}. 
For the mUED model, the cross section is taken at leading order from \textsc{Herwig++}. 

\subsection{Standard Model event samples}
\label{simulbg}

\begin{table*}[ht]
\begin{center}
\scriptsize
\begin{tabular}{l l c c c c }
\hline\hline
Physics process & Generator& Cross-section & PDF set & Tune\\
&& normalisation & & \\
\hline\hline
$W(\rightarrow \ell\nu)$ + jets  & \sherpa 1.4.1 \cite{sherpa} & NNLO \cite{DYNNLO1,DYNNLO2} &NLO CT10 \cite{Lai:2010vv} & \sherpa default \\
$Z/\gamma^{*}(\rightarrow \ell \ell)$ + jets &&&&\\
($m_{\ell\ell}>40$~\GeV) & \sherpa 1.4.1 &NNLO \cite{DYNNLO1,DYNNLO2} &NLO CT10 & \sherpa default\\
$t\bar{t}$ & {\sc powheg-box} r2129 & NNLO+NNLL &NLO CT10 &\sc{Perugia2011C}\\
& \cite{PowhegBOX1,PowhegBOX2,PowhegBOX3} & \cite{ttbarxsec1,ttbarxsec2}& & \cite{pythiaperugia}\\
Single-top & &&&\\
($t$-channel) & \acermc 3.8 \cite{Kersevan:2004yg} & NNLO+NNLL \cite{Kidonakis:2011} & CTEQ6L1 \cite{Pumplin:2002vw} & \sc{Perugia2011C}\\
Single-top &&&&\\
($s$-channel and $Wt$) & {\sc powheg-box} r1556 & NNLO+NNLL \cite{Kidonakis:2010a,Kidonakis:2010b} &NLO CT10 & \sc{Perugia2011C}\\ 
$t+Z$ & \madgraph5 1.3.28 \cite{Alwall:2011uj}  & LO  &CTEQ6L1 & AUET2\cite{AUET2}\\
$t\bar{t}+W(W)/Z$  & \madgraph5 1.3.28  & NLO \cite{Campbell:2012,Lazopoulos:2008} & CTEQ6L1 & AUET2 \\
\multicolumn{4}{l}{{\bf Single-lepton analyses:}}\\
$WW$, $WZ$ and $ZZ$ & \sherpa 1.4.1  & NLO \cite{diboson1,diboson2} &NLO CT10 & \sherpa default\\
$W\gamma$ and $Z\gamma$    & \sherpa 1.4.1 & LO &NLO CT10 & \sherpa default\\
\multicolumn{4}{l}{{\bf Dilepton analyses:}}\\
$WW$, $WZ$ and $ZZ$ & {\sc powheg-box} r1508 \cite{Nason:2013ydw} & NLO \cite{diboson1,diboson2} &NLO CT10 & AUET2 \\\hline
\multicolumn{4}{l}{{\bf Hard-lepton analyses:}}\\
Drell--Yan  & \sherpa 1.4.1  & NNLO \cite{Melnikov:2006kv} &NLO CT10 & \sherpa default\\
($8<m_{\ell\ell}<40$~\GeV) &&&&\\
\multicolumn{4}{l}{{\bf Soft-lepton analyses:}}\\
$Z/\gamma^{*}(\rightarrow \ell \ell)$ + jets  &\alpgen 2.14~\cite{Mangano:2002ea} &NNLO \cite{Melnikov:2006kv} & CTEQ6L1 & AUET2\\
 ($10<m_{\ell\ell}<60$~\GeV) & & & &\\
\hline\hline
\end{tabular}
\caption{Simulated background event samples used in this paper (where $\ell = e, \mu, \tau$): the corresponding generators, cross-section normalisation, 
	PDF set and underlying event tune are shown. More details (e.g. concerning the parton showers) can be found in the text. 
}
\label{tab:MC}
\end{center}
\end{table*}

The simulated event samples for the SM backgrounds are summarised in table \ref{tab:MC}, along with the PDFs and tunes used. Further samples are
also used to compute systematic uncertainties, as explained in section \ref{sec:systuncert}. 
The Drell--Yan samples used in the hard-lepton analyses have a filter which selects events at generation level by requiring the leptons to satisfy 
\pT$^{\ell_1(\ell_2)}>9 (5)$ \GeV~and $|\eta_{\ell_{1,2}}|<2.8$. This filter prevents its use in the soft-lepton analyses which
   use \alpgen samples with a lepton \pT~cut at 5 \GeV.
The \alpgen \cite{Mangano:2002ea} samples are generated with the MLM matching scheme and with $0 \le N_{\rm{parton}} \le 5$; for these samples   
\herwig6.520 \cite{Corcella:2000bw} is used for simulating the parton shower and fragmentation processes in combination
with \jimmy \cite{Butterworth:1996zw} for underlying-event simulation. 
\pythia6.426 is used for the \madgraph5, \acermc \cite{Kersevan:2004yg} 
and all \powheg \cite{PowhegBOX1,PowhegBOX2,PowhegBOX3} samples except for the diboson \powheg samples, which use \pythia8.163 \cite{Pythia8}. 
The \powheg diboson samples have dilepton filters which increase the number of Monte Carlo events available for the dilepton analyses.
\sherpa uses its own parton shower and fragmentation, and the \sherpa $W$+jets and $Z/\gamma^{*}$+jets samples are generated with massive
$b/c$-quarks to improve the treatment of the associated production of $W/Z$ bosons with heavy flavour. 

\subsection{Detector simulation}
\label{simul}
The detector simulation is performed either with a full ATLAS detector simulation \cite{:2010wqa}
based on \textsc{Geant4} \cite{Agostinelli:2002hh} or a fast simulation based on the parameterisation of the performance
of the ATLAS electromagnetic and hadronic calorimeters \cite{atlfast} and on \textsc{Geant4} elsewhere.
All simulated samples are generated with
a range of minimum-bias interactions (simulated using \pythia8 with the MSTW2008LO PDF set~\cite{Sherstnev:2007nd} and the A2 tune \cite{a2tune}) overlaid on the
hard-scattering event to account for the
multiple $pp$ interactions in
the same bunch crossing (pile-up). The overlay
also treats the impact of pile-up on bunch crossings
other than the bunch crossing in which the event occurred.
Event-level weights are applied to the simulated samples to account for
differences between data and simulation for
the lepton trigger, identification and reconstruction efficiencies, and for the efficiency and misidentification rate
of the algorithm used to identify jets containing $b$-hadrons ($b$-tagging).

\section{Trigger and data collection}
\label{sec:trigger}
The data used in this paper
were collected in 2012, during which the instantaneous
luminosity of the LHC reached $7.7 \times 10^{33}$ cm$^{-2}\rm{s}^{-1}$.
The average number of expected interactions per bunch crossing ranged from approximately 6 to 40, with a mean of 21.
After applying data-quality requirements related to the beam and
detector conditions, the total
integrated luminosity is 20.1~\ifb~in the soft-lepton channel and 20.3~\ifb~in the hard-lepton channel; the
integrated luminosities differ as these channels use different trigger requirements. 
The uncertainty on the integrated luminosity is $\pm$2.8\%.
It is derived, following the same methodology as that detailed in ref.~\cite{2011lumi}, from a preliminary
calibration of the luminosity scale derived from beam-separation scans performed in November 2012.

In the hard single-electron channel the L1 decision is based on electron
requirements only, while electron and \met~requirements are used at the HLT. The trigger thresholds on HLT objects are 24~\GeV~for
the electron and 35~\GeV~for~\met.  The~\met~trigger 
is fully
efficient for \met$>80$~\GeV. The electron trigger selects events containing one or more electron candidates, based on the presence
of an energy cluster in the electromagnetic calorimeter, with a shower shape consistent with that of an electron, and has no explicit electron isolation requirement except a loose one at L1.
For electrons with $\pT>25$ \GeV, the trigger efficiency increases from 70\% to close to 100\% as the electron \pT~increases from 24 to 30 \GeV.

In the hard single-muon channel the L1 decision is based on muon and jet requirements only, while 
the HLT also includes requirements on \met. The trigger thresholds on HLT objects are at 24~\GeV~for the muon, 65~\GeV~for the jet and 40~\GeV~for~\met.
The muon trigger selects events containing one or more muon candidates based on the hit patterns in the MS and ID, and has no muon isolation requirement. 
The combined trigger reaches its maximal efficiency of approximately $70\%~(90\%)$ for a muon in the barrel (end-cap) for muons satisfying \pT$>$25~\GeV, \met$>100$~\GeV~and
fully calibrated jets with $\pt>80$~\GeV.

In the hard two-lepton channel, a combination of single-lepton and dilepton triggers is used with different \pT~requirements on the electron(s) and muon(s). 
The maximal trigger efficiency is reached when requiring the leading lepton to have $\pT>14$ \GeV~in $ee$ and $\mu\mu$ events or $\pT^{e}(\pT^{\mu})>10(18)$ \GeV~ in $e\mu$ events.
If both leptons are in the barrel (end-cap), these plateau efficiencies are approximately 96\%, 88\% and 80\% (91\%, 92\% and 82\%) for $ee$, $e\mu$ and $\mu\mu$ events, 
respectively.

Since the thresholds in the single-lepton and dilepton triggers are too high to be suitable for the soft-lepton event selections,
this channel relies on a \met$>80$~\GeV~trigger  
which is fully efficient for events with a jet with \pt$>$80~\GeV~and \met$>150$~\GeV.

\section{Object reconstruction}
\label{objectreco}

In this section, the final-state object reconstruction and selection requirements are described. The preselection described below identifies candidate objects.
Some objects are also required to pass a tighter selection before they are used in the event selection.
The event selection criteria and the various signal regions are described in detail in section \ref{sec:EventSelection}.

\subsection{Object preselection}

The primary vertex of the event \cite{PV} is required to be consistent with the beam-spot envelope. When more than one such vertex is found, the vertex with
the largest summed $|\pt|^{2}$ of the associated tracks is chosen.

Jets are reconstructed from three-dimensional calorimeter energy clusters using the anti-$k_{\rm{t}}$ algorithm
\cite{Cacciari:2008gp, Cacciari:2005hq} with a radius parameter $R = $ 0.4. Jets
arising from detector noise, cosmic rays or other non-collision sources are rejected, as described in ref.~\cite{JES}.
To take into account the differences in calorimeter response between electrons/photons and hadrons, each cluster is classified,
prior to the jet reconstruction, as coming from an electromagnetic or hadronic shower
on the basis of its shape \cite{CSCbook}. The jet energy is then
corrected at cluster level by weighting electromagnetic and hadronic energy deposits with correction factors derived from Monte Carlo simulation.
A correction is applied to subtract the expected contamination from pile-up: it is calculated as the product of the jet area in the $(\eta, \phi)$ space and the average energy density of the event \cite{jetPU}.
A further calibration, relating the response of the calorimeter to true jet energy \cite{JES,JES2}, is then applied, with a residual correction to account for differences between the data in situ measurements and the Monte Carlo simulation. 
Once calibrated, the ``preselected'' jets are required to have \pt~$>$ 20~\GeV~and $|\eta|<2.5$. 

Electrons are reconstructed from clusters in the electromagnetic
calorimeter matched to tracks in the ID \cite{electronref}.  
The ``preselected'' electrons are required to pass a variant of the ``medium'' selection 
of ref.~\cite{electronref}, which was modified in 2012 to reduce the impact 
of pile-up. These electrons must have $|\eta|<2.47$ and \pt~$>$ 7 (10) GeV in the soft(hard)-lepton channel.
As each electron can also be reconstructed as a jet, electrons with $\Delta R(e,\rm{jet})<0.2$ are kept, where $\Delta R = \sqrt{(\Delta \eta)^2 + (\Delta \phi)^2}$, 
and the jet is discarded in order to resolve the ambiguity; for $0.2<\Delta R(e,\rm{jet})<0.4$, the electron is discarded and the jet is kept; 
for $\Delta R(e,\rm{jet})>0.4$ both the electron and the jet are kept.
The electrons are also required to be
well separated from the preselected muons described below, with $\Delta R(e,\mu)>0.01$. 
If two preselected electrons are found to have an angular separation  
$\Delta R(e,e)<0.05$, only the higher-\pT~electron is kept. 
Finally, any event containing a preselected electron
in the transition region between the barrel and end-cap electromagnetic calorimeters, $1.37 < |\eta| < 1.52$, is rejected. 

Muons are identified either as a combined track in the MS
and ID systems, or as an ID track matched to a  MS segment
\cite{muonrefnew}.  Requirements on the
quality of the ID track are identical to those in ref.~\cite{ATLAS_1lep}.
``Preselected'' muons in the soft(hard)-lepton channel are required to have \pt~$>$ 6~(10)~\GeV, 
$|\eta| < 2.40$ and $\Delta R(\mu,\rm{jet})>0.4$ with respect to the closest preselected jet.

The missing transverse momentum is computed from the transverse momenta of identified electrons, photons, jets and muons, 
and from all calorimeter clusters within $|\eta|<4.5$ not associated with such objects \cite{Aad:2012re}. 

\subsection{Signal object selection}

For the final selection of events used to define the various signal regions, some objects are required to pass more stringent
 requirements, which are described below.

``Signal'' jets have a higher threshold than preselected jets, with $\pT>25$~\GeV. 
Signal jets with $|\eta|<2.4$ are further required to be associated with the hard-scattering
process by demanding that at least 25\% of the scalar sum of the \pT~of all tracks associated
with the jet comes from tracks associated with the primary vertex in the event. This jet vertex fraction requirement
is applied in order to remove jets which come from pile-up~\cite{ATLAS-CONF-2013-083}; it is not applied
to jets with \pT~greater than 50 \GeV~nor to the $b$-tagged
jets (see below), since the probability of a pile-up jet satisfying either of these requirements is negligible. 

Signal jets containing $b$-hadrons are identified using the neural-network-based algorithm MV1 described in ref.~\cite{MV1}, which uses 
information about track impact parameters and reconstructed secondary vertices.
The presence of $b$-jets is vetoed in the hard dilepton signal regions and in some of the soft-lepton signal regions in order to reduce the \ttbar~background. 
In the single-lepton channels, there is no requirement on $b$-jets in the event selection, 
but they are used in the background estimation, as explained in section \ref{bkgestimate}.
The tightness of the selection criteria used in the $b$-tagging is optimised for each channel.
In all signal regions except for the soft dimuon signal region, the chosen criteria give an inclusive
$b$-tagging efficiency of 60\% in a simulated sample of \ttbar~ events; in the soft dimuon signal region they are chosen to give an inclusive efficiency of 80\%.  
For a $b$-jet efficiency of 60\% (80\%), the algorithm provides a rejection factor of approximately 585 (25) for light-quark 
and gluon jets, and of approximately 8 (3) for charm jets \cite{ATLAS-CONF-2012-043}.

The ``signal'' electrons are required to be isolated, and the isolation requirement depends on the electron transverse momentum.
For $\pT<25$~\GeV~($\pT\ge25$ \GeV), the scalar sum of the \pt~of tracks within a cone of size $\Delta R = 0.3$ (0.2) around the electron,  
excluding the electron itself, is required to be less than 16\% (10\%) of the electron \pt.
For $\pT<25$~\GeV, the distance $|z_{0}\sin{\theta}|$ must be $\le0.4$ mm, where $z_{0}$ is the longitudinal impact parameter with respect 
to the primary vertex. 
For $\pT\ge25$ \GeV, $|z_{0}|$ is required to be $\le2$ mm.
Finally, for electrons with $\pT<25$~\GeV, the significance of the distance of closest approach of the electron to the 
primary vertex in the transverse plane must be $|d_0/\sigma_{d_0}|<5$,  
while for electrons with $\pT\ge 25$~\GeV, the distance of closest approach itself must be $|d_0|\le1$ mm. 

Isolation is also required in the ``signal'' muon definition.
For $\pT<25$~\GeV,  the scalar sum of the \pt~of tracks within a cone of size $\Delta R = 0.3$ around the muon candidate, excluding the muon itself, is 
required to be less than 12\% of the muon \pt, while for $\pT\ge25$ \GeV, the same sum within a cone of size $\Delta R = 0.2$ is required to be less than 1.8 \GeV.
Muons with $\pT<25$~\GeV~are required to have $|z_{0}\sin{\theta}|$ less than 1 mm and $|d_0/\sigma_{d_0}|$ less than 3.

With the lepton selection described above, the combined isolation and identification efficiency measured in simulated $t\bar{t}$ events for electrons (muons) is 56\% (72\%) at $\pT=20$~\GeV~and 84\% (82\%) at $\pT=100$~\GeV.

\section{Event selection}\label{sec:EventSelection}

Events selected by the triggers are required to have a primary vertex with at least five associated tracks with $\pt>400$ \MeV. 
An event is rejected if it contains any preselected jet which fails to satisfy the quality 
criteria which are designed to suppress non-collision backgrounds and detector noise \cite{jetcleaning1,jetcleaning2}, or any
preselected muon with $|z_0|>1.0$ mm~and $|d_0|>0.2$~mm in order to remove cosmic-ray muons. These selection criteria remove $\mathcal{O}(2\%)$
of data events.

This analysis is based on a number of signal regions (SR), each designed to
maximise the sensitivity to different SUSY topologies
in terms of the chosen discriminating variables. As described in
detail in section~\ref{bkgestimate}, a number of control regions (CR) are constructed to
constrain the dominant backgrounds. These control regions are designed to
have a high purity, a small statistical uncertainty in terms of the background process
of interest and to contain only a small fraction of the potential SUSY signal. 
Because of these requirements, the CRs are not necessarily
close to the SRs in terms of the main discriminating variables. As described
in Section~\ref{sec:bkgfit}, validation regions (VR), closer to the SRs than the CRs, are
used to verify the compatibility between data and MC. Figures \ref{fig:softSR}--\ref{fig:razor_regions}
illustrate these concepts, respectively for the soft-lepton, hard
single-lepton and hard dilepton analyses.

\subsection{Signal regions}\label{sec:sr}
 The selection criteria used to define the various signal regions in this paper are 
 summarised in table \ref{tab:softlepSR} for the soft-lepton signal regions, in table \ref{tab:hardlepSR} for the hard single-lepton signal regions and in 
 table \ref{tab:hard2lSR} for the hard dilepton signal regions. 

The soft and hard single-lepton signal regions are designed with lower jet multiplicities to cover squark pair production and with higher jet multiplicities to cover 
gluino pair production. 
The soft single-lepton channel focuses on 
models with a compressed mass spectrum, with the 3-jet inclusive selection being defined to make the analysis sensitive to squark pair production in the case where there is a large mass gap between the squark and the LSP. 
The soft dimuon channel is optimised for mUED searches. The hard dilepton channel targets gluino and first- 
and second-generation squark production, as well as mUED searches; it is not designed to search for signal events in which a real $Z$ boson is present. The correspondence between the analysis channels and the various models probed is summarised in table \ref{tab:SRvsModels}.

\begin{table}[hbt]
\begin{center}
\small
\begin{tabular}{|l|c|c|c|c|}
\hline
 & \multicolumn{3}{c|}{\bf Single-bin (binned) soft single-lepton} & {\bf Soft dimuon} \\\hline
 & {\bf 3-jet} & {\bf 5-jet} & {\bf 3-jet inclusive} & {\bf 2-jet} \\\hline\hline
$N_{\ell}$ & \multicolumn{3}{c|}{1 electron or muon} & 2 muons\\\hline
\ptl [\GeV]   & \multicolumn{3}{c|}{ [7,25] for electron, [6,25] for muon} & [6,25]\\\hline
Lepton veto & \multicolumn{4}{c|}{ No additional electron or muon with \pt$>$ 7 \GeV~or 6 \GeV, respectively}  \\\hline
$m_{\mu\mu}$ [\GeV] & $-$ & $-$ & $-$ & [15,60] \\\hline\hline
$N_{\mathrm{jet}}$  & [3,4] & $\geq$ 5 & $\geq$ 3 & $\geq 2$ \\\hline
\pt$^{\mathrm{jet}}$[\GeV] & $>$ 180, 25, 25 & $>$ 180, 25, 25, 25, 25 & $>$ 130, 100, 25 & $>$ 80, 25 \\\hline
$N_{b\mathrm{-tag}}$ &  $-$ &$-$ & 0 & 0\\\hline
\hline
\met\ [\GeV]  & $>$400 & $>$300 & \multicolumn{2}{c|}{$>$ 180} \\\hline
\mt\ [\GeV]   & \multicolumn{2}{c|}{ $>$ 100 } & $>$120 & $>40$ \\\hline
\met/$m_{\mathrm{eff}}^{\mathrm{incl}}$ & \multicolumn{2}{|c|}{$> 0.3$ $(0.1)$} & $>0.1$  & $> 0.3$ \\\hline
$\Delta R_{\mathrm{min}}(\mathrm{jet},\ell)$ & $>1.0$ & $-$ & $-$ & $>1.0$ ($\mathrm{2^{nd}}$ muon)\\\hline
Binned variable & \multicolumn{3}{c|}{(\met/$m_{\mathrm{eff}}^{\mathrm{incl}}$ in 4 bins)} & $-$ \\\hline
Bin width & \multicolumn{3}{c|}{(0.1, $4^{\mathrm{th}}$ is inclusive)} & $-$ \\\hline
\end{tabular}
\caption{Overview of the selection criteria for the soft single-lepton and dimuon signal regions.
	For each jet multiplicity in the single-lepton channel, two sets of requirements are defined, corresponding to single-bin and binned signal regions (see the text at the end of Section \ref{sec:sr}).
	The requirements of the binned signal region are shown in parentheses when they differ from those of the single-bin signal region. The variables $\Delta R_{\mathrm{min}}(\mathrm{jet},\ell)$, \mt~and $m_{\mathrm{eff}}^{\mathrm{incl}}$ are defined in equations \ref{eq:rmin}, \ref{eq:mt} and \ref{eq:meffinc}, respectively.
}
\label{tab:softlepSR}
\end{center}
\end{table}

\begin{table}[hbt]
\begin{center}
\small
\begin{tabular}{|l|c|c|c|}
\hline
 & \multicolumn{3}{c|}{\bf Single-bin (binned) hard single-lepton} \\\hline
 & {\bf 3-jet} & {\bf 5-jet} & {\bf 6-jet} \\\hline\hline
$N_{\ell}$ & \multicolumn{3}{c|}{1 electron or muon} \\\hline
\ptl [\GeV]   & \multicolumn{3}{c|}{ $>$~25 }\\\hline
Lepton veto & \multicolumn{3}{c|}{ \pt$^{\mathrm{2^{nd} lepton}}<$ 10 \GeV} \\\hline
$N_{\mathrm{jet}}$  & $\geq$ 3 & $\geq$ 5 & $\geq$ 6 \\\hline
\pt$^{\mathrm{jet}}$[\GeV] & $>$ 80, 80, 30 & $>$ 80, 50, 40, 40, 40 & $>$ 80, 50, 40, 40, 40, 40\\\hline
Jet veto & (\pt$^{\mathrm{5^{th} jet}}< 40$ \GeV) & (\pt$^{\mathrm{6^{th} jet}}< 40$ \GeV) & $-$ \\\hline
\met\ [\GeV]  & $>$500 (300) & $>$300&  $>$350 (250) \\\hline
\mt\ [\GeV]   & $>$ 150  & $>$ 200 (150) & $>$ 150 \\\hline
\met/$m_{\mathrm{eff}}^{\mathrm{excl}}$ & $>$ 0.3 & $-$ & $-$  \\\hline
$m_{\mathrm{eff}}^{\mathrm{incl}}$ [\GeV] & \multicolumn{2}{c|}{ $>$ 1400 (800)} & $>$ 600 \\\hline
Binned variable & \multicolumn{2}{c|}{($m_{\mathrm{eff}}^{\mathrm{incl}}$ in 4 bins)} & (\met~in 3 bins) \\\hline  
Bin width & \multicolumn{2}{c|}{(200 \GeV, $4^{\mathrm{th}}$ is inclusive)} & (100 \GeV, $3^{\mathrm{rd}}$ is inclusive) \\\hline     
 
\end{tabular}
\caption{Overview of the selection criteria for the hard single-lepton signal regions.
	For each jet multiplicity, two sets of requirements are defined, corresponding to single-bin and binned signal regions (see the text at the end of Section \ref{sec:sr}).
	The requirements of the binned signal region are shown in parentheses when they differ from those of the single-bin signal region. The variables \mt~and $m_{\mathrm{eff}}^{\mathrm{incl}}$ are defined in equations \ref{eq:mt} and \ref{eq:meffinc}, respectively, while $m_{\mathrm{eff}}^{\mathrm{excl}}$ is defined in the text.
}
\label{tab:hardlepSR}
\end{center}
\end{table}

\begin{table}[hbt]
\begin{center}
\small
\begin{tabular}{|l|c|c|c|c|c|}
\hline
 & \multicolumn{4}{c|}{\bf Single-bin (binned) hard dilepton} \\\hline
 & \multicolumn{2}{c|}{{\bf Low-multiplicity ($\le2$-jet) }} & \multicolumn{2}{c|}{{\bf 3-jet}}  \\\hline\hline
 & {\bf $ee/\mu\mu$} & {\bf $e\mu$} & {\bf $ee/\mu\mu$} & {\bf $e\mu$}\\\hline\hline
$N_{\ell}$ & \multicolumn{4}{c|}{$2$, $2$ of opposite sign or $\ge 2$} \\\hline
\ptl [\GeV]   & \multicolumn{4}{c|}{ $>$14,10 } \\\hline
$N_{\ell\ell}$ with 81$<m_{\ell\ell}<$101 \GeV & 0 & $-$ & 0 & $-$ \\\hline
$N_{\mathrm{jet}}$  & \multicolumn{2}{c|}{$\leq$ 2} & \multicolumn{2}{c|}{$\geq$ 3} \\\hline
\pt$^{\mathrm{jet}}$[\GeV] & \multicolumn{2}{c|}{$>$ 50,50} & \multicolumn{2}{c|}{$>$ 50, 50, 50} \\\hline
$N_{b\mathrm{-tag}}$ &\multicolumn{4}{c|}{0} \\\hline
\hline
$R$   & \multicolumn{2}{c|}{$>$0.5} & \multicolumn{2}{c|}{$>$0.35} \\\hline
$M_{R}^{'}$ [\GeV] & \multicolumn{2}{c|}{$>600$ ($>400$ in 8 bins)} & \multicolumn{2}{c|}{$>800$ ($>800$ in 5 bins)} \\\hline
$M_{R}^{'}$ bin width [\GeV] & \multicolumn{4}{c|}{(100, the last is inclusive )} \\\hline
\end{tabular}
\caption{Overview of the selection criteria for the hard dilepton signal regions. The requirements on the number and charge of the leptons 
depend on the model probed (see the text). 
For each jet multiplicity, two sets of requirements are defined, corresponding to single-bin and binned signal regions (see the text at the end of Section \ref{sec:sr}).
The requirements of the binned signal region are shown in parentheses when they differ from those of the single-bin signal region. The variables $M_{R}^{'}$ and $R$ are defined in equations \ref{eq:mr} and \ref{eq:r}, respectively.  
}
\label{tab:hard2lSR}
\end{center}
\end{table}

\begin{figure}[htb]
\centering
\includegraphics[width=0.49\textwidth]{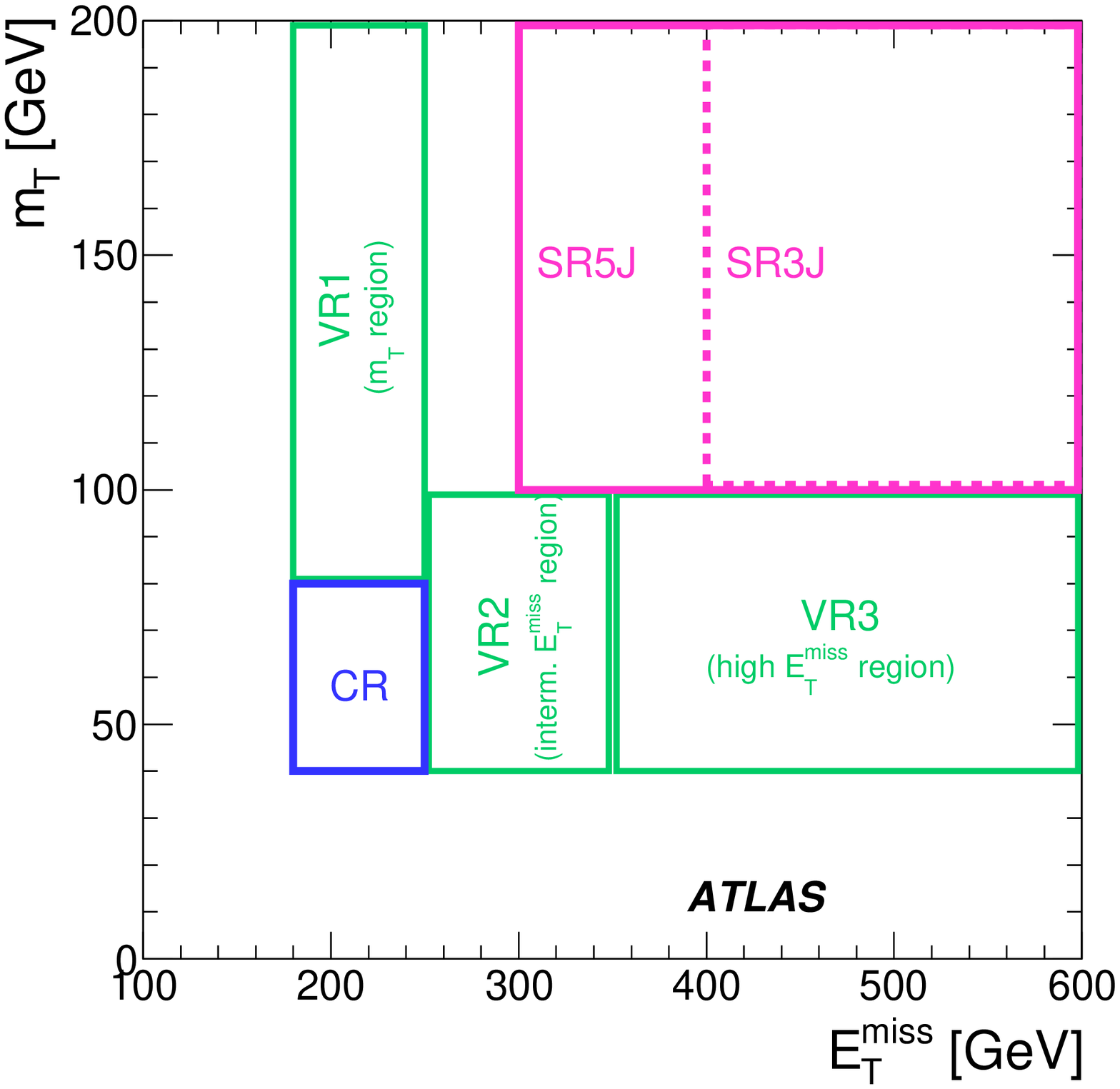}
\includegraphics[width=0.49\textwidth]{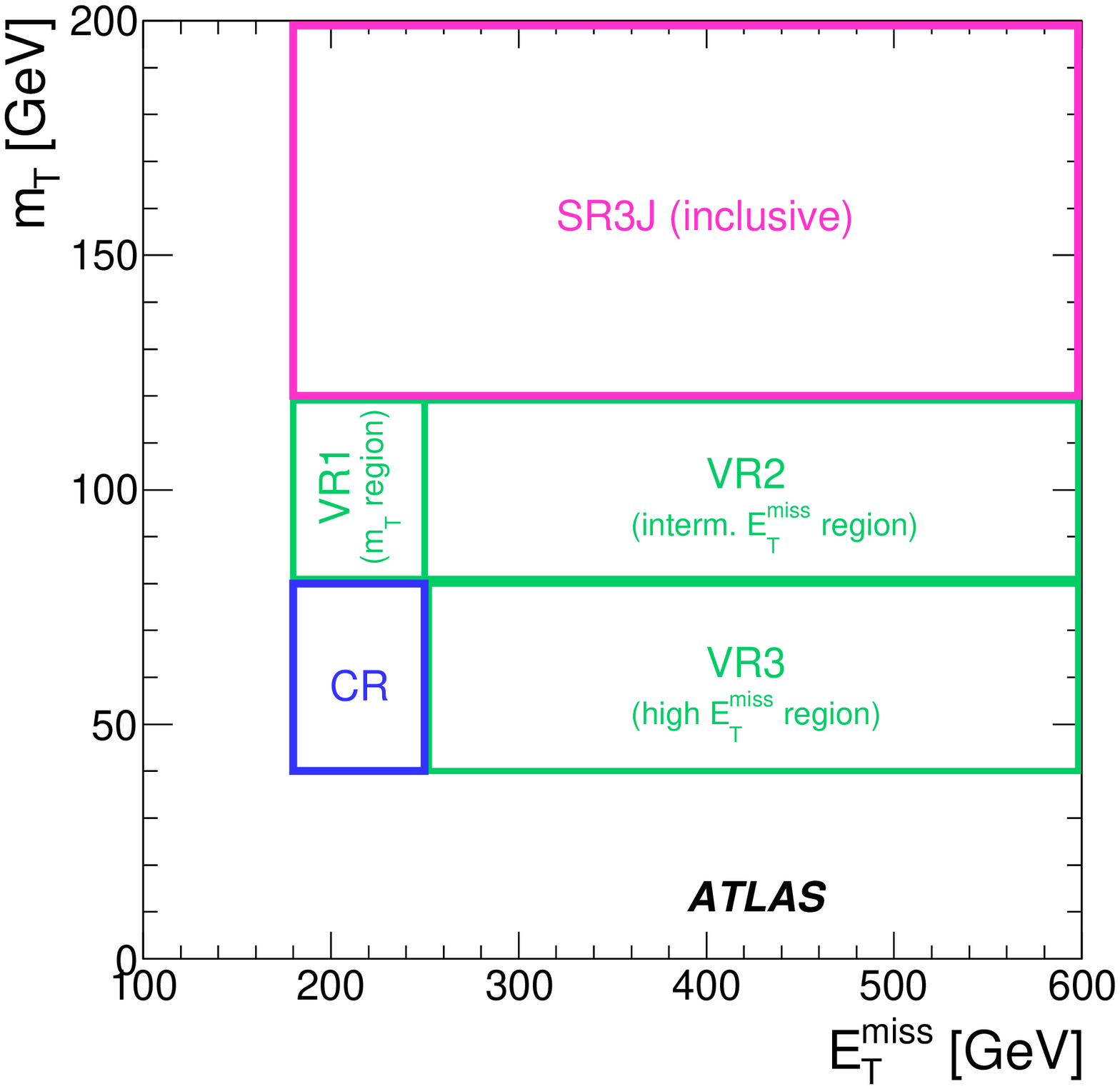}
\includegraphics[width=0.49\textwidth]{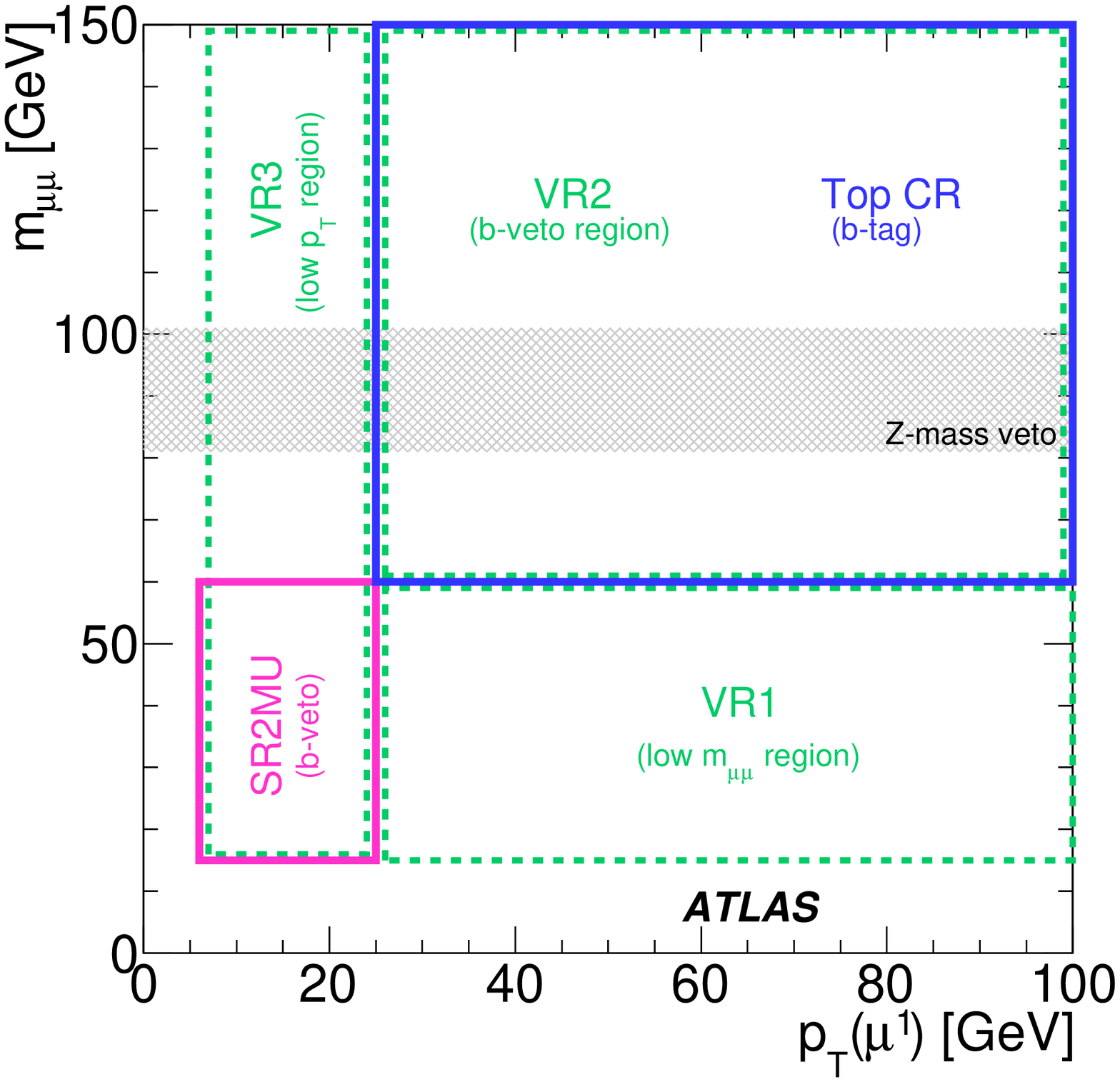}
\caption{
Graphical illustration of the soft lepton signal regions (SR) used in this paper. The soft single-lepton signal regions are shown in the plane of transverse mass
\mt~(see equation \ref{eq:mt}) versus missing transverse momentum \met: the 3- and 5-jet regions are depicted
in the upper left plot while the 3-jet inclusive region is shown in the upper right plot; the soft dimuon signal region is shown in the bottom plot in the plane of the dimuon mass,$m_{\mu\mu}$, versus the \pT~of the
leading muon, $p_{\text{T}}(\mu^1)$. The control regions (CR) and validation regions (VR) described in sections \ref{bkgestimate} and \ref{sec:bkgfit}, respectively, are also shown.
} \label{fig:softSR}
\end{figure}
\begin{figure}[htb]
\centering
\includegraphics[width=0.49\textwidth]{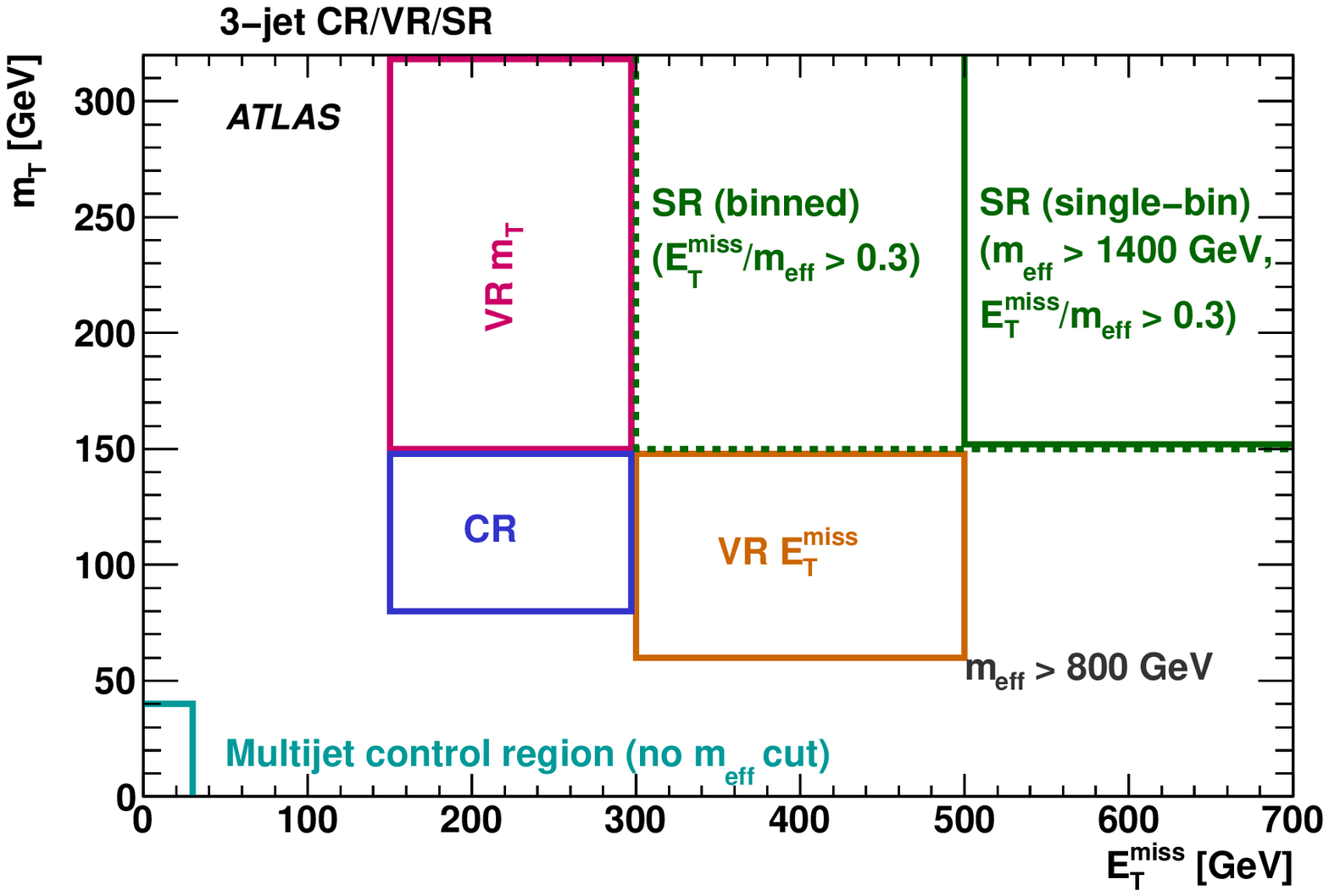}
\includegraphics[width=0.49\textwidth]{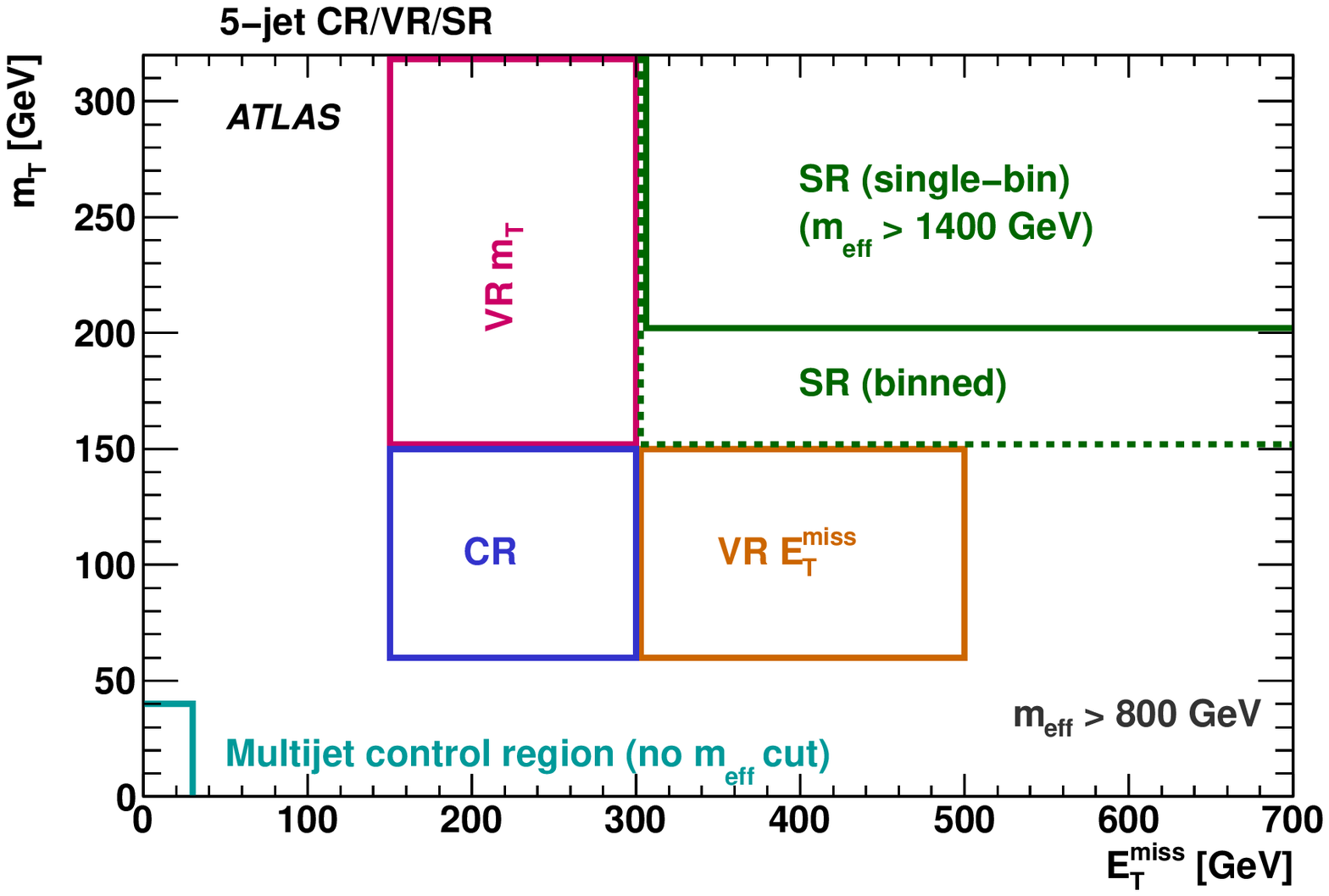}
\includegraphics[width=0.49\textwidth]{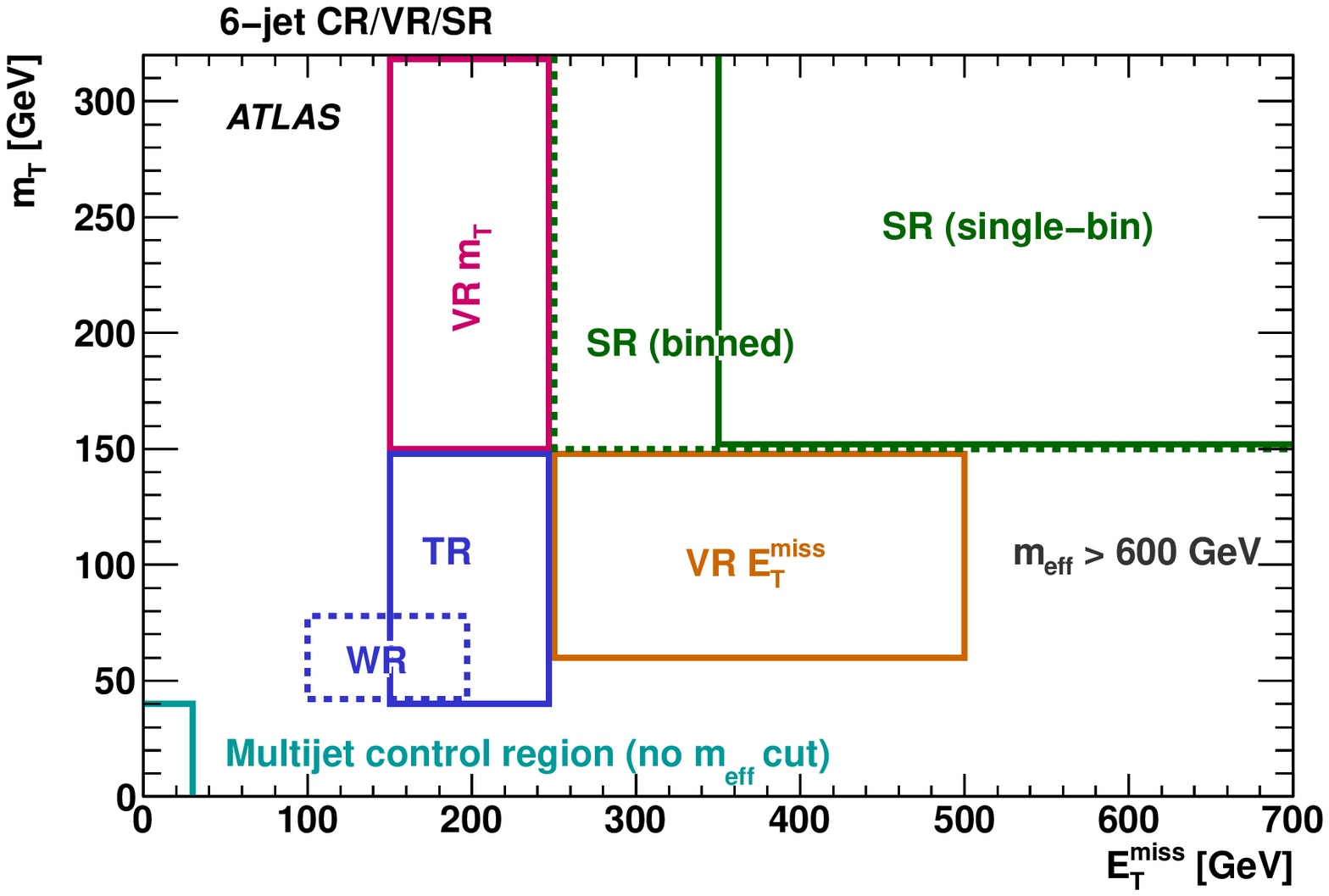}
\caption{
Graphical illustration of the hard single-lepton 3-jet (top left), 5-jet (top right) and 6-jet (bottom) signal regions (SR) used in this paper, shown in the plane of transverse mass
\mt~(see equation \ref{eq:mt}) versus missing transverse momentum \met. 
The control regions (CR) and validation regions (VR) described in sections \ref{bkgestimate} and \ref{sec:bkgfit}, respectively, are also shown.
} \label{fig:hardSR}
\end{figure}
\begin{figure}[htb]
\centering
   \begin{center}
     \includegraphics[width=0.47\textwidth]{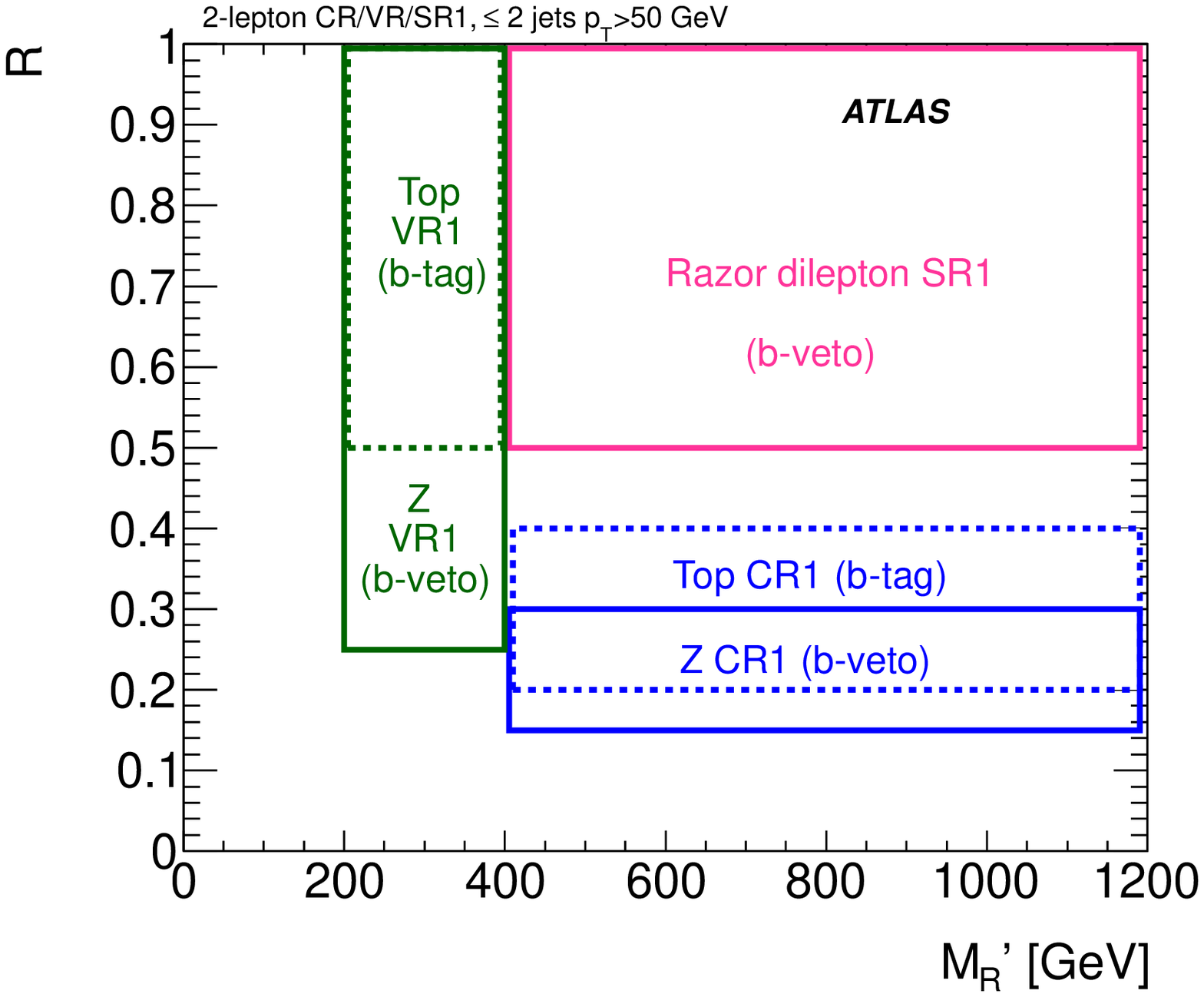}
     \includegraphics[width=0.47\textwidth]{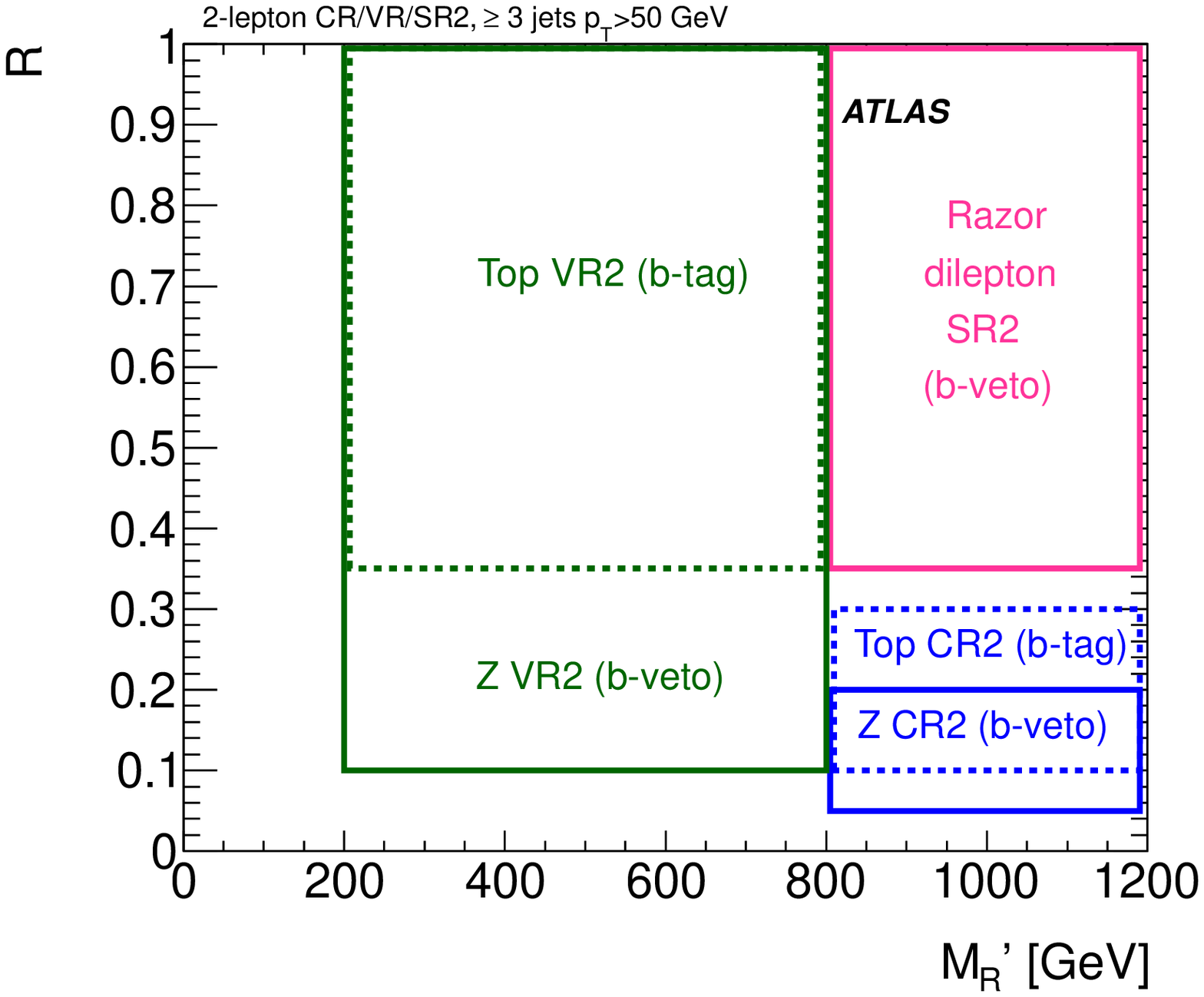}
   \end{center}
   \caption{
     \label{fig:razor_regions}
Graphical illustration of the hard dilepton signal regions (SR) used in this paper. The low-multiplicity (left) and 3-jet (right)
hard dilepton signal regions are shown in the plane of $R$-frame mass $M_R'$ versus razor variable $R$ (see equations \ref{eq:mt} and \ref{eq:r}). 
The control regions (CR) and validation regions (VR) described in sections \ref{bkgestimate} and \ref{sec:bkgfit}, respectively, are also shown.
}
\end{figure}

\begin{table}[hbt]
\begin{center}
\small
 \begin{tabular}{|l|c|c|c|c|}
    \hline
      {\bf Model}  & \multicolumn{2}{c|}{\bf Soft} & \multicolumn{2}{c|}{\bf Hard} \\
                                                       & {\bf single-lepton} & {\bf dimuon} & {\bf single-lepton} & {\bf dilepton}  \\\hline\hline
      mSUGRA/CMSSM  & & & \checkmark & \\\hline
      bRPV mSUGRA/CMSSM  & & & \checkmark & \\\hline
      nGM  & & & \checkmark & \\\hline
      NUHMG  & & & \checkmark & \\\hline
      mUED & & \checkmark &  & \checkmark \\\hline
      $\gluino\gluino$ production, $\gluino\rightarrow tc\ninoone$ &   & & \checkmark & \\\hline
      $\gluino\gluino$ production, $\gluino\rightarrow \ttbar\ninoone$ &   & & \checkmark & \\\hline
      $\gluino\gluino$ production, $\gluino\rightarrow qqW\ninoone$  & \checkmark & &  \checkmark & \\\hline
      $\squark\squark$ production, $\squark\rightarrow qW\ninoone$  & \checkmark & &  \checkmark & \\\hline
      $\gluino\gluino$ production, $\gluino\rightarrow qq(\ell\ell/\ell\nu/\nu\nu)\ninoone$  & & &  \checkmark & \checkmark  \\\hline
      $\squark\squark$ production, $\squark\rightarrow q(\ell\ell/\ell\nu/\nu\nu)\ninoone$ & & & & \checkmark \\\hline
      $\gluino\gluino$ production, $\gluino\rightarrow qqWZ\ninoone$ &  & & \checkmark & \\\hline
 \end{tabular}
 \caption{Analysis channels used to probe each of the models described in section \ref{simulsignal}.
    }
 \label{tab:SRvsModels}
\end{center}
\end{table}

The following variables, derived from the kinematic properties of the
objects,  are used in the event selection.

The minimum angular separation between the signal lepton $\ell$ and all preselected jets,
\begin{equation}
     \Delta R_{\mathrm{min}}(\mathrm{jet},\ell)=\min{(\Delta R(\mathrm{jet}_{1},\ell),\Delta R(\mathrm{jet}_{2},\ell),...,\Delta R(\mathrm{jet}_{n},\ell))},
          \label{eq:rmin}
\end{equation}
is used to reduce the background coming from misidentified or non-prompt leptons in the soft-lepton signal region with three jets and in the soft dimuon signal region. 
In the latter case, the subleading signal muon 
is used to compute $\Delta R_{\mathrm{min}}$. 
As the expected signal jet multiplicity grows, the $\Delta R_{\mathrm{min}}$ requirement starts to impair the signal acceptance; this requirement is hence not 
applied in the soft-lepton 5-jet and 3-jet inclusive signal regions (see table~\ref{tab:softlepSR}). 

The dilepton mass $m_{\ell\ell}$ for leptons of the same flavour and opposite charge is required to be outside the $Z$ boson mass window
in the soft and hard dileptonic channels in order to reject background events in which a real $Z$ boson decays to leptons.

The transverse mass ($m_{\rm{T}}$) of the lepton ($\ell$)
and ${\boldsymbol p}_{\mathrm{T}}^\mathrm{miss}$ is defined as
\begin{equation}
    m_{\mathrm{T}} =
\sqrt{2 p_{\mathrm{T}}^{\ell} \met
  (1-\cos[\Delta\phi(\vec{\ell},{\boldsymbol p}_{\mathrm{T}}^\mathrm{miss})])},
       \label{eq:mt}
\end{equation}
and is used in all signal regions to reject events containing a $W\rightarrow\ell\nu$ decay, except in the hard dilepton
signal regions where this background is expected to be small.
In the soft dimuon channel, the transverse mass is defined using the subleading muon.

The inclusive effective mass ($m_{\rm{eff}}^{\rm{inc}}$) is the scalar sum of
the \pt~of the lepton(s), the jets and \met:
\begin{equation}
    m_{\mathrm{eff}}^{\mathrm{inc}} = 
    \sum_{i=1}^{{N}_{\ell}}p_{\mathrm{T},i}^{\ell} +  \sum_{j=1}^{{N}_\mathrm{jet}}
    p_{\mathrm{T},j} + \met
         \label{eq:meffinc}
\end{equation}
where the index $i$ identifies all the signal leptons and the index $j$ all the signal jets in the event. 
The inclusive effective mass is correlated with the overall mass scale of the 
hard scattering and provides good discrimination against SM backgrounds, 
without being too sensitive to the details of the SUSY decay cascade. It is used in the hard single-lepton channel.

The ratio \met/$m_{\rm{eff}}^{\mathrm{inc}}$ is used in the soft-lepton signal regions; it reflects the
change in  the \met~resolution as a function of the calorimeter activity in the event. In the hard single-lepton channel, 
a similar ratio is computed, \met/$m_{\rm{eff}}^{\rm{excl}}$, where 
the exclusive effective mass, $m_{\rm{eff}}^{\rm{excl}}$, is defined in a similar way to $m_{\rm{eff}}^{\rm{inc}}$, with
the exception that only the three leading signal jets are considered.
This variable is used to remove events with large \met~coming from a poorly reconstructed jet.

Razor variables~\cite{razor} are used in the hard dilepton signal region. They are a set of kinematic variables that exploit the symmetry in the visible portion of sparticle 
decays when sparticles are produced in pairs. The final-state jets and leptons are grouped into two ``mega-jets'', where 
all visible objects from one side of the di-sparticle decay are collected together to create 
a single four-vector, representing the decay products of a single sparticle.
The mega-jet construction involves iterating over all possible combinations of the four-vectors of the visible reconstructed objects,
with the favoured combination being that which minimises the sum of the squared masses of the mega-jet four-vectors.
Using this mega-jet configuration, with some simplifying assumptions (e.g. symmetric sparticle production), 
the rest frame of the sparticles (the so-called ``$R$-frame'' described in ref.~\cite{razor}) can be reconstructed, 
and a characteristic mass $M_R'$ can be defined in this frame:

\begin{equation}
	M_{R}' = \sqrt{ ( j_{1,E} + j_{2,E} )^2 - ( j_{1,{\rm L}} + j_{2,{\rm L}} )^2 } ,
	     \label{eq:mr}
\end{equation}

\noindent where $j_{i,{\rm L}}$ denotes the longitudinal momentum, and $j_{i,E}$ the energy in the $R$-frame, of the mega-jet $i$. 
The transverse information of the event is contained in another variable, $M^R_{\rm T}$. 
In the di-sparticle decay there are two mega-jets, each with associated \met\ from the escaping LSPs. 
Assigning half of the missing transverse momentum per event to each of the LSPs, $M^R_{\rm T}$ is defined as

\begin{equation}
	M_{\rm T}^R = \sqrt{\frac{ |{\boldsymbol p}_{\mathrm{T}}^\mathrm{miss}|( |\vec{j}_{1,{\rm T}}| + |\vec{j}_{2,{\rm T}}| ) -  {\boldsymbol p}_{\mathrm{T}}^\mathrm{miss} \cdot ( \vec{j}_{1,{\rm T}} + \vec{j}_{2,{\rm T}} ) }{2}} ,
	     \label{eq:mtr}
\end{equation}

\noindent where ${j}_{i,{\rm T}}$ denotes the transverse momentum of the mega-jet $i$. 

Finally, the razor variable is defined as:
\begin{equation}
	R = \frac{M_{\rm T}^R}{M_{R}'} .
	     \label{eq:r}
\end{equation}
For SM processes, $R$ tends to have a low value, while it is approximately uniformly distributed between zero and one for SUSY-like signal events. 
Thus $R$ can be used as a discriminant between signal and background.
A selection using $R$ is made to reduce background processes before a search for new physics phenomena is performed using the distribution of the variable $M_R'$. 

In order to have signal regions which are orthogonal to each other in lepton multiplicity, a veto is placed on the presence of a
second lepton in the hard and soft single-lepton channels. Following this veto, all signal regions are orthogonal except the 
inclusive and exclusive soft single-lepton signal regions and the soft dilepton and hard dilepton signal regions.  A veto on the 
third lepton in the soft dimuon channel is placed to reduce the fake lepton contribution (see section~\ref{bkgestimate}).
For the hard dilepton channel, the requirements on the number of leptons and on their charge depends on the model probed: in the case
of models where only two leptons are expected in the final state, events containing any additional lepton are vetoed. Furthermore,
for models with squark pair production followed by one-step decays, only events with opposite-sign dilepton pairs are selected. 

In all the search channels except the soft dimuon channel, two sets of requirements are optimised for each jet multiplicity: one single-bin signal region
optimised for discovery reach, which is also used to place limits on the visible cross section, and one signal region which is binned
in an appropriate variable in order to exploit the expected shape of the distribution of signal events when placing model-dependent limits.
The binned variables are: \met/$m_{\mathrm{eff}}^{\mathrm{incl}}$ in the soft single-lepton regions (four bins of width 0.1), 
$m_{\mathrm{eff}}^{\mathrm{incl}}$ in the hard single-lepton 3-jet and 5-jet signal regions (four bins of 200 \GeV), \met~in the hard single-lepton 6-jet signal region 
(three bins of 100 \GeV) and $M_R'$ in the hard dilepton signal regions (eight bins of 100 \GeV~in the low-multiplicity signal region and five bins of 100 \GeV~in the 3-jet signal region). In all regions, 
the last bin is inclusive. The binned signal regions can differ from the single-binned signal regions in that some requirements may be relaxed. The binned hard single-lepton
signal regions are made orthogonal in jet multiplicity to one another by placing a jet veto, as can be seen in table~\ref{tab:hardlepSR}, in order to allow their statistical combination and have finer-grained
requirements than in a single combined signal region.

\section{Background estimation}
\label{bkgestimate}
The dominant background in all the analyses presented here is top quark pair production. The $W$+jets and $Z$+jets backgrounds are also 
important in the single-lepton and hard dilepton channels, respectively.
These backgrounds are estimated using control regions
optimised to be enriched in SM events from the background process of interest,while containing only a small contribution from the signal of interest, as described below.
The normalisation of the simulation for these background processes is obtained simultaneously in all control regions for each signal region using the fit described in section \ref{sec:bkgfit}. 
The simulation is thus used only to extrapolate the results to the signal region, 
and is therefore not affected by potentially large theoretical uncertainties on the total expected rates in specific regions of phase space.
The control regions are chosen to be kinematically close to the signal regions in order to minimise the theoretical
uncertainties related to the extrapolation, while containing enough events to avoid compromising the background estimate with 
a large statistical uncertainty.

Events with ``fake'' or non-prompt leptons can also mimic the signal if they have sufficiently large \met.
A jet can be misidentified as a lepton (fake lepton), or a real lepton can arise as a decay product of $b$- or $c$-hadrons in jets
but can still be sufficiently isolated (non-prompt lepton).
Such lepton-like objects are collectively referred to as fake leptons in this paper.

\subsection{Backgrounds from \ttbar~ and $W/Z$+jets}

The control regions used in the soft single-lepton and soft dimuon channels are illustrated in figure~\ref{fig:softSR} 
and summarised precisely in table \ref{tab:softlepCR}. The soft single-lepton control regions 
are built using events with lower \met~and~\mt~values than in the
 signal regions by requiring $180<$\met$<250$ \GeV~and $40<$\mt$<80$ \GeV, and by removing the requirement 
on \met/$m_{\mathrm{eff}}^{\mathrm{incl}}$. The $W$+jets and \ttbar~background components in these control 
regions are separated by a requirement on the number of $b$-tagged 
signal jets. Events in the \ttbar~control region are defined by requiring that at least one signal jet is $b$-tagged; 
otherwise, they are associated with the $W$+jets control region.  

In the soft dimuon analysis, the \ttbar~control region is defined by requiring the leading muon to have \pt~$>$25~\GeV\ instead 
of pt~$<$25~\GeV. The veto on $b$-tagged jets is reversed to require at least one $b$-tagged signal jet among the 
three leading jets and the requirement on \met/$m_{\mathrm{eff}}^{\mathrm{incl}}$ is removed.
The dimuon mass is required to be higher than in the signal region, $m_{\mu\mu}>60$ \GeV, and at least 10 GeV away from the $Z$ boson mass. 

The hard single-lepton control regions are defined by lowering the requirements on \met~and \mt~and by removing the \met/$m_{\mathrm{eff}}^{\mathrm{excl}}$
requirement in the 3-jet region. 
Table \ref{tab:hardlepCR} lists the control region requirements which differ from the signal region selections for the hard single-lepton channel. 
These various control regions are illustrated in figure~\ref{fig:hardSR}.
The different regions are kept orthogonal by vetoing on the presence of a fifth (sixth) jet in the 3 (5)-jet control region.
To increase the number of events, the \pt~requirements on subleading jets are also lowered with respect to the signal regions.
Finally, the $W$+jets and \ttbar~components of these control regions are separated by a requirement on the number of
signal jets which are $b$-tagged, considering the first three leading jets. In order to enhance the $W$+jets contribution 
over the \ttbar~contribution in the 6-jet $W$+jets control region, the \mt~and \met~requirements are lowered in this region with respect to the 
6-jet \ttbar~control region. 

As summarised in table \ref{tab:hard2lepCR} and illustrated in figure~\ref{fig:razor_regions},  
the control regions for the hard dilepton channel are defined at lower values
of the $R$ variable for events in which there are exactly two leptons of opposite sign in order to enhance the 
background processes. The $Z$+jets and \ttbar~components of these control regions are separated 
by a requirement on the number of signal jets which are $b$-tagged. The control regions are binned in the discriminating variable
$M_{R}^{'}$ in order to use the same 
shape information as in the signal regions. 

The \ttbar~control regions in all channels include a small fraction of at most 11\% of $Wt$ events; this background is not
normalised by the fit in the control region, but evaluated directly from simulation, as are other lower-rate background 
processes involving top quarks (see section \ref{otherBG}).

Figures \ref{fig:1softCR}--\ref{fig:2softCR} show the \mt~and $m_{\mu\mu}$~distributions, prior to the upper \mt~and lower $m_{\mu\mu}$ cuts,
in the soft single-lepton and soft dimuon control regions, respectively. Figures \ref{fig:1hardCR1} and \ref{fig:1hardCR2} show the \met~
distribution, prior to the upper \met~cut, in the hard single-lepton control regions. Figure \ref{fig:2hardCR2} shows the $R$ distribution in the hard dilepton control regions. 
All these distributions are shown after the fitting procedure is applied to adjust the MC normalisation, as described in 
section \ref{sec:bkgfit}. 
For illustration, examples of expected signal distributions are also shown in these figures.
The fraction of events in the control regions coming from the background of interest, hereafter called purity, is 
given in the caption of these figures. 
As the normalisation factors are obtained in a simultaneous fit to all control regions for a given 
signal region, the cross-contamination of the control regions with different processes is taken into account and lower purity in some regions
does not degrade significantly the accuracy of the background estimation.
The agreement between the data and the SM background estimate is reasonable within the statistical and systematic uncertainties. 
The systematic uncertainties shown do not include an uncertainty on the cross sections of the backgrounds 
that are normalised using the fitting procedure, but do include the relevant theoretical uncertainties on the 
extrapolation of the background normalisation obtained from each CR to the relevant SR (see section~\ref{sec:systuncert}).
The results of the fit, in particular the signal region predictions, are further discussed in section~\ref{sec:results}.

\begin{table}[hbt]
\begin{center}
\small
\begin{tabular}{|l|c|c|c|c|}
\hline
 & \multicolumn{3}{c|}{\bf Soft single-lepton} & {\bf Soft dimuon} \\\hline
 & {\bf 3-jet} & {\bf 5-jet} & {\bf 3-jet inclusive} & {\bf 2-jet} \\\hline
 & \multicolumn{3}{c|}{\bf $W$+jets / $t\bar{t}$} & $t\bar{t}$ \\\hline\hline
\ptl [\GeV]   & \multicolumn{3}{c|}{ [7,25] (electron) , [6,25] (muon)} & $>$25,6\\\hline
$m_{\mu\mu}$ [\GeV] & \multicolumn{3}{c|}{$-$} & $>60$, $|m_{\mu\mu}-m_Z|>10$  \\\hline\hline
$N_{b\mathrm{-tag}}$ &  \multicolumn{3}{c|}{ 0 / $\ge1$ } & $\ge1$\\\hline
\met\ [\GeV]  &  \multicolumn{3}{c|}{[180,250]} &  $>180$ \\\hline
\mt\ [\GeV]   & \multicolumn{3}{c|}{ [40,80] } & $>40$ \\\hline
$\Delta R_{\mathrm{min}}(\mathrm{jet},\ell)$ & $>1.0$ & $-$ & $-$ & $>1.0$  \\\hline
\end{tabular}
\caption{Overview of the selection criteria for the CR used in the soft single-lepton and soft dimuon channels: only the criteria which differ from the corresponding signal region selections in at least one CR are shown (see figure~\ref{fig:softSR} for an illustration of the above CRs).}
\label{tab:softlepCR}
\end{center}
\end{table}

\begin{table}[hbt]
\begin{center}
\small
\begin{tabular}{|l|c|c|c|}
\hline
 & \multicolumn{3}{c|}{\bf Hard single-lepton} \\\hline
 & {\bf 3-jet} & {\bf 5-jet} & {\bf 6-jet} \\\hline\hline
 & \multicolumn{3}{c|}{\bf $W$+jets / $t\bar{t}$} \\\hline
\pt$^{\mathrm{jet}}$[\GeV] & $>$ 80, 80, 30 & $>$ 80, 50, 30, 30, 30 & $>$ 80, 50, 30, 30, 30, 30\\\hline
Jet veto & \pt$^{\mathrm{5^{th} jet}}< 30$ \GeV & \pt$^{\mathrm{6^{th} jet}}< 30$ \GeV & $-$ \\\hline
$N_{b\mathrm{-tag}}$ & \multicolumn{3}{c|}{0 / $\geq 1$} \\\hline
\met\ [\GeV]  & \multicolumn{2}{c|}{[150,300]} &  [100,200] / [150,250] \\\hline
\mt\ [\GeV]   & [80,150]  & [60,150] &  [40,80] / [40,150] \\\hline
\met/$m_{\mathrm{eff}}^{\mathrm{excl}}$ & $-$ & $-$ & $-$  \\\hline
\end{tabular}
\caption{Overview of the selection criteria for the $W$+jets and \ttbar~CR used in the hard single-lepton channel: only the criteria which differ from the corresponding signal
	region selections in at least one CR are shown (see figure~\ref{fig:hardSR} for an illustration of the above CRs). 
}
\label{tab:hardlepCR}
\end{center}
\end{table}

\begin{table}[hbt]
\begin{center}
\small
\begin{tabular}{|l|c|c|c|c|c|c|}
\hline
 & \multicolumn{6}{c|}{\bf Hard dilepton} \\\hline
 & \multicolumn{3}{c|}{{\bf Low-multiplicity}} & \multicolumn{3}{c|}{{\bf 3-jet}} \\\hline\hline
 & \multicolumn{2}{c|}{{\bf $ee/\mu\mu$}} & {\bf $e\mu$} & \multicolumn{2}{c|}{{\bf $ee/\mu\mu$}} & {\bf $e\mu$} \\\hline
 & $Z$ CR & \ttbar~CR & \ttbar~CR &  $Z$ CR & \ttbar~CR & \ttbar~CR  \\\hline\hline
$N_{\ell}$ & \multicolumn{6}{c|}{$2$ of opposite sign} \\\hline
$N_{b\mathrm{-tag}}$ & 0 &\multicolumn{2}{c|}{1} & 0 &\multicolumn{2}{c|}{1}\\\hline
$R$   & [0.15,0.3] & \multicolumn{2}{c|}{[0.2,0.4]} & [0.05,0.2] & \multicolumn{2}{c|}{[0.1,0.3]} \\\hline
$M_{R}^{'}$ [\GeV] & \multicolumn{3}{c|}{[400,1200]} & \multicolumn{3}{c|}{[800,1600]} \\\hline
$M_{R}^{'}$ bin width [\GeV] & \multicolumn{3}{c|}{100} & \multicolumn{3}{c|}{200} \\\hline

\end{tabular}
\caption{Overview of the selection criteria for the $Z$+jets and \ttbar~CR used in the hard dilepton channel: only the criteria which differ from the corresponding signal
	region selections in at least one CR are shown (see figure~\ref{fig:razor_regions} for an illustration of the above CRs).
}
\label{tab:hard2lepCR}
\end{center}
\end{table}

\begin{figure}[ht]
\centering
\includegraphics[width=0.49\textwidth]{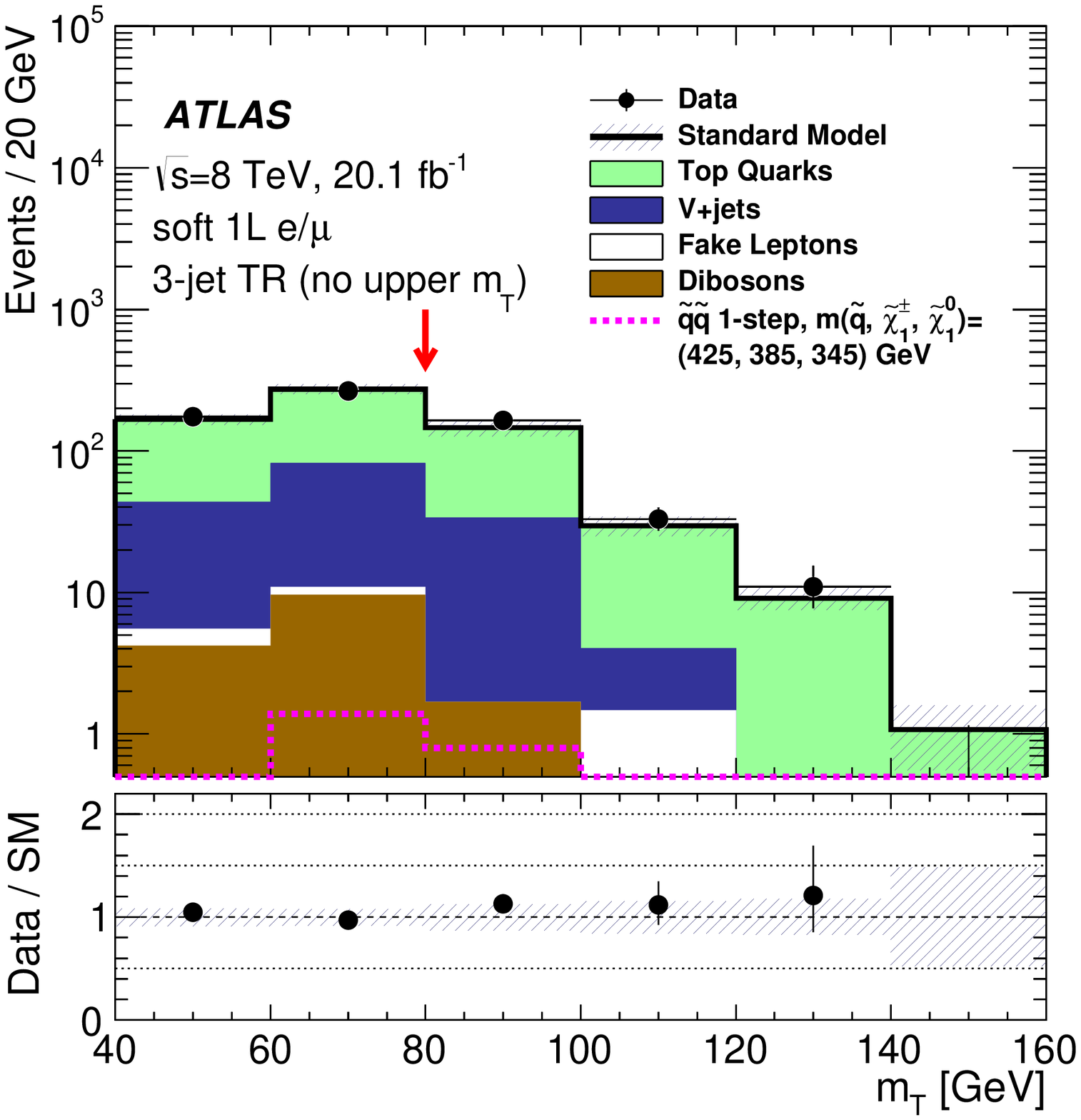}
\includegraphics[width=0.49\textwidth]{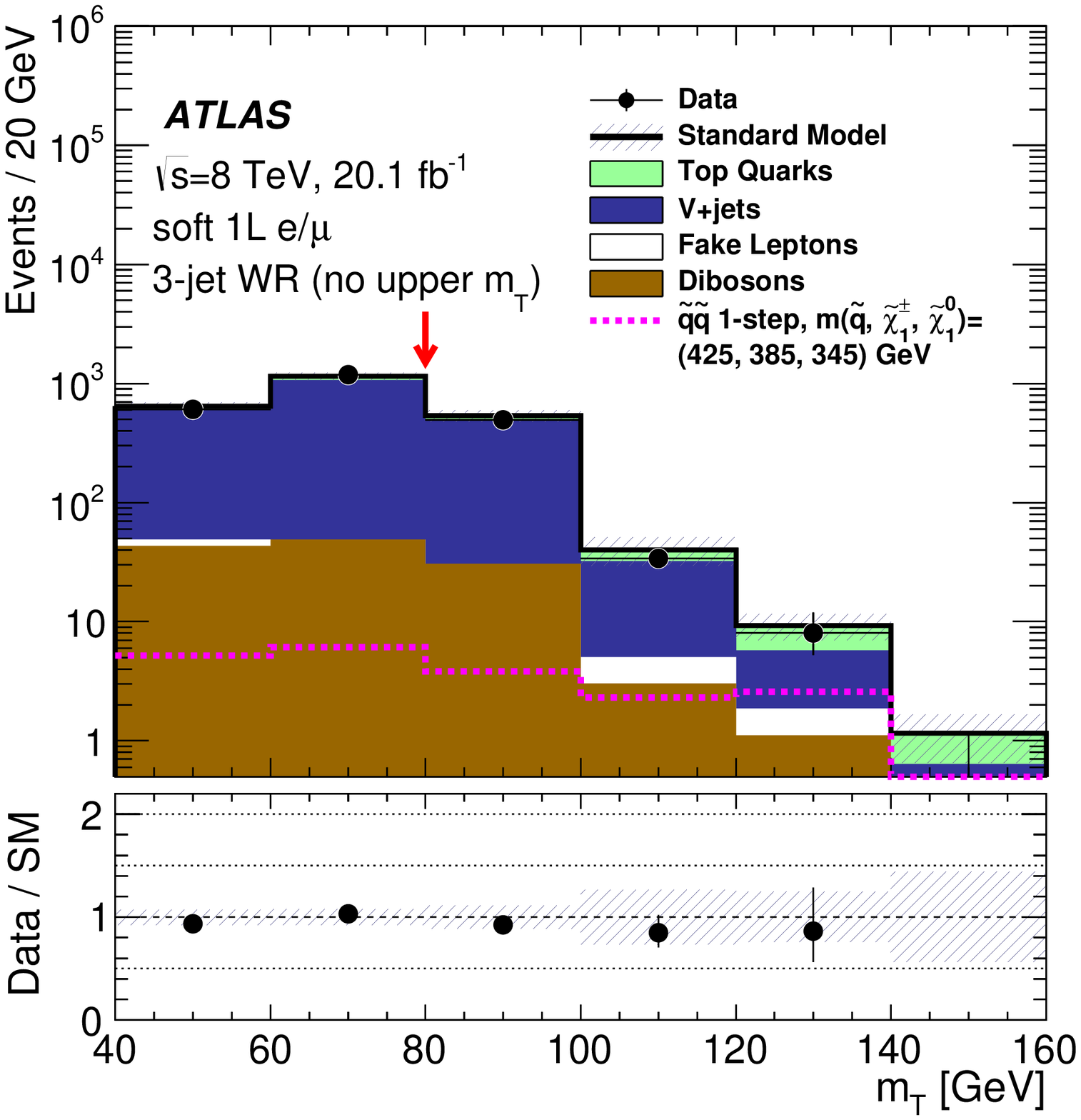}
\includegraphics[width=0.49\textwidth]{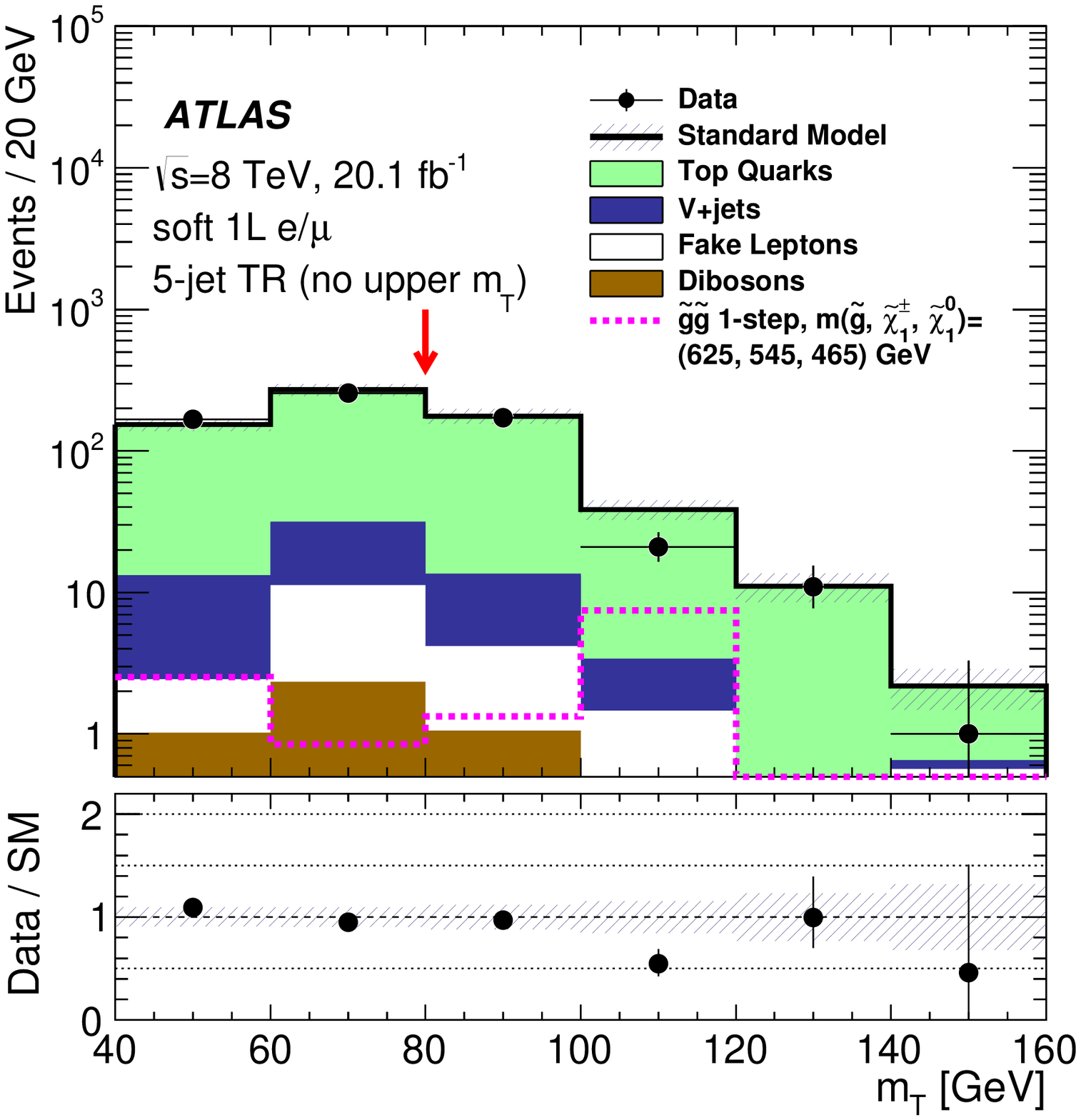}
\includegraphics[width=0.49\textwidth]{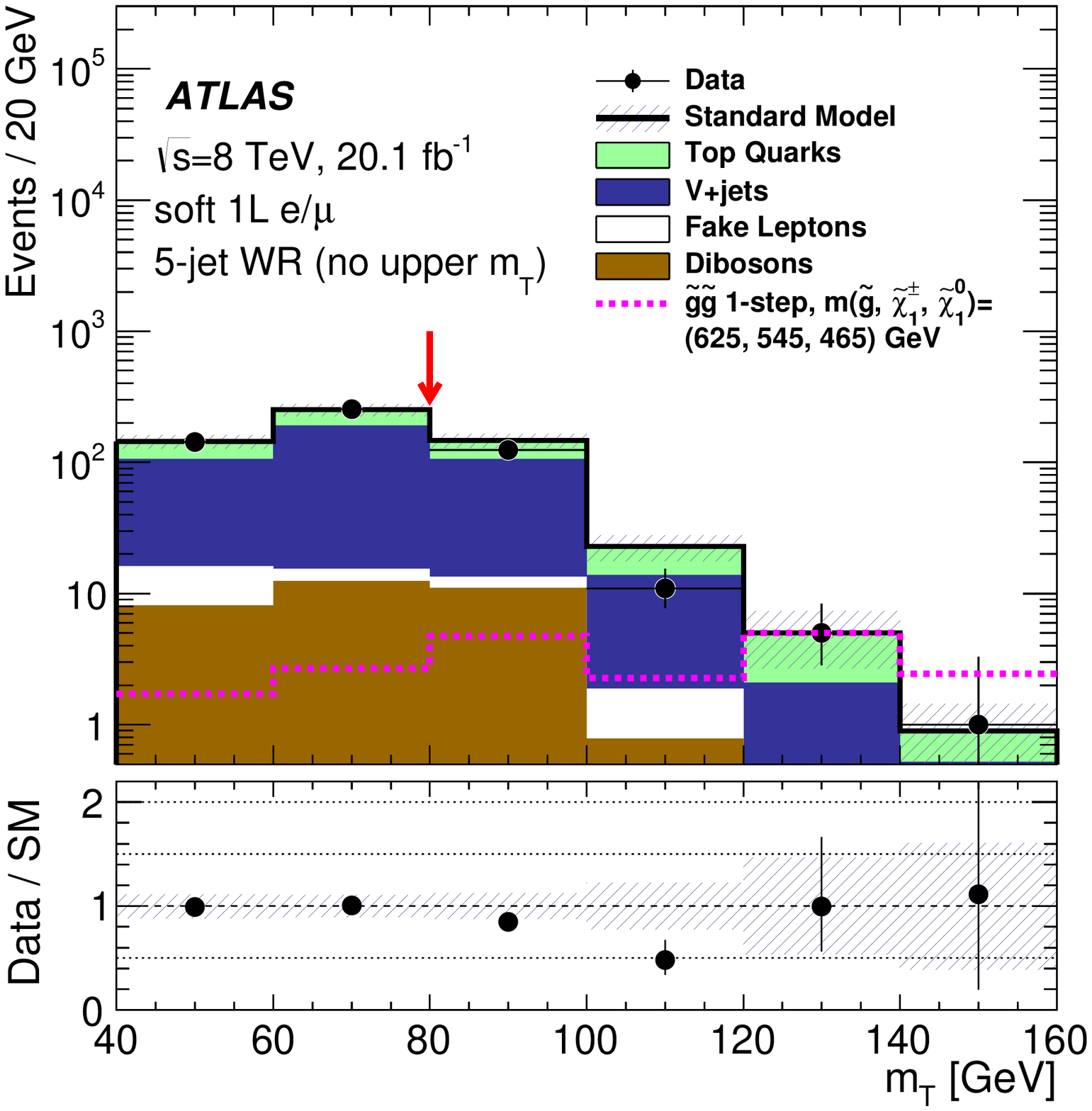}
\caption{Distribution of the transverse mass \mt~in the 3-jet (top) and 5-jet (bottom) \ttbar~(left) and
$W$+jets (right) control regions used in the soft single-lepton channel. The upper \mt~cut, indicated by the arrow, is not applied in these
distributions. The purity in the 
background of interest is 56\% (87\%) for the 3-jet \ttbar~($W$) control region and 82\% (66\%) for the 5-jet \ttbar~($W$) control region.
The ``Data/SM'' plots show the ratio of data to the summed Standard Model expectation, which is derived from the fit described in section \ref{sec:bkgfit}. The uncertainty band on the Standard Model expectation shown here
combines the statistical uncertainty on the simulated event samples with the relevant systematic uncertainties (see text).
The last bin includes the overflow. The ``Top Quarks'' label includes all top-quark-related backgrounds, while ``V+jets''
includes $W$+jets, $Z$+jets and other Drell-Yan backgrounds such as $Z\rightarrow\tautau$ and $\gamma^*/Z$ outside the $Z$ pole region.
For illustration, the expected signal distribution is shown 
for first- and second-generation squark pair production with $m_{\tilde{q}}=425$ \GeV, $m_{\tilde{\chi}^{\pm}_{1}}$=385 \GeV~and $m_{\tilde{\chi}^0_1}=345$ \GeV~(top), 
and for gluino pair production with 
$m_{\tilde{g}}=625$ \GeV, $m_{\tilde{\chi}^{\pm}_{1}}$=545 \GeV~and 
$m_{\tilde{\chi}^0_1}=465$ \GeV~(bottom).
} \label{fig:1softCR}
\end{figure}
\begin{figure}[ht]
\centering
\includegraphics[width=0.49\textwidth]{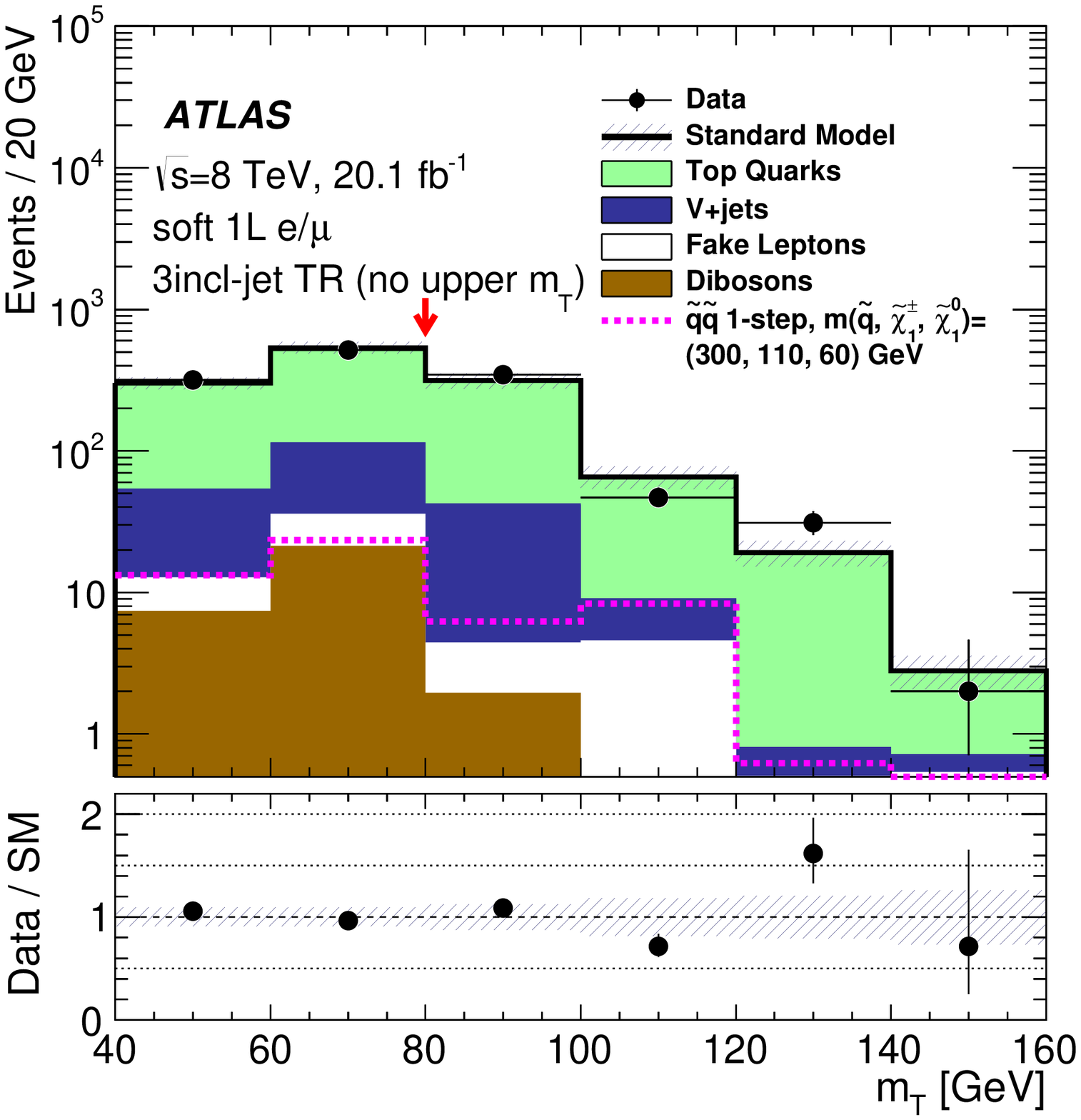}
\includegraphics[width=0.49\textwidth]{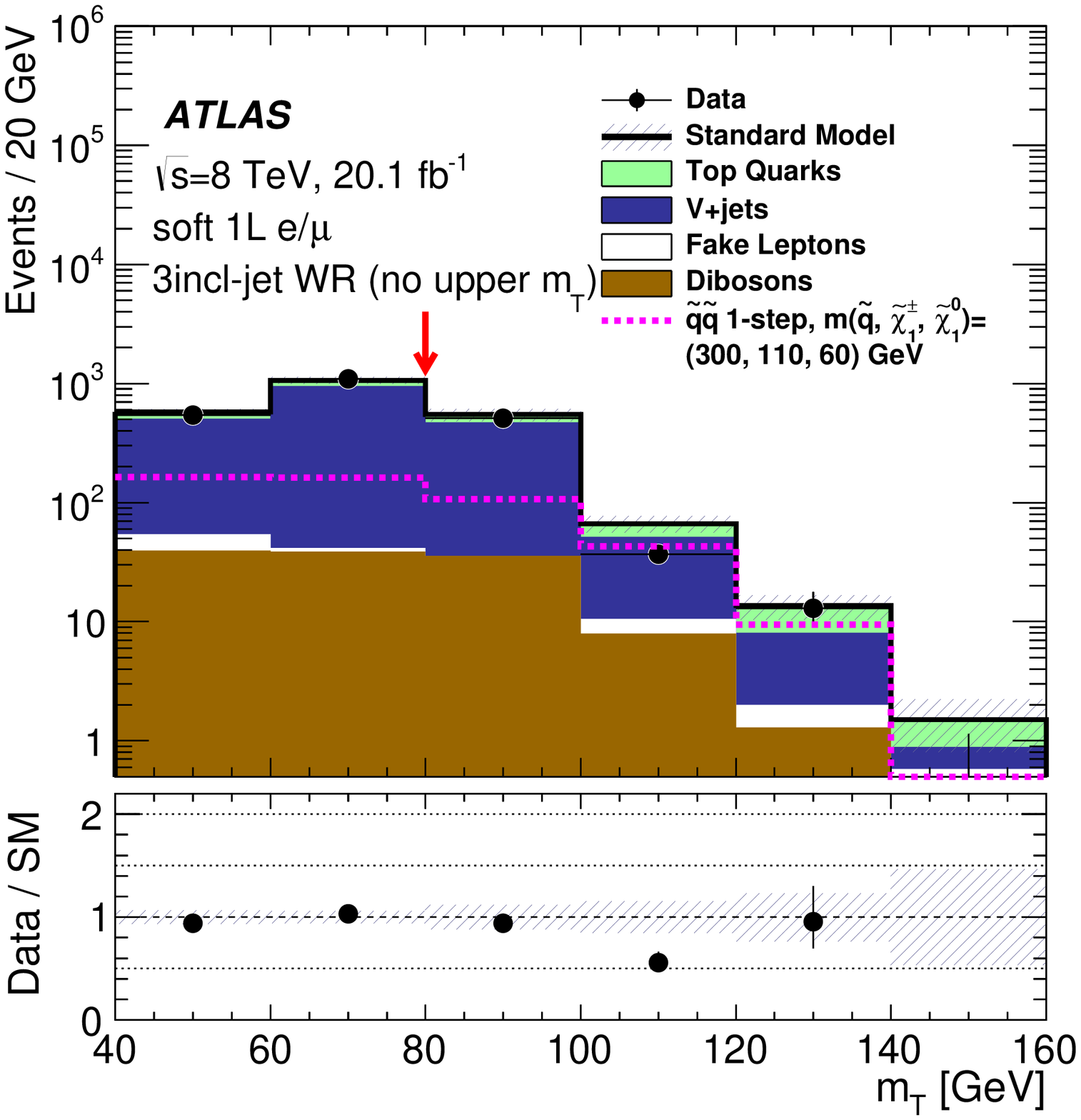}
\caption{Distribution of the transverse mass \mt~in the 3-jet inclusive \ttbar~(left) and $W$+jets (right) control regions used in the soft single-lepton channel. 
	The upper \mt~cut, indicated by the arrow, is not applied in these
	distributions. The purity in the background of interest is 70\% (82\%) for \ttbar~($W$) control region.
The ``Data/SM'' plots show the ratio of data to the summed Standard Model expectation, which is derived from the fit described in section \ref{sec:bkgfit}. The uncertainty band on the Standard Model expectation shown here
combines the statistical uncertainty on the simulated event samples with the relevant systematic uncertainties (see text).
The last bin includes the overflow.  The ``Top Quarks'' label includes all top-quark-related backgrounds, while ``V+jets'' 
includes $W$+jets, $Z$+jets and other Drell-Yan backgrounds such as $Z\rightarrow\tautau$ and $\gamma^*/Z$ outside the $Z$ pole region.
For illustration, the expected signal distribution is shown
for first- and second-generation squark pair production with 
$m_{\tilde{q}}=300$ \GeV, $m_{\tilde{\chi}^{\pm}_{1}}$=110 \GeV~and $m_{\tilde{\chi}^0_1}=60$ \GeV.
} \label{fig:1softCR3jinc}
\end{figure}

\begin{figure}[ht]
\centering
\includegraphics[width=0.49\textwidth]{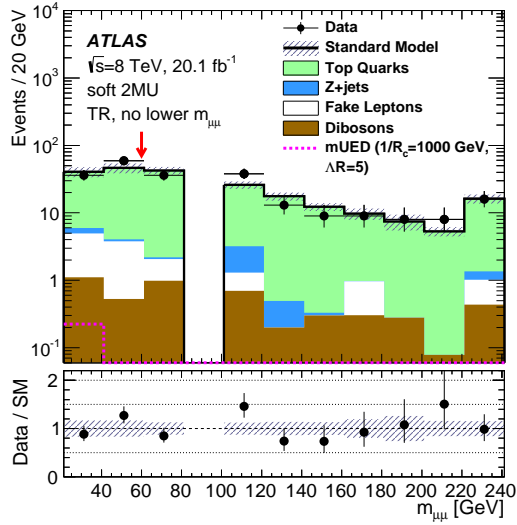}

\caption{Distribution of the dimuon invariant mass $m_{\mu\mu}$~in the \ttbar~control region used in the soft dimuon channel.
	The lower $m_{\mu\mu}$ cut, indicated by the arrow, is not applied in this distribution.
The purity in \ttbar~is 83\% for this region.	
The ``Data/SM'' plots show the ratio of data to the summed Standard Model expectation, which is derived from the fit described in section \ref{sec:bkgfit}. The uncertainty band on the Standard Model expectation shown here
combines the statistical uncertainty on the simulated event samples with the relevant systematic uncertainties (see text).
The last bin includes the overflow. The ``Top Quarks'' label includes all top-quark-related backgrounds, while ``V+jets'' 
includes $W$+jets, $Z$+jets and other Drell-Yan backgrounds such as $Z\rightarrow\tautau$ and $\gamma^*/Z$ outside the $Z$ pole region.
For illustration, the expected signal distribution of the mUED model point
with $R_{\mathrm{c}}^{-1}$ = 1000 \GeV~and $\Lambda R_{\mathrm{c}}$=5 is also shown.
} \label{fig:2softCR}
\end{figure}
\begin{figure}[ht]
\centering
\includegraphics[width=0.49\textwidth]{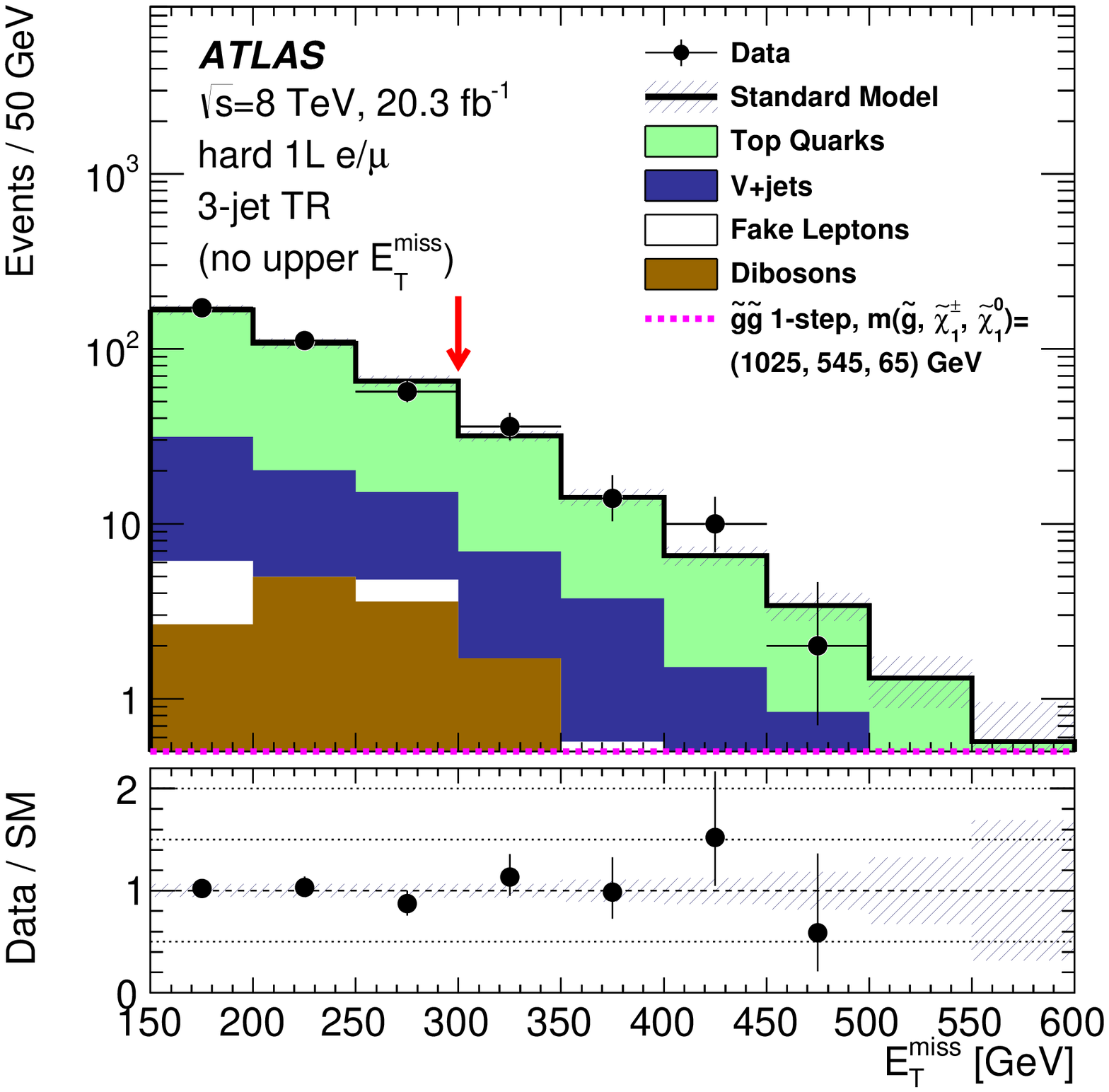}
\includegraphics[width=0.49\textwidth]{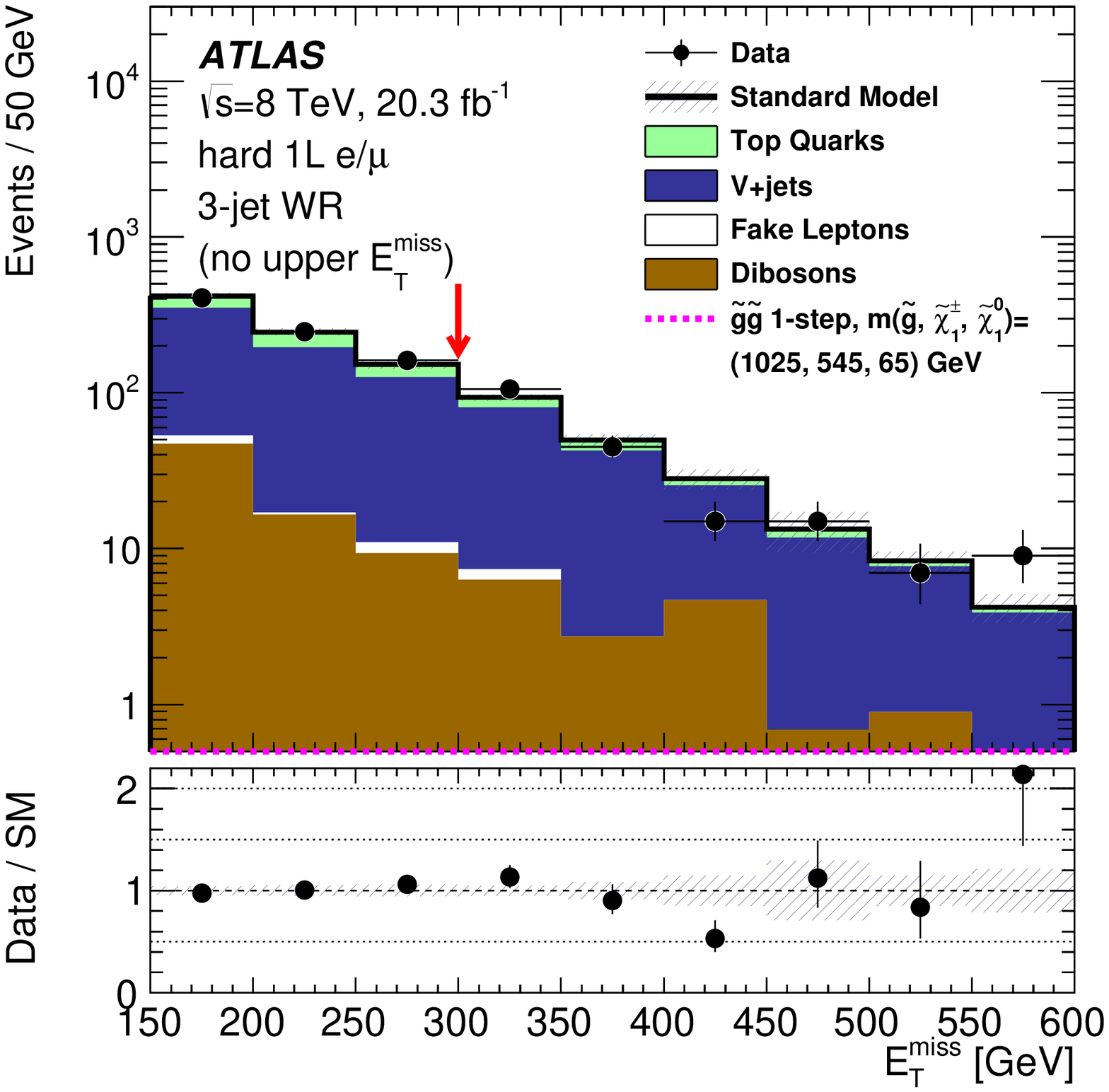}
\includegraphics[width=0.49\textwidth]{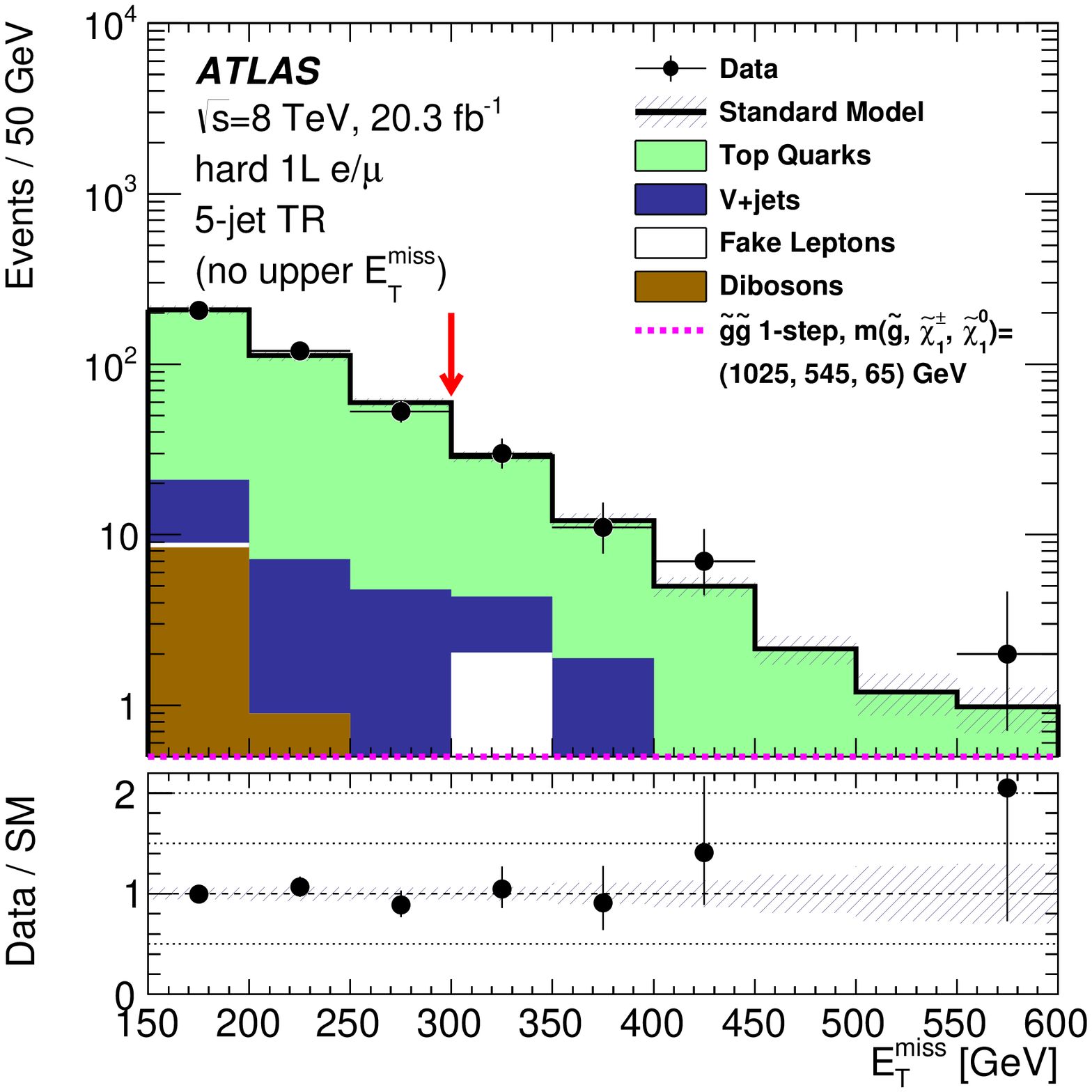}
\includegraphics[width=0.49\textwidth]{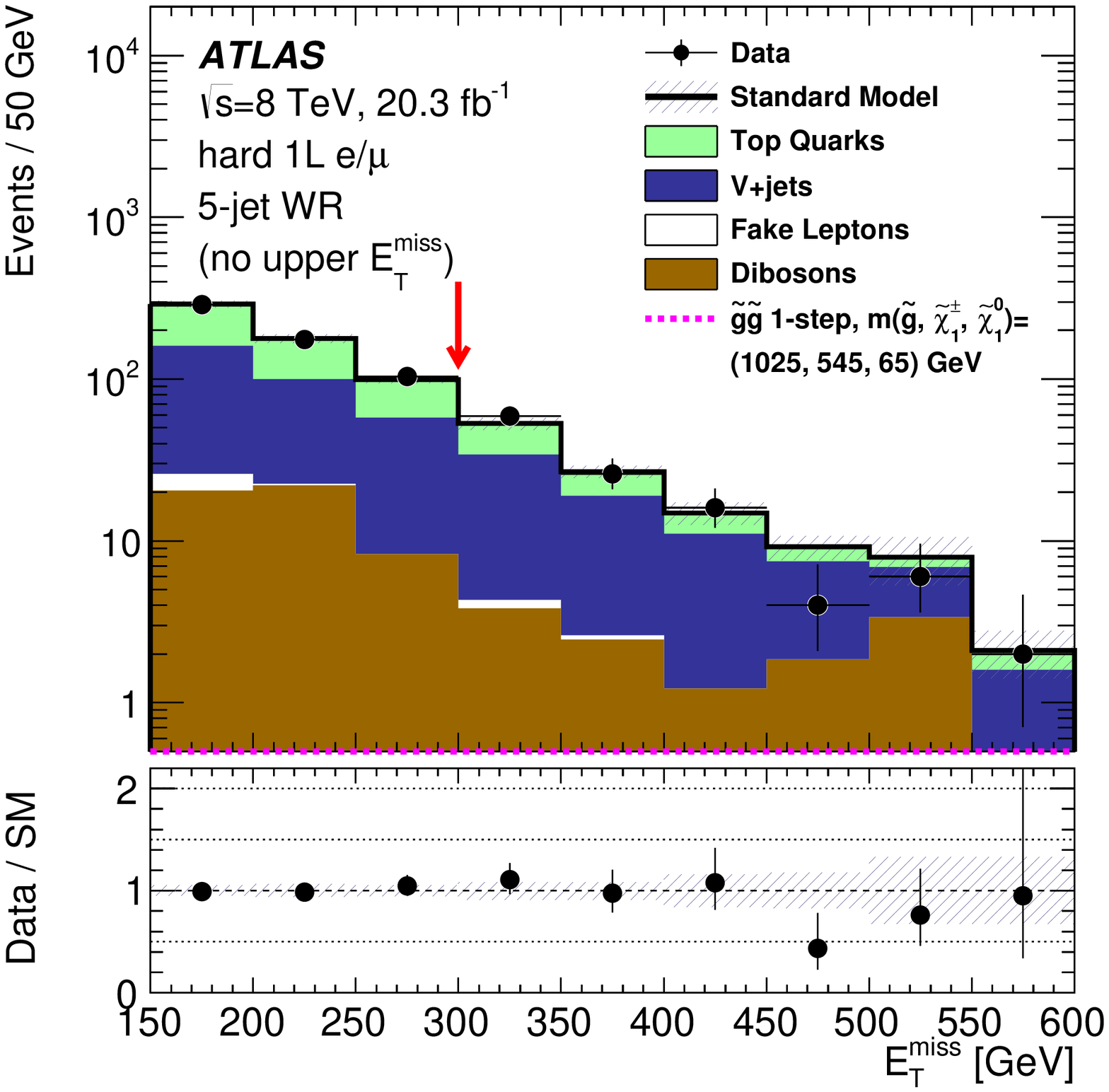}
\caption{Distribution of the missing transverse momentum \met~in the 3-jet (top) and 5-jet (bottom) \ttbar~(left) and
$W$+jets (right) control regions used in the hard single-lepton channel.
The upper \met~cut, indicated by the arrow, is not applied in these distributions.
The purity in the background of interest is 66\% (72\%) for the 3-jet \ttbar~($W$) control region and 81\% (45\%) for the
5-jet \ttbar~($W$) control region; the 5-jet $W$ control region is cross-contaminated by \ttbar~events at the level of 40\%.
The ``Data/SM'' plots show the ratio of data to the summed Standard Model expectation, which is derived from the fit described in section \ref{sec:bkgfit}. The uncertainty band on the Standard Model expectation shown here
combines the statistical uncertainty on the simulated event samples with the relevant systematic uncertainties (see text).
The last bin includes the overflow. The ``Top Quarks'' label includes all top-quark-related backgrounds, while ``V+jets'' 
includes $W$+jets, $Z$+jets and other Drell-Yan backgrounds such as $Z\rightarrow\tautau$ and $\gamma^*/Z$ outside the $Z$ pole region. For illustration, the expected signal
distributions are shown for gluino pair production with
 $m_{\tilde{g}}=1025 \GeV, m_{\tilde{\chi}^{\pm}_{1}}=545 \GeV$ and $m_{\tilde{\chi}^0_1}=65 \GeV$.
} \label{fig:1hardCR1}
\end{figure}
\begin{figure}[ht]
\centering
\includegraphics[width=0.49\textwidth]{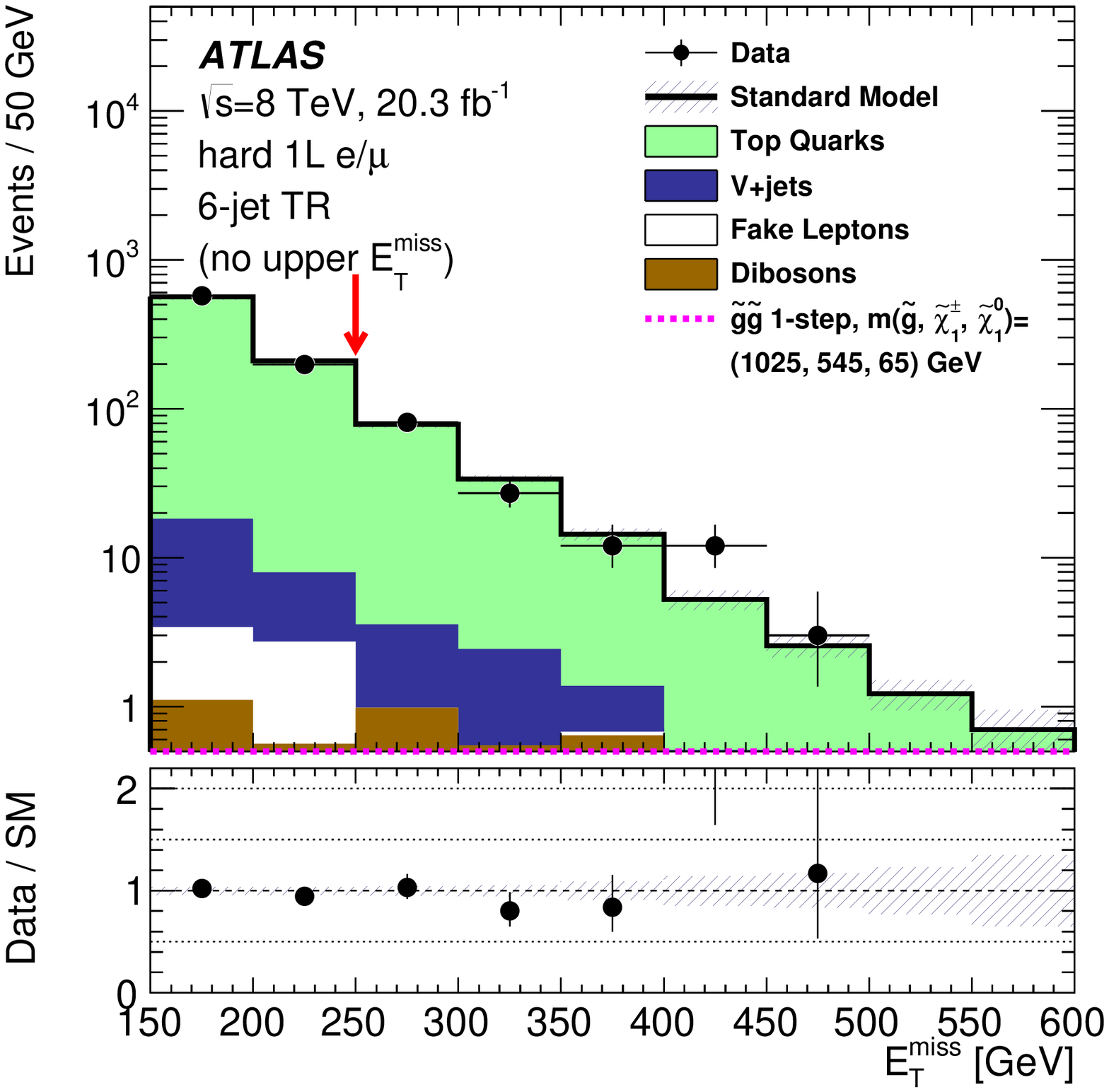}
\includegraphics[width=0.49\textwidth]{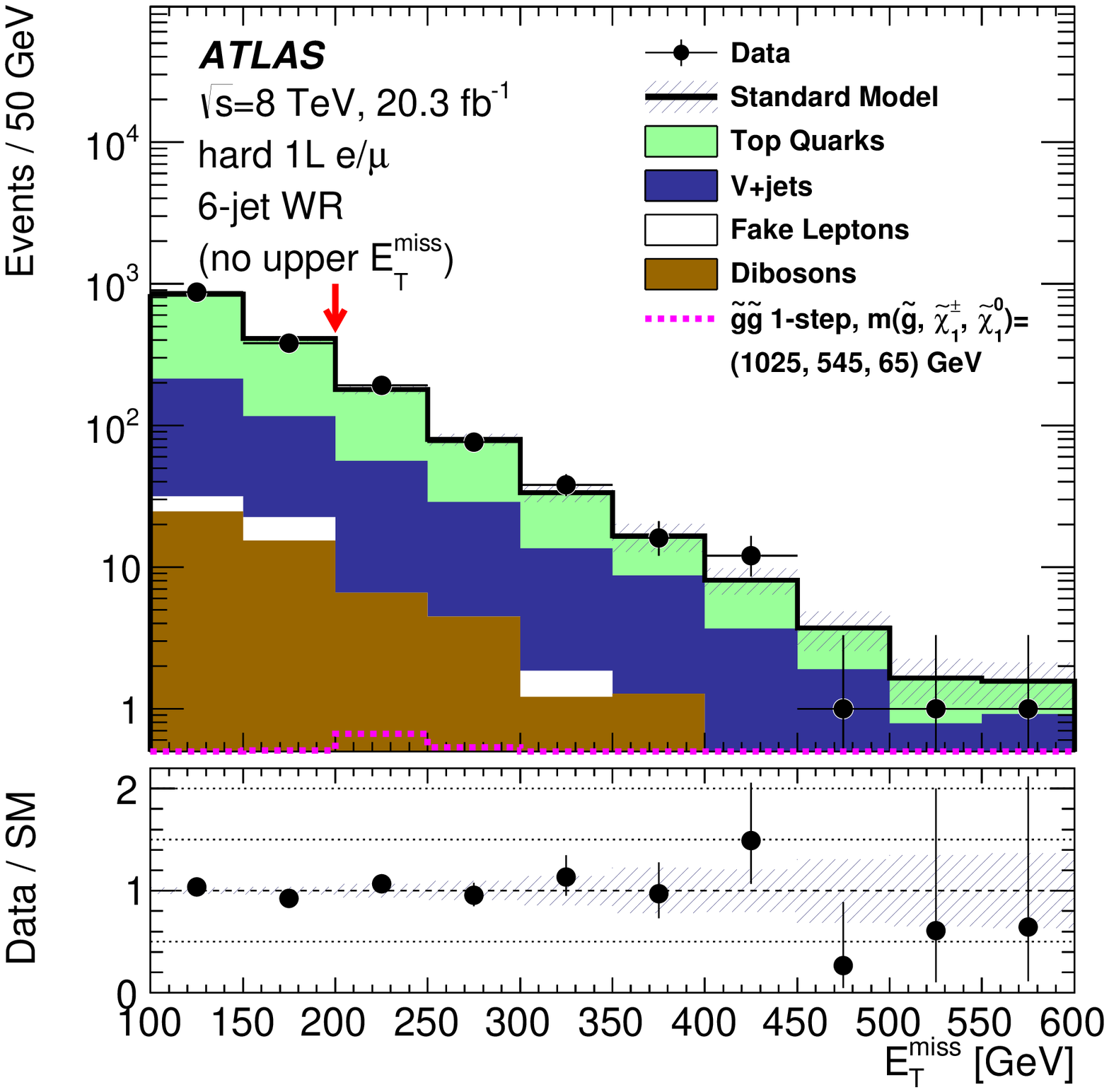}
\caption{Distribution of the missing transverse momentum \met~in 6-jet \ttbar~(left) and $W$+jets (right) control regions used in the hard single-lepton channel. 
	The upper \met~cut, indicated by the arrow, is not applied in these distributions.
	The purity in the background of interest is 90\% (21\%) for the \ttbar~($W$) control region; the $W$ control region is cross-contaminated 
by \ttbar~events at the 70\% level.
The ``Data/SM'' plots show the ratio of data to the summed Standard Model expectation, which is derived from the fit described in section \ref{sec:bkgfit}. The uncertainty band on the Standard Model expectation shown here
combines the statistical uncertainty on the simulated event samples with the relevant systematic uncertainties (see text). 
The last bin includes the overflow. The ``Top Quarks'' label includes all top-quark-related backgrounds, while ``V+jets'' 
includes $W$+jets, $Z$+jets and other Drell-Yan backgrounds such as $Z\rightarrow\tautau$ and $\gamma^*/Z$ outside the $Z$ pole region. For illustration, the expected signal
distributions are shown for gluino pair production 
 with $m_{\tilde{g}}=1025 \GeV, m_{\tilde{\chi}^{\pm}_{1}}=545 \GeV$ and $m_{\tilde{\chi}^0_1}=65 \GeV$.
} \label{fig:1hardCR2}
\end{figure}
\begin{figure}[ht]
\centering
\includegraphics[width=0.42\textwidth]{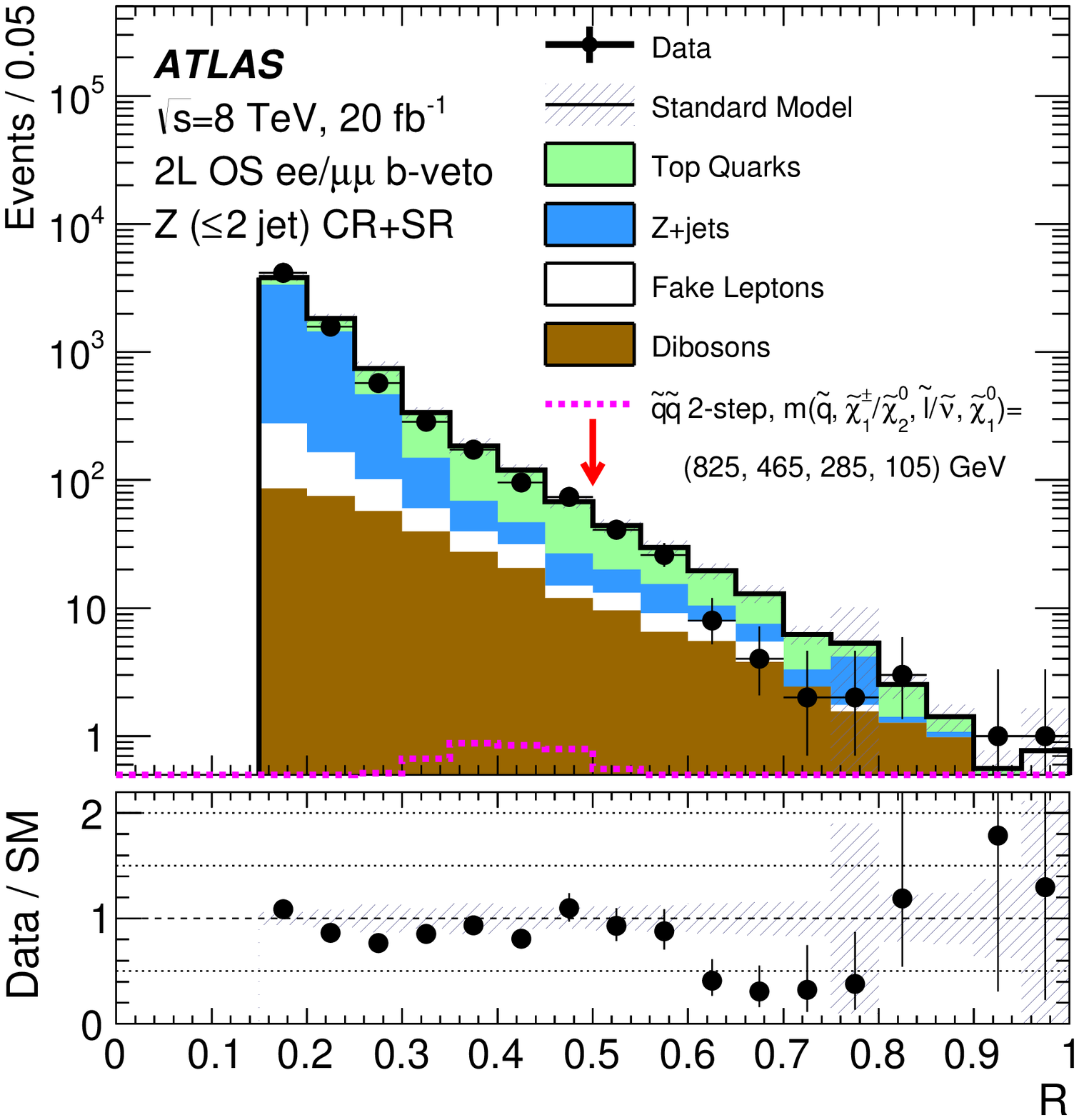}
\includegraphics[width=0.42\textwidth]{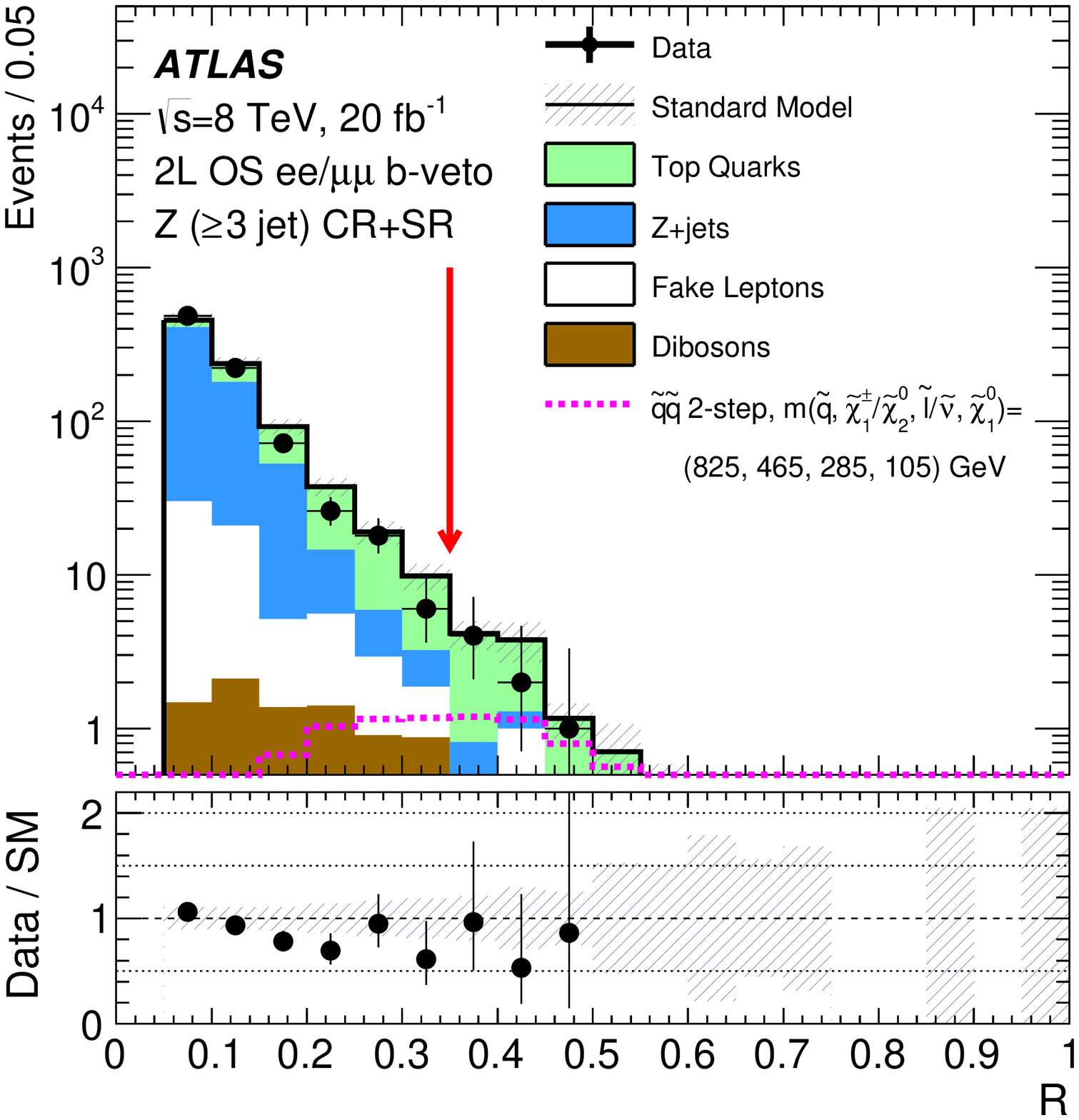}
\includegraphics[width=0.42\textwidth]{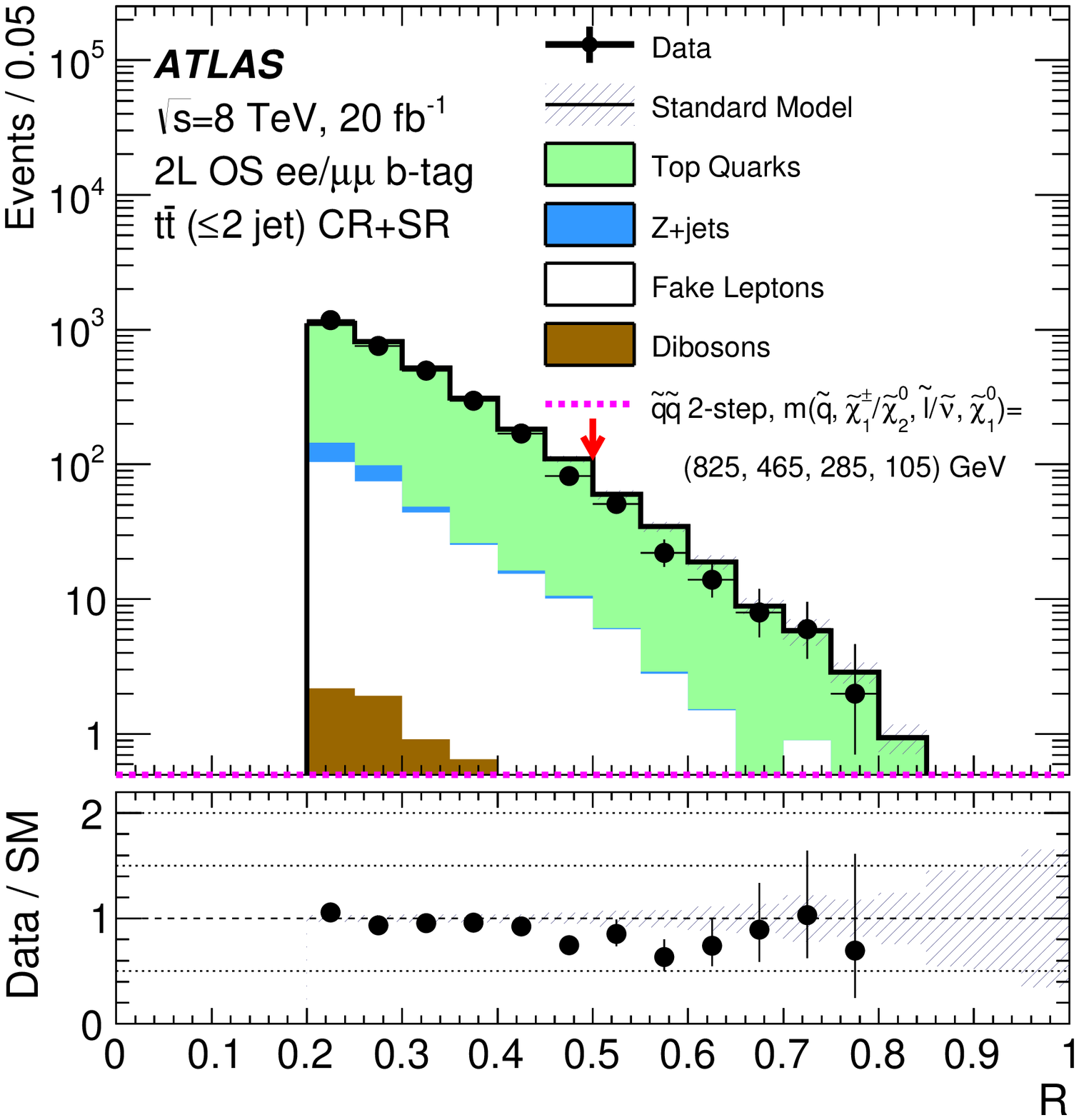}
\includegraphics[width=0.42\textwidth]{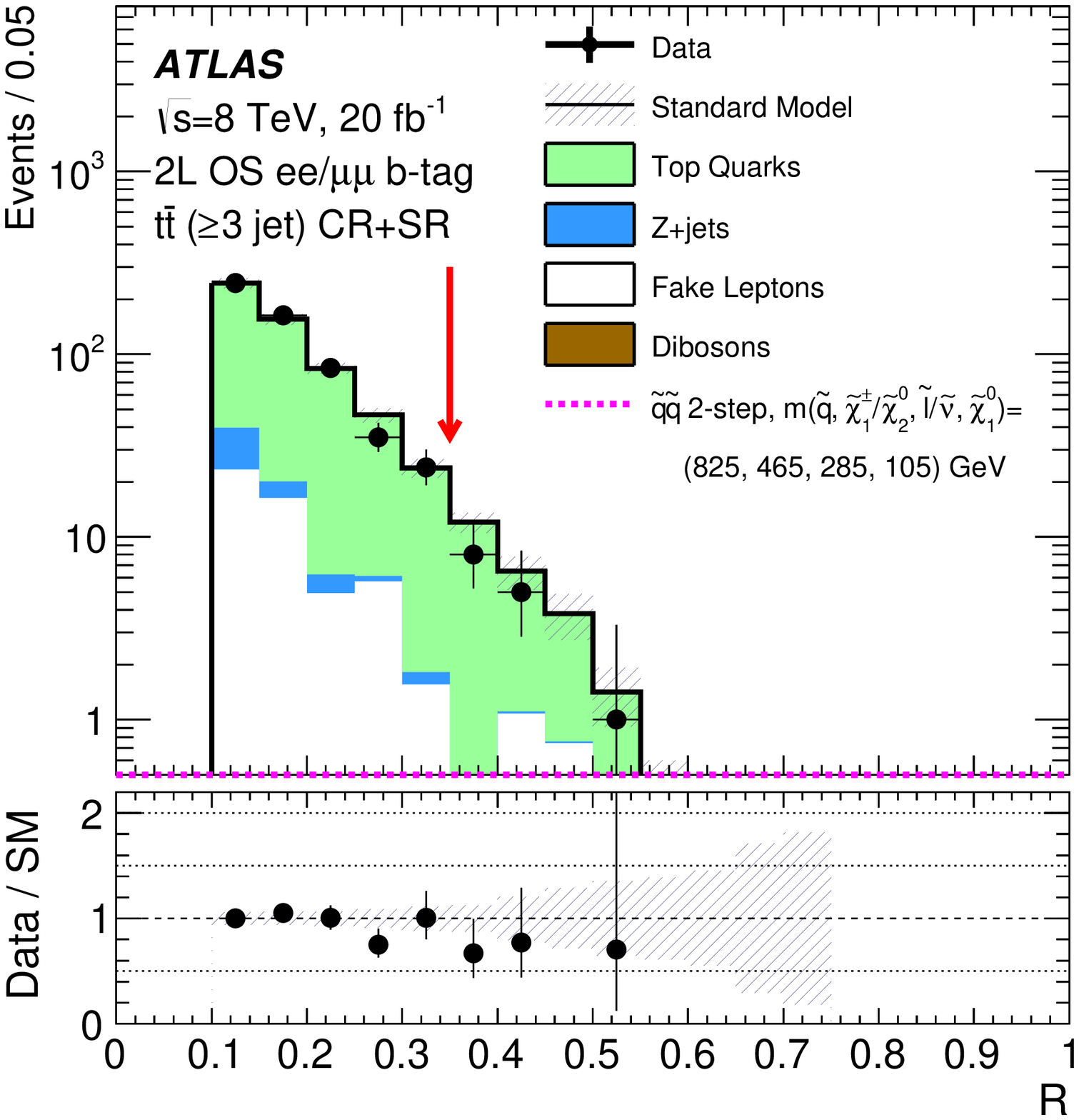}
\includegraphics[width=0.42\textwidth]{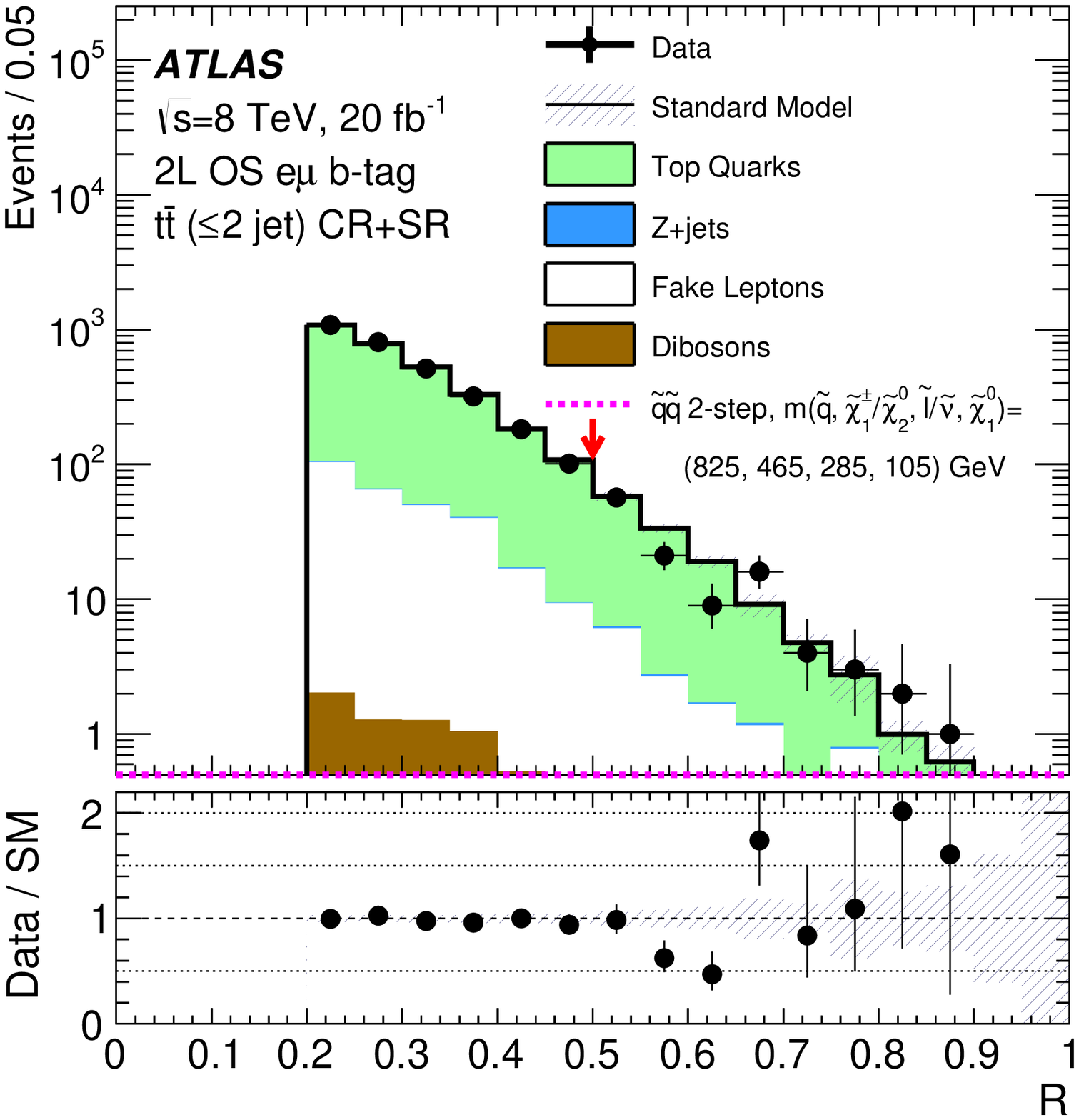}
\includegraphics[width=0.42\textwidth]{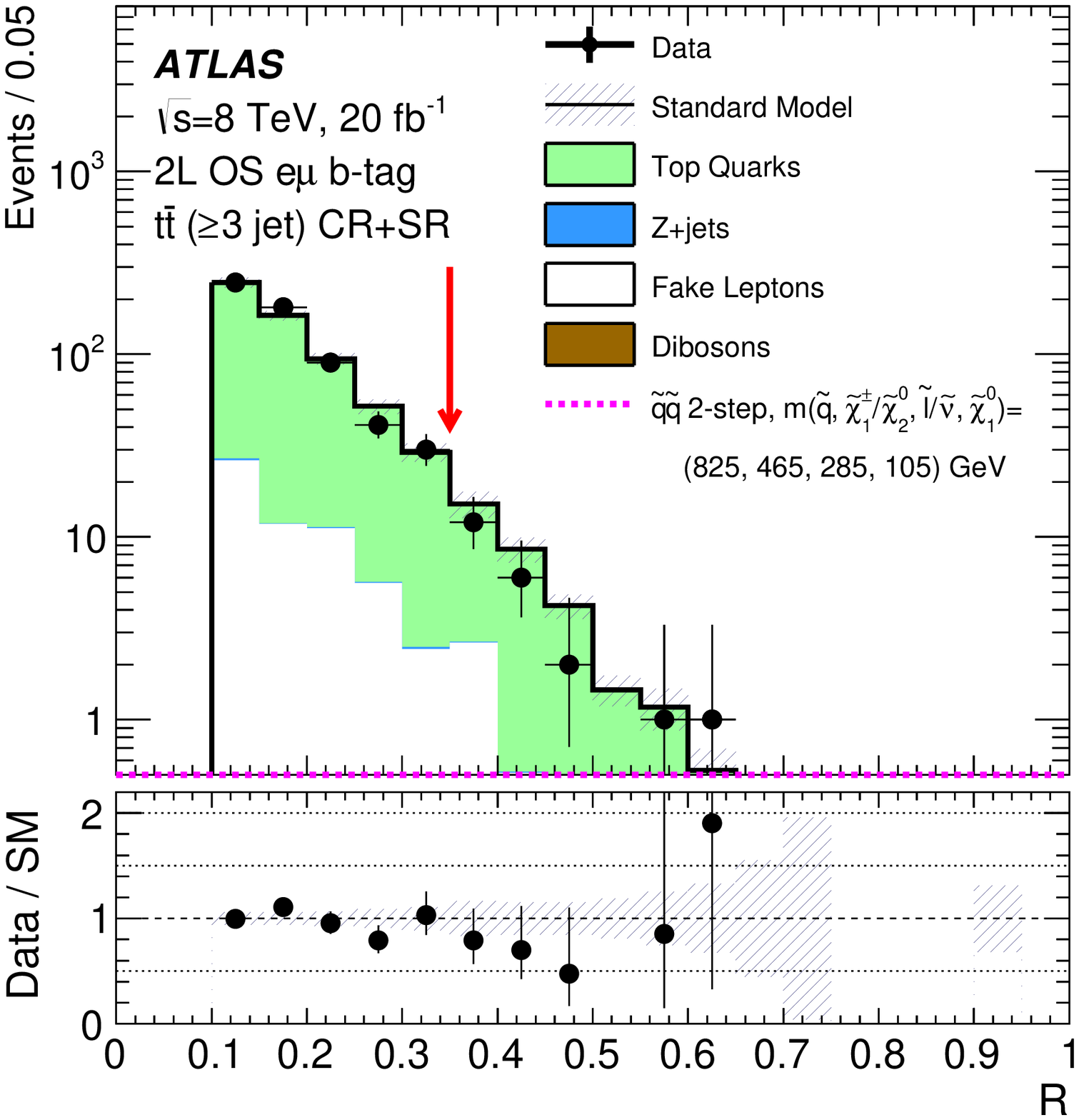}
\caption{Distribution of the razor variable $R$~in the low-multiplicity (left) and 3-jet (right) $Z$ (top) control region and in the \ttbar~ control region with same-flavour (middle) 
	or opposite-flavour (bottom) leptons used in the hard dilepton channel. 
	The upper cut on $R$ (illustrated by the arrow) separating signal from control regions is not applied in these distributions.
	The purity in the background of interest is 85\% and 75\% for
	the \ttbar~and $Z$ control regions, respectively.
The ``Data/SM'' plots show the ratio of data to the summed Standard Model expectation, which is derived from the fit described in section \ref{sec:bkgfit}. The uncertainty band on the Standard Model expectation shown here
combines the statistical uncertainty on the simulated event samples with the relevant systematic uncertainties (see text). The ``Top Quarks'' label includes all top-quark-related backgrounds, while ``V+jets'' 
includes $W$+jets, $Z$+jets and other Drell-Yan backgrounds such as $Z\rightarrow\tautau$ and $\gamma^*/Z$ outside the $Z$ pole region. For illustration, the expected signal
distributions are shown for squark pair production with $m_{\tilde{q}}=825 \GeV, m_{\tilde{\chi}^{\pm}_{1}/\tilde{\chi}^{0}_{2}}=465 \GeV, m_{\tilde{\ell}/\tilde{\nu}}=285 \GeV$ and $m_{\tilde{\chi}^0_1}=105 \GeV$.
} \label{fig:2hardCR2}
\end{figure}

\clearpage
\subsection{Fake-lepton background}
\label{sec:misidlep}
The multijet and $Z(\rightarrow\nu\nu)$+jets processes are 
important sources of fake-lepton background in the single-lepton analyses, while $W(\rightarrow\ell\nu$)+jets 
and \ttbar~production (where one of the leptons comes from a $W$ boson decay and the other is from a $b$-hadron decay)
are expected to dominate the fake-lepton background in the soft dimuon and hard dilepton analyses. 

The fake-lepton background in the signal region as well as in the control regions, where it is more significant, 
is estimated entirely from the data by means of a matrix
method described in ref. \cite{matrixmethod} and briefly summarised below; the procedure is applied separately for electrons and muons.

In this method, the process creating the fake lepton is
enhanced in a control sample where preselected leptons are used instead of the signal leptons, and all signal or control region criteria are applied.
If $N_\mathrm{pass}$ and $N_\mathrm{fail}$ are the number of
events found passing or failing the signal lepton selection  
in this control sample, then the number of events with a fake 
lepton in a single-lepton signal or control region is given by:
\begin{equation}
N_\mathrm{pass}^\mathrm{fake}=\frac{N_\mathrm{fail} - (1/\epsilon^\mathrm{real} - 1)N_\mathrm{pass}}{1/\epsilon^\mathrm{fake} - 1/\epsilon^\mathrm{real}}
\end{equation}
where $\epsilon^\mathrm{real}$ is the relative identification efficiency (from preselection to selection) for prompt
leptons and $\epsilon^\mathrm{fake}$ is the probability to misidentify jets or non-prompt leptons as prompt leptons.
For dileptonic signal or control regions, the estimation of 
this background is based on the same principle, this time using a four-by-four matrix 
to take into account the different fake combinations for the leading and subleading lepton: one prompt and one fake lepton, two fake leptons,
or two prompt leptons.

The relative identification efficiency $\epsilon^\mathrm{real}$ is obtained 
from data in bins of \pt~and $\eta$~using a tag-and-probe method in $Z \rightarrow \ell\ell$ events.
The value of $\epsilon^\mathrm{fake}$
is estimated in control regions enriched in multijet events. 
The multijet control region is composed of 
events with at least one preselected lepton and at least one signal jet with
\pt~$>$ 60 GeV, \mt~$<$ 40 GeV and \met~$<$ 30 GeV. 
Since the control region is defined at low \met~values, the triggers described in section \ref{sec:trigger}
cannot be used. Instead, a combination of prescaled single-lepton triggers and unprescaled dilepton
triggers is used. The prompt lepton contribution is subtracted from this multijet control region using MC simulation and
$\epsilon^\mathrm{fake}$ is given by the fraction of preselected leptons passing the signal lepton selection in this region.
The events are split into two samples depending on whether they have at least one $b$-tagged jet or none,
in order to allow $\epsilon^\mathrm{fake}$ to vary as a function of the fake-lepton source. The value of $\epsilon^\mathrm{fake}$ 
 is also extracted in bins of \pt~and $\eta$. The same relative efficiencies are used across all analysis channels.

\subsection {Other backgrounds}
\label{otherBG}
All other backgrounds are estimated from simulation, using the most
accurate theoretical cross sections available. These include single-top ($t$-channel, $s$-channel, $Wt$), $t+Z$ , dibosons ($WW$, $WZ$, $ZZ$, $W\gamma$, $Z\gamma$), $t\bar{t}+W$, $t\bar{t}+Z$ and $t\bar{t}+WW$ production. In the soft dimuon channel, the Drell--Yan and $Z$+jets backgrounds 
are also estimated from simulation as there is no dedicated control region to evaluate their contribution.

\clearpage
\section{Systematic uncertainties}
\label{sec:systuncert}

Systematic uncertainties have an impact on 
the extrapolation factors used to derive
the dominant background yields in the signal regions; they also impact the fake-lepton estimation and
the yields of the backgrounds estimated from simulation only, which in turn affect the normalisation 
of the dominant backgrounds in the control regions. Finally, systematic uncertainties also affect the expected signal yields.  

\subsection{Experimental uncertainties}

The following detector-related systematic uncertainties are taken
into account.

Uncertainties on the lepton identification, momentum/energy scale and resolution are estimated from samples of $Z \rightarrow \ell^{+}\ell^{-}$,
$J/\psi \rightarrow \ell^{+}\ell^{-}$ and $W \rightarrow \ell\nu$ decays~\cite{electronref,muonrefnew}.
The jet energy scale (JES) uncertainty depends on \pT~and $\eta$ as well as on the event topology and flavour composition of the jet. 
It has been obtained from simulation and a combination of 
test beam and in-situ measurements from $pp$ collisions, as described in refs.~\cite{JES,JES2}. 
The jet energy resolution (JER) uncertainty has also 
been estimated using in-situ measurements \cite{atlas-jer}. 
The JES and JER uncertainties arising from the high luminosity and pile-up in 2012 data are also taken into account.
These jet and lepton uncertainties are propagated to the \met~calculation, which also includes an uncertainty coming from the energy deposits 
which are not associated with a well-identified physics object~\cite{Aad:2012re}. 

The jet vertex fraction requirement in the identification of signal jets has an associated uncertainty which is assessed by varying the threshold from 
its nominal value of 0.25 down to 0.21 and up to 0.28 (see ref.~\cite{jvfuncert}). 

Uncertainties associated with the $b$-tagging efficiency are derived 
from data-driven measurements in \ttbar\ events~\cite{ATLAS-CONF-2014-004,ATLAS-CONF-2012-043}, while uncertainties associated with the probability of mistakenly 
$b$-tagging a jet which does not contain a $b$-hadron are determined using dijet samples \cite{ATLAS-CONF-2012-040}. 

Uncertainties (1--5\%) are also assigned to the various trigger efficiencies, based on studies comparing their efficiencies
as measured in data to those predicted by simulation. 

Finally, uncertainties are also assigned to the fake-lepton background estimation, including
a statistical uncertainty on the number of events in the control samples and an uncertainty on the relative identification efficiency obtained by 
comparing its value as measured in \ttbar~events to the value obtained in $Z\rightarrow\ell\ell$ events. In addition, a 20\% uncertainty is 
assigned to the subtraction of the $W/Z+$jet backgrounds from the control samples used to estimate the relative misidentification efficiency.
In the soft dimuon case, the misidentification efficiency is assigned an additional 30\% uncertainty which covers the difference 
obtained when varying the origin of the misidentified prompt leptons (light jets or heavy-flavour jets).

\subsection{Theoretical uncertainties on the background estimation}

The following theory-related uncertainties are taken into account. 
For backgrounds with free normalization, the uncertainties affect only the change in the prediction in the VR or SR relative to the yield predicted in the CR.

The uncertainty due to the factorisation and renormalisation scales is computed by varying these scales up and down by a factor of two with respect to the  nominal setting. PDF uncertainties 
are computed following the PDF4LHC recommendations \cite{Botje01}. 

For the $W/Z$+jets and \ttbar~$+W$ samples, an uncertainty derived by comparing samples generated with different numbers of partons is applied 
to cover the impact of the finite number of partons generated in the
nominal samples. For the \ttbar, single-top and diboson backgrounds, a parton shower modelling uncertainty is evaluated by comparing \pythia to \herwigpjimmy.

The uncertainty related to the choice of Monte Carlo generator for the \ttbar~background is derived by comparing \powhegpjimmy to \alpgenpjimmy samples. 
For the \ttbar~and single-top backgrounds, uncertainties on the emission of initial- and final-state radiation (ISR/FSR) are also accounted for by the use 
of \textsc{AcerMC}+\pythia samples generated with different tunes. 

The small contribution to the single-top background from $t+Z$ is assigned a total cross-section uncertainty of 50\%, while
the other single-top production channels are assigned uncertainties of 3.9\% ($s$-channel \cite{Kidonakis:2010a}), $[+3.9,-2.2]$\% 
($t$-channel \cite{Kidonakis:2011}), and 6.8\% ($Wt$-channel \cite{Kidonakis:2010b}).
Samples using diagram subtraction and diagram removal schemes are compared to incorporate interference effects between single-top 
and \ttbar~production at NLO.
For the $Wt$ channel, a Monte Carlo generator uncertainty is derived from the comparison between \powheg and \mcatnlo samples.  
For \ttbar~$+W$ a total cross-section uncertainty of 22\%, derived from ref.~\cite{Campbell:2012},  is applied.

The $WW$, $WZ$ and $ZZ$ backgrounds are estimated purely from MC simulation and assigned a 5\%, 5\% and 7\% cross-section uncertainty, respectively, as estimated from 
missing higher order corrections, and from uncertainties on the PDF and on the value of the strong coupling constant. In addition to this, a generator uncertainty is applied, 
derived from comparing the nominal \textsc{powheg} diboson samples used in the dileptonic signal regions to
samples generated with a\mcatnlo\cite{aMCatNLO}, in both cases hadronised with \textsc{Pythia}. 

For the diboson, single-top and \ttbar$+W$ backgrounds, the theoretical uncertainties, with the exception of those on the cross section and PDF, 
are combined and applied as an uncertainty envelope. In the case of the small \ttbar~$+Z$ and \ttbar~$+WW$ backgrounds a total cross-section uncertainty of 50\% is applied.

Finally, a statistical uncertainty corresponding to the finite size of the MC samples used is also taken into account.

\subsection{Dominant uncertainties on the background estimation}

The backgrounds in the signal region are estimated using a fit procedure that is described in section \ref{sec:bkgfit}. 
The dominant uncertainties, after this procedure is applied, are reported in table \ref{tab:dominantUncert}.

\begin{table}[hbt]
\begin{center}
\small
\begin{tabular}{|l|c|c|c|c|}
    \hline
    Source & \multicolumn{4}{c|}{Relative systematic uncertainty (\%)}  \\\hline\hline
    & \multicolumn{3}{c|}{\bf Binned soft single-lepton} & {\bf Soft dimuon} \\
	   & {\bf 3-jet} & {\bf 5-jet} & {\bf 3-jet inclusive} & {\bf 2-jet} \\\hline
    Total systematic uncertainty & 20 & 24 & 17 & 43 \\  
    Lepton identification    & {$-$}  & {$-$}  & {5} & {$-$} \\
    JER       & {6}  &$-${} & {$-$}& {$-$}\\
    JES (flavour composition)   & {$-$} & {$-$} & {$-$}& {5}  \\
    Fake leptons  & {10} & {6} & {5}   & {40}  \\
    \ttbar~MC generator  & {11}  & {9} & {7} & {8} \\
    \ttbar~parton shower  & {$-$} & {19} & {$-$} & {$-$}\\
    \ttbar~scales, ISR and FSR          & {$-$} & {$-$} & {9} &  {5}  \\
    \ttbar~normalisation & {$-$} & {7} & {$-$} & {$-$} \\
    MC statistics & {8} &{$-$} & {6}  & {7} \\

    \hline

    & \multicolumn{3}{c|}{\bf Binned hard single-lepton} &\multicolumn{1}{c}{} \\
    & {\bf 3-jet} & {\bf 5-jet} & {\bf 6-jet} &\multicolumn{1}{c}{}\\\cline{1-4}
    Total systematic uncertainty & 9 & 22 & 24 \\
    \ttbar~MC generator                                     & {$-$} & {9}  & {23}&\multicolumn{1}{c}{}\\
    \ttbar~parton shower                 & $-$  & {17}   &$-$ &\multicolumn{1}{c}{} \\
    \ttbar~scales, ISR and FSR  & {$-$} & {7}      &$-$ &\multicolumn{1}{c}{} \\
    \ttbar~normalisation                 &  {5}   & {6}    &$-$ &\multicolumn{1}{c}{}  \\
    MC statistics 				    & {$-$}  & {5}      & 5 & \multicolumn{1}{c}{}  \\

    \cline{1-4}\hline
     & \multicolumn{4}{c|}{\bf Binned hard dilepton} \\
     & \multicolumn{2}{c|}{{\bf Low-multiplicity ($\le2$-jet) }} & \multicolumn{2}{c|}{{\bf 3-jet}}  \\
     & {\bf $ee/\mu\mu$} & {\bf $e\mu$} & {\bf $ee/\mu\mu$} & {\bf $e\mu$}\\\hline
    Total systematic uncertainty  & 11 & 11 & 23 & 18 \\
    $b$-tagging                   & {7} & {6} & {11} & {11}       \\
    JES (in-situ measurement)  &  {$-$}   &  {$-$}   & {$-$}     & {5} \\
    Fake leptons          & {5} & {$-$}    &  {$-$}    & {$-$}      \\
    MC statistics      & {6} &  {$-$}   &  {$-$}    & {$-$}      \\

    \hline
 \end{tabular}
 \caption{The main sources of systematic uncertainty on the SM background estimates for the various signal regions are shown and their value given as relative 
	 uncertainties (in \%) on the signal region event yields. The values are only shown if the 
	 relative uncertainty is at least 5\%.
    }
 \label{tab:dominantUncert}
\end{center}
\end{table}

The theoretical uncertainties related to the \ttbar~background are dominant for all the soft and hard single-lepton signal regions.
The fake-lepton background uncertainty is the main uncertainty for the soft dimuon signal region.
This uncertainty is also important in the other soft-lepton signal regions, reaching 5--10\% of the total event yield.
For the hard dilepton signal regions, the dominant uncertainties are those related to $b$-tagging.

\subsection{Theoretical uncertainties on the signal expectation}

The mUED model cross sections are based on a calculation at LO in QCD, and the events are generated with a leading order MC event generator.
No theoretical uncertainties on the acceptance are considered for this case.

Several theoretical uncertainties on the acceptance for the remaining signal models are taken into account. 
These uncertainties are estimated using \madgraph5+\pythia6 
samples for which the following parameters are varied up and down in turn by a factor of two: the \madgraph scale used to determine the event-by-event 
renormalisation and factorisation scale, the \madgraph parameter used to determine the scale for QCD radiation, the
\pythia parameter which controls the QCD scale value used for final-state radiation (the upward variation of this parameter is by a factor of 1.5) 
and the \madgraph parameter used for jet matching. This results in an uncertainty of $\mathcal{O}(5\%$--$25\%)$; it is larger for smaller mass 
differences in the decay cascade and for higher jet multiplicity regions.
The uncertainty on the modelling of initial-state radiation plays an important role
for small mass differences in the decay cascade in the simplified models.


For all models but mUED, the NLO+NLL cross-section uncertainty is taken from an envelope of cross-section predictions using 
different PDF sets and factorisation and renormalisation scales, as described in ref.~\cite{Kramer:2012bx}.
These uncertainties grow from 15\% at low ($\sim 500$~\GeV) squark and gluino masses up to 25\% at $m_{\squark}=900$~\GeV\ (30\% at $m_{\gluino}=1150$~\GeV.

\section{Background fit}
\label{sec:bkgfit}
As discussed in section~\ref{bkgestimate}, the background in the signal region is estimated with a fit based on the profile likelihood method \cite{Cowan:2010js} using the HistFitter \cite{histfitter} framework. 
The inputs to the fit, for each of the signal regions, are as follows:
\begin{enumerate}
  \item The number of events observed in each of the control regions, and the corresponding number of events expected from simulation. 
  \item The extrapolation factors (obtained from the simulation) which relate the number of predicted $W/Z$+jets or \ttbar~events in their associated 
  control region to that predicted in the signal region.
  \item The number of fake-lepton events in each region obtained with the data-driven method.
  \item The number of events predicted by the simulation in each region for the other backgrounds. 
\end{enumerate}

The numbers of observed and predicted events in each of these regions are described using Poisson probability density functions. 
There are two free parameters considered per signal region: a normalisation scale for the $W$+jets (or $Z$+jets) background and another scale  
for the \ttbar~ background. 
The other backgrounds are allowed to vary in the fit within their respective uncertainties. 
The systematic uncertainties (see section \ref{sec:systuncert}) and the MC statistical uncertainties on the expected values are included in the fit as nuisance parameters which are 
constrained by a Gaussian function with a width corresponding to the size of the uncertainty considered and a Poissonian function, respectively.
Correlations between these parameters are also taken into account. 

The product of the various probability density functions forms the likelihood, which
the fit maximises by adjusting the input parameters and the nuisance parameters described above.  
The fit may introduce a negative correlation between the \ttbar~and $W$+jets (or $Z$+jets) normalisation scales.
The relative uncertainty on the individual contributions may therefore increase, but the sum of the contributions is estimated more precisely: 
the total background relative uncertainty may then be smaller than the sum in quadrature of the individual components.

The background fit results are cross-checked in validation regions defined to be kinematically close to the signal regions but orthogonal to 
both the control and signal regions.
The data in the validation regions are not used to constrain the fits; they are
only used to compare the results of the fits to statistically independent observations.
The criteria used to define the validation regions are summarised in tables \ref{tab:softVR}--\ref{tab:hard2VR}. 
The validation regions are also illustrated in figures \ref{fig:softSR}--\ref{fig:razor_regions}.
As shown in table \ref{tab:softVR}, the soft single-lepton channel uses three validation regions to probe the \mt~and \met~extrapolations of each 
control region (3-jet, 5-jet or 3-jet inclusive $W$+jets and \ttbar~regions); there are therefore eighteen validation regions in the soft single-lepton channel.
In the soft dimuon case, shown in the same table, three validation regions are defined to cross-check the lepton \pt~and  
$m_{\mu\mu}$~extrapolations and the extrapolation from requiring one $b$-jet in the control region to none in the signal region. 
In the hard single-lepton channel, each signal region is associated with two validation regions in order to probe the \met~and 
the \mt~extrapolations independently, as shown in table~\ref{tab:hardVR}. Finally, as shown in table~\ref{tab:hard2VR}, in the hard dilepton channel, 
a total of twelve validation regions are defined to verify the extrapolation in the razor variable $R$ between the various control and signal regions.
\begin{table}[tb]
\begin{center}
\small
\tabcolsep=0.11cm
\begin{tabular}{|l|c|c|c|c|c|c|}
\hline
         & \multicolumn{3}{c|}{\bf Soft single-lepton} & \multicolumn{3}{c|}{\bf Soft dimuon} \\\hline
	 & { \mt } & { Interm. \met} & { High \met} & { Low $m_{\mu\mu}$} & { $b$-veto} & { Low \pt}  \\
         & { region} &{ region}      & { region}    & { region}      & { region } & { region}  \\\hline\hline
\ptl [\GeV]   & \multicolumn{3}{c|}{[10,25] (electron), [6,25] (muon)} & $>$25 & $>$25 & [6,25] \\
\hline
$N_{b\mathrm{-tag}}$  & \multicolumn{3}{c|}{$-$} & $-$ & 0 & $\geq 1$ \\
\hline
\met\ [\GeV]  & [180,250] & [250,350] ($>250$) & $>$350 ($>250$) & \multicolumn{3}{c|}{$>180$}  \\
\hline
\mt\ [\GeV]   & $>$80 ([80,120]) & [40,100] ([80,120]) & [40,100] & \multicolumn{3}{c|}{$>40$}  \\\hline
$m_{\mu\mu}$ [\GeV] & \multicolumn{3}{c|}{$-$} & $<60$ & $>60$ & $-$ \\\hline
\end{tabular}
\caption{Validation region definitions for the soft single-lepton and dilepton channels (see figure~\ref{fig:softSR}). 
	Only the variables for which the selection differs 
from the respective control region (see table~\ref{tab:softlepCR}) in at least one validation region are shown. When the 3-jet inclusive selection differs in \met~or in \mt~ from the 3-jet or 5-jet validation regions,
 the values are shown in parentheses.}
\label{tab:softVR}
\end{center}
\end{table}

\begin{table}[hbt]
\begin{center}
\small
\begin{tabular}{|l|c|c|c|c|c|c|}
\hline
 & \multicolumn{6}{c|}{\bf Hard single-lepton} \\\hline
 & \multicolumn{2}{c|}{\bf 3-jet} & \multicolumn{2}{c|}{\bf 5-jet} & \multicolumn{2}{c|}{\bf 6-jet} \\\hline\hline
 & {\bf \met} & {\bf \mt} & {\bf \met} & {\bf \mt} & {\bf \met} & {\bf \mt} \\
 & {\bf region} & {\bf region} & {\bf region} & {\bf region} & {\bf region} & {\bf region} \\\hline
\pt$^{\mathrm{jet}}$[\GeV] & \multicolumn{2}{c|}{$>$ 80, 80, 30} & \multicolumn{2}{c|}{$>$ 80, 50, 40, 40, 40} & \multicolumn{2}{c|}{$>$ 80, 50, 40, 40, 40, 40}\\\hline
Jet veto & \multicolumn{2}{c|}{\pt$^{\mathrm{5^{th} jet}}< 40$ \GeV} & \multicolumn{2}{c|}{\pt$^{\mathrm{6^{th} jet}}< 40$ \GeV} & \multicolumn{2}{c|}{$-$} \\\hline
$N_{b\mathrm{-tag}}$ & \multicolumn{6}{c|}{$-$} \\\hline
\met\ [\GeV]  & [300,500] & [150,300] & [300,500] & [150,300] & [250,500] & [150,250] \\\hline
\mt\ [\GeV]   & [60,150]  & [150,320] & [60,150]  & [150,320] & [60,150]  & [120,320] \\\hline
\end{tabular}
\caption{Validation region definitions for the hard single-lepton channel (see figure~\ref{fig:hardSR}).
 Only the variables for which the selection differs from the respective control region (see table~\ref{tab:hardlepCR}) in at least one validation region are shown.}
\label{tab:hardVR}
\end{center}
\end{table}

\begin{table}[hbt]
\begin{center}
\small
\begin{tabular}{|l|c|c|c|c|c|c|}
\hline
 & \multicolumn{6}{c|}{\bf Hard dilepton} \\\hline
 & \multicolumn{3}{c|}{{\bf Low-multiplicity}} & \multicolumn{3}{c|}{{\bf 3-jet}} \\\hline\hline
 & \multicolumn{2}{c|}{{\bf $ee/\mu\mu$}} & {\bf $e\mu$} & \multicolumn{2}{c|}{{\bf $ee/\mu\mu$}} & {\bf $e\mu$} \\\hline
 & $Z$ region & \ttbar~region & \ttbar~region &  $Z$ region & \ttbar~region & \ttbar~region  \\\hline\hline
$N_{\ell}$ & \multicolumn{6}{c|}{$2$ or $\geq 2$} \\\hline
$N_{b\mathrm{-tag}}$ & 0 &\multicolumn{2}{c|}{1} & 0 &\multicolumn{2}{c|}{1}\\\hline
$R$   & [0.25,1.0] & \multicolumn{2}{c|}{[0.5,1.0]} & [0.1,1.0] & \multicolumn{2}{c|}{[0.35,1.0]} \\\hline
$M_{R}^{'}$ [\GeV] & \multicolumn{3}{c|}{[200,400]} & \multicolumn{3}{c|}{[200,800]} \\\hline
$M_{R}^{'}$ bin width [\GeV] & \multicolumn{3}{c|}{50} & \multicolumn{3}{c|}{100} \\\hline
\end{tabular}
\caption{Validation region definitions for the hard dilepton channel (see figure~\ref{fig:razor_regions}). Only the criteria which differ from the respective control
region selections (see table~\ref{tab:hard2lepCR})  in at least one validation region are shown.
}
\label{tab:hard2VR}
\end{center}
\end{table}

The comparison of observed versus 
predicted event counts in the validation regions as obtained from the background-only fit 
are summarised in figure \ref{fig:pullplot}. 
More details on the
normalisation factors found after the fit are given in section~\ref{fitresults}.
Good agreement is seen between the predicted and observed yields in all regions.                
The systematic uncertainties are correlated across multiple
control regions in each channel, see section \ref{sec:systuncert}, and are thus correlated for the validation regions in each channel.
One of the main sources of systematic uncertainty in the validation regions is the theoretical uncertainty on the \ttbar~estimation related to the 
MC generator choice. The conservative uncertainty obtained from the comparison of \alpgen to \powheg is correlated across channels 
and could explain the somewhat small absolute values obtained for the pulls shown in this figure. Furthermore, the 3-jet inclusive soft single-lepton validation regions are not independent 
of the other soft single-lepton validation 
regions; the maximum overlap of $\sim 40$\% occurs between the high-\met~3-jet inclusive and 3-jet validation regions.  

\begin{figure}[!ht]
\centering
\includegraphics[width=0.98\textwidth]{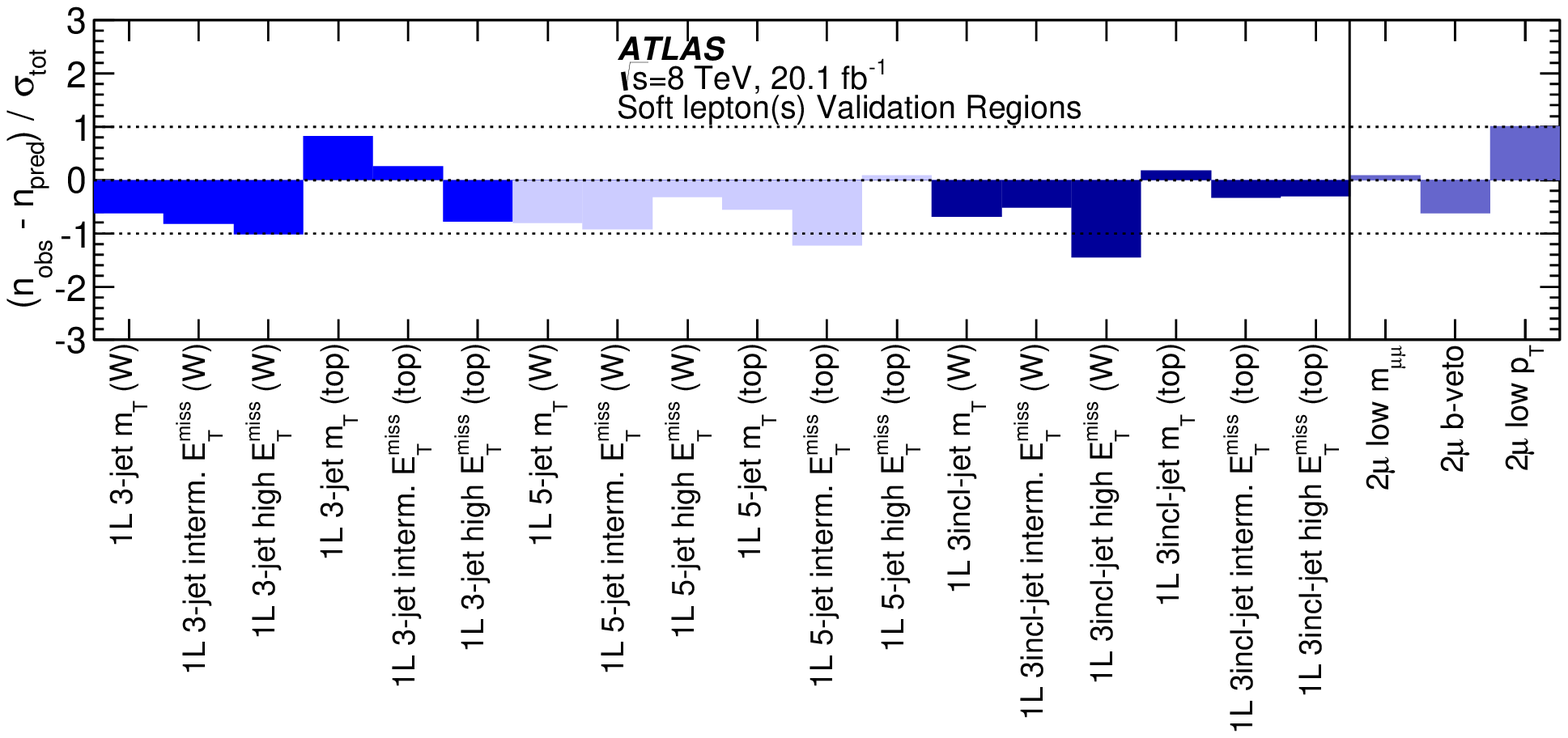}
\includegraphics[width=0.98\textwidth]{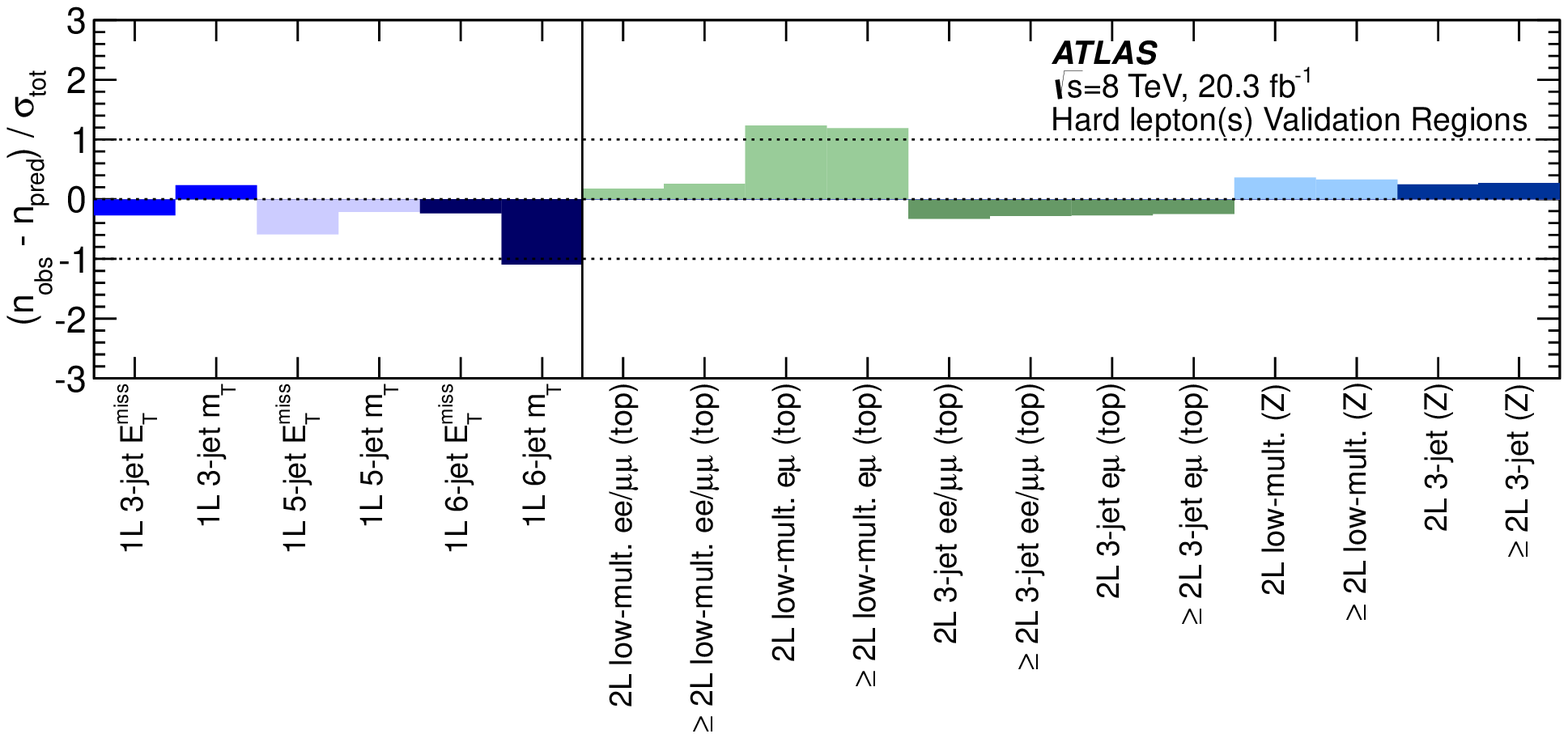}
\caption{Summary of the fit results: for each validation region (see tables \ref{tab:softVR}--\ref{tab:hard2VR}), 
the difference between the observed ($n_\mathrm{obs}$) and predicted ($n_\mathrm{pred}$) number of events, divided by the total 
(statistical and systematic) uncertainty on the prediction ($\sigma_{\mathrm{tot}}$), is shown. Soft-lepton (top) and hard-lepton (bottom) validation regions are shown.
The hard single-lepton validation regions are a mixture of $W$+jets and top processes. All single-lepton and soft dimuon validation regions, except the 3-jet inclusive soft-lepton validation regions, are statistically independent. The hard dilepton validation regions with exactly two or at least two leptons largely overlap.
The systematic uncertainties are partially correlated across multiple 
control regions defined for each of the four main channels (see section \ref{sec:systuncert}) and are thus correlated for the validation regions of each channel.
} \label{fig:pullplot}
\end{figure}

\section{Results and interpretation}
\label{sec:results}
\subsection{Background fit results and limits on the visible cross section}\label{fitresults}
The distributions of $\met/\meff$ in the soft single-lepton signal regions, 
of \met~in the soft dimuon signal region, of $m_{\mathrm{eff}}^{\mathrm{incl}}$ (\met) in the binned hard single-lepton 3-jet and 5-jet (6-jet) signal regions, 
and of $M_R'$ in the hard dilepton inclusive signal regions are shown in figures \ref{fig:SRplots_isr}--\ref{fig:InclSR_MR}.  The results of the fit in 
these signal regions are shown in tables \ref{table11nominal.results.systematics.in.logL.fit.table.results.yields}--\ref{table.results.yields.2l}.
The normalisation factors of the dominant backgrounds after the fit are found to be $\mathcal{O}(0.7$--$0.9)$ for \ttbar~(with the lowest value found in the soft dimuon 
channel), $\mathcal{O}(0.5$--$0.9)$ for $W$+jets (with the lowest values found in the high jet-multiplicity signal regions of the hard single-lepton 
channel) and $\mathcal{O}(0.9)$ for $Z$+jets. 
The ranges of values quoted for the $W/Z$+jets and \ttbar~background normalisation factors indicate that the event generators used 
(see table~\ref{tab:MC}) predict significantly larger rates for these processes in certain control regions than is observed in data. 
This observation is compatible with the differential cross-section measurements published by ATLAS for these processes with looser 
selection requirements~\cite{ttdiffxs1,ttdiffxs2,ZjetsXsec,WjetsXsec,RjetsXsec}. The inclusion of the systematic uncertainties as
nuisance parameters in the fit has a negligible effect on the central value of the fitted normalisation factors.

The number of events observed in all the signal regions
presented in this paper is consistent with the post-fit SM
expectations. The signal regions are not all independent: some
CRs are common to several SRs (e.g. binned and unbinned). The binned and unbinned signal regions overlap, 
and the 3-jet inclusive soft
single-lepton signal region overlaps with the other soft single-lepton signal
regions. The data and Monte Carlo distributions of the variables used in the
extrapolation from control to signal regions are in agreement within 
uncertainties.

\begin{figure}[hbt]
  \begin{center}
    \includegraphics[width=0.49\textwidth]{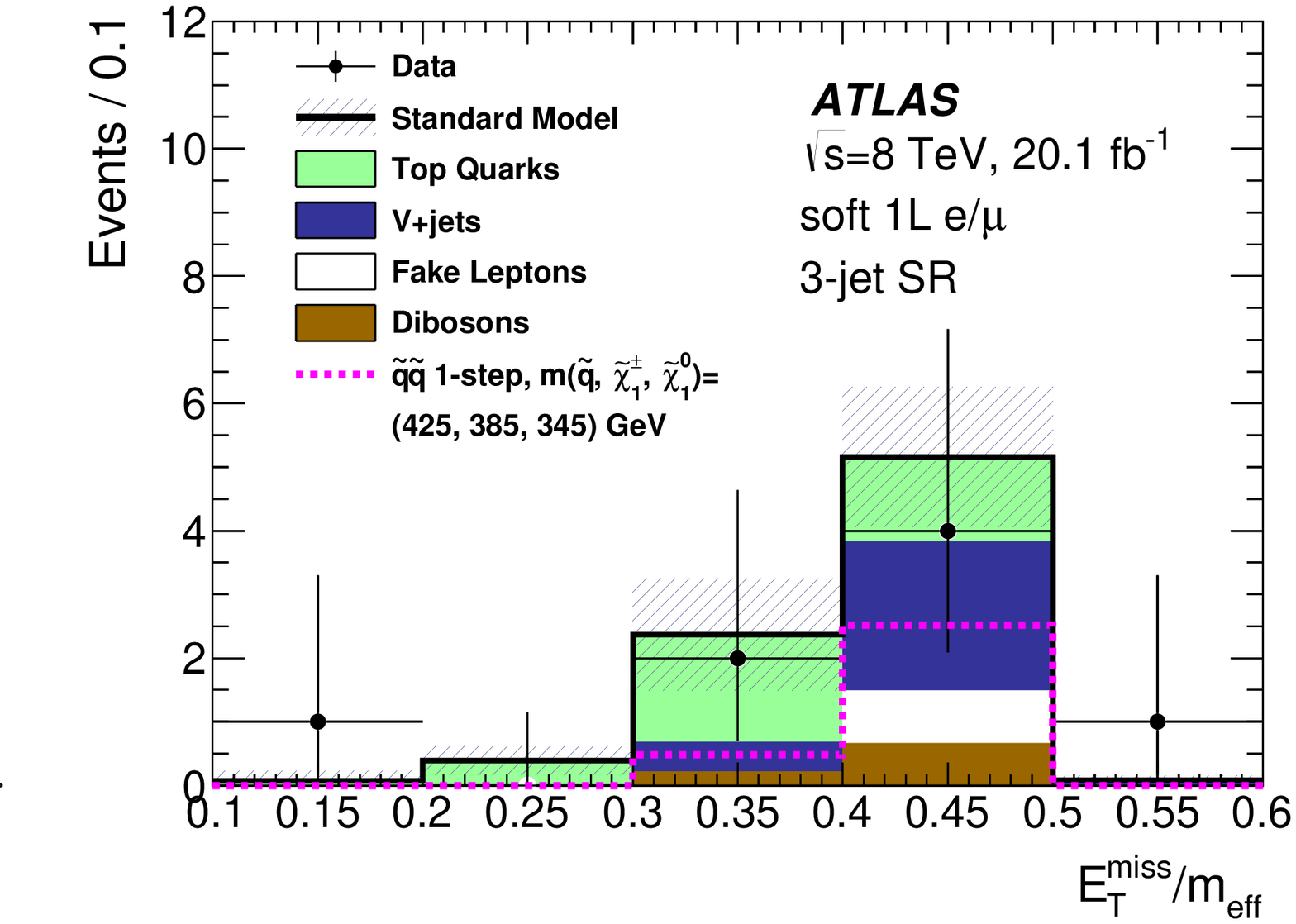}
    \includegraphics[width=0.49\textwidth]{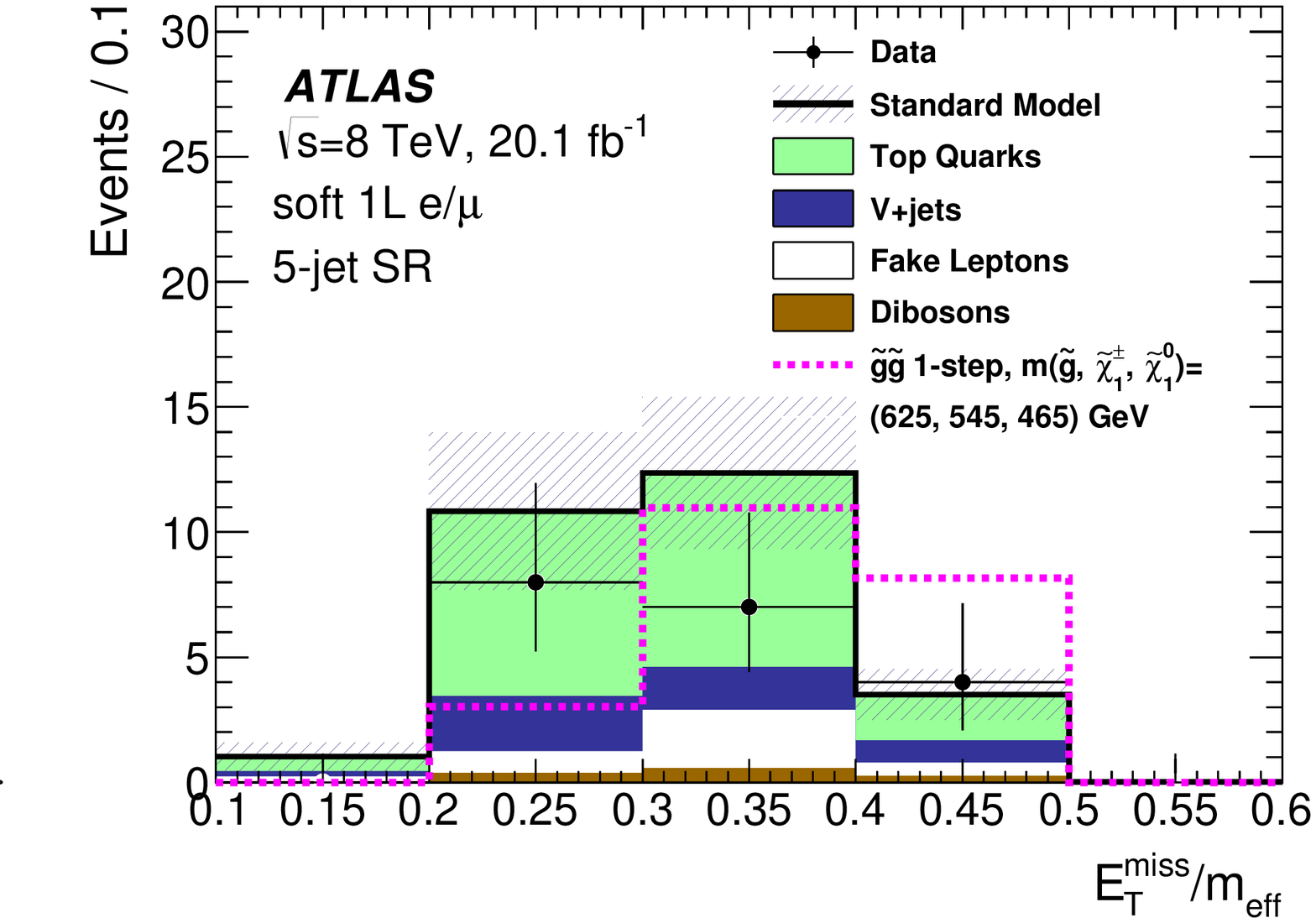}
    \includegraphics[width=0.49\textwidth]{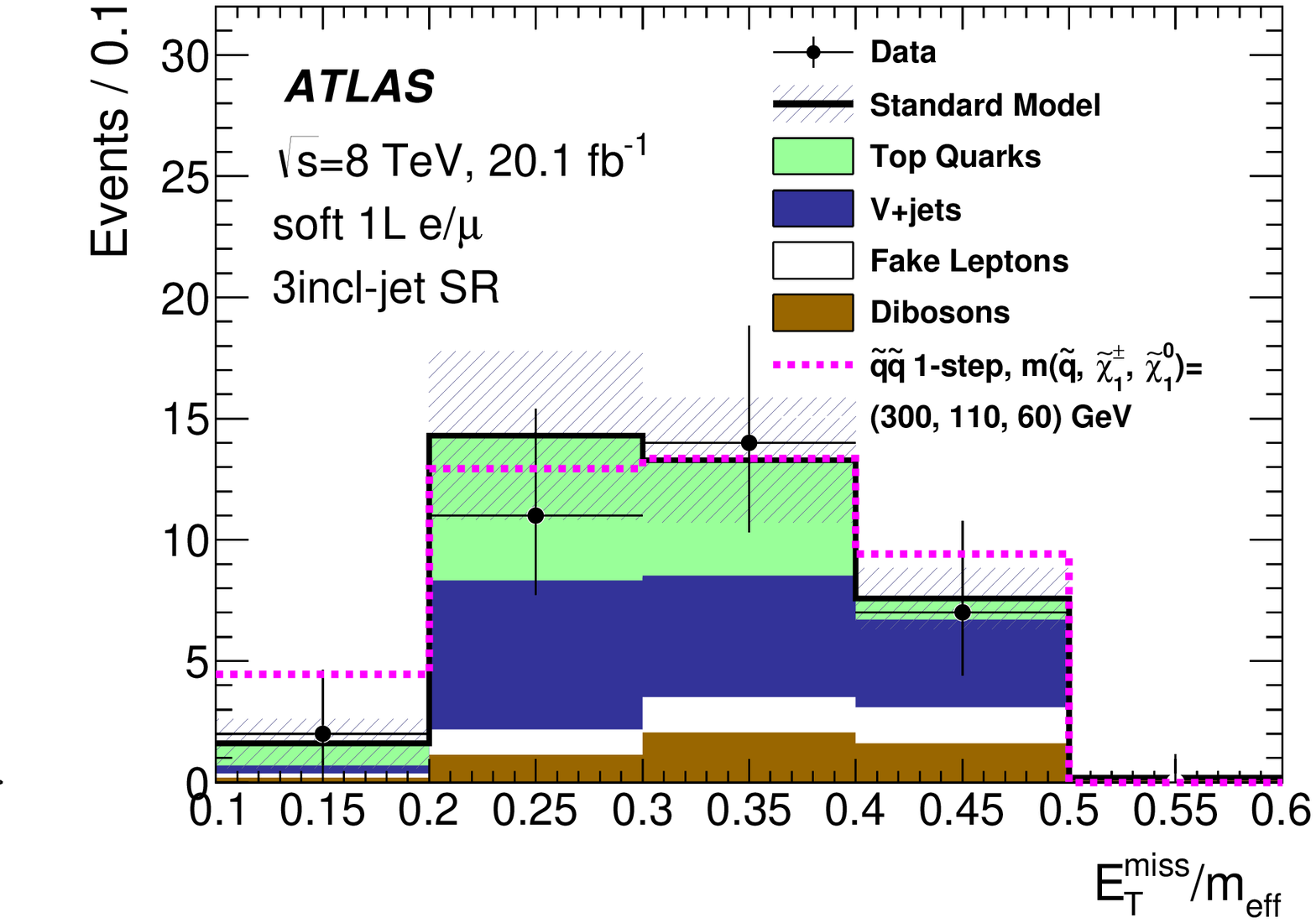}
    \includegraphics[width=0.49\textwidth]{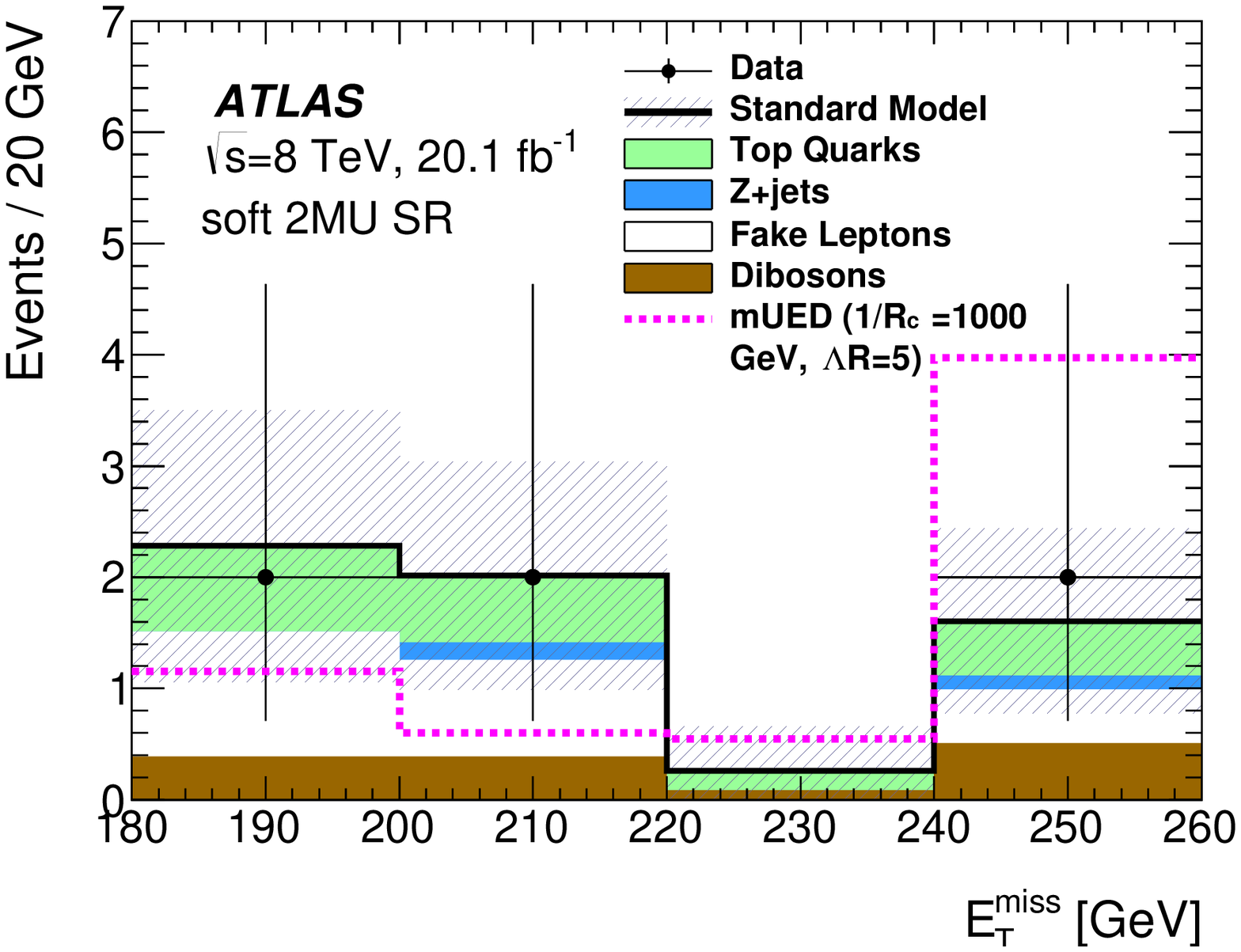}
  \end{center}
  \caption{Distribution of the ratio of the missing transverse momentum to the effective mass $\met/\meff$~in soft single-lepton 3-jet (upper left), 5-jet (upper right) and 3-jet inclusive (bottom left) signal regions and distribution of the missing transverse momentum \met~in the soft dimuon signal region (bottom right). The Standard Model expectation is derived from the fit. 
The uncertainty band on the Standard Model expectation shown here combines the statistical uncertainty on the simulated event samples and the 
 systematic uncertainties. The last bin includes the overflow. 
 The ``Top Quarks'' include \ttbar, single top, $t+Z$, \ttbar+$W$ and \ttbar+$WW$, while ``V+jets'' 
 includes $W$+jets, $Z$+jets and Drell--Yan contributions. For illustration, the expected signal
distributions are shown for three signal benchmark points: $(m_{\squark} , m_{\chinoonepm} , m_{\ninoone} ) = (425 , 385 , 345) \GeV$, $(m_{\gluino} , m_{\chinoonepm} , m_{\ninoone} )=(625 , 545 , 465) \GeV$ and $(m_{\squark} , m_{\chinoonepm} , m_{\ninoone} ) = (300 , 110 , 60) \GeV$ in the single-lepton channel and for
the mUED model point with $R_{\mathrm{c}}^{-1}$ = 1000 \GeV~and $\Lambda R_{\mathrm{c}}$=5 in the dimuon channel.
}
  \label{fig:SRplots_isr}
\end{figure}

\begin{figure}[htb]
\centering
\includegraphics[width=0.49\textwidth]{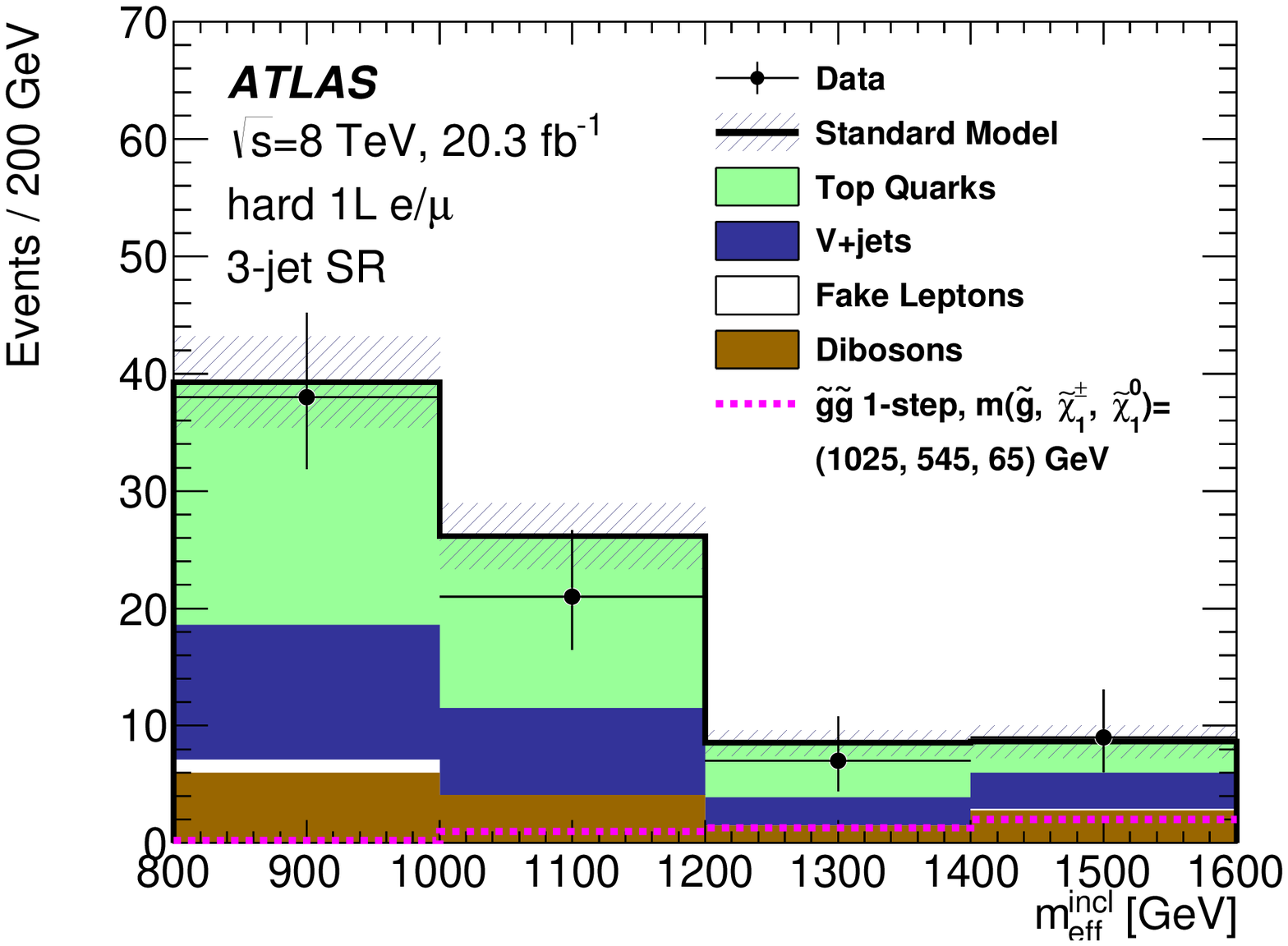}
\includegraphics[width=0.49\textwidth]{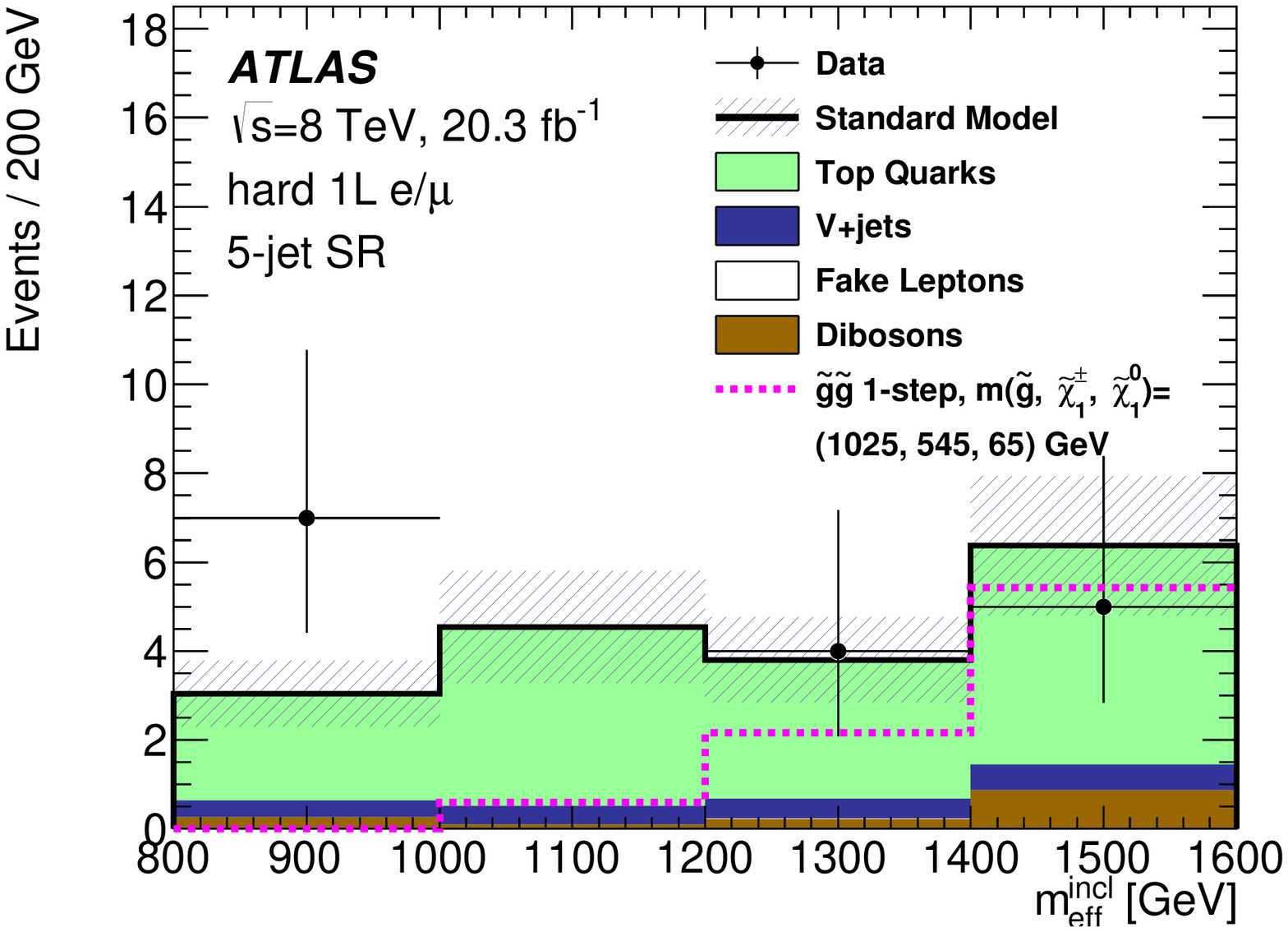}
\includegraphics[width=0.49\textwidth]{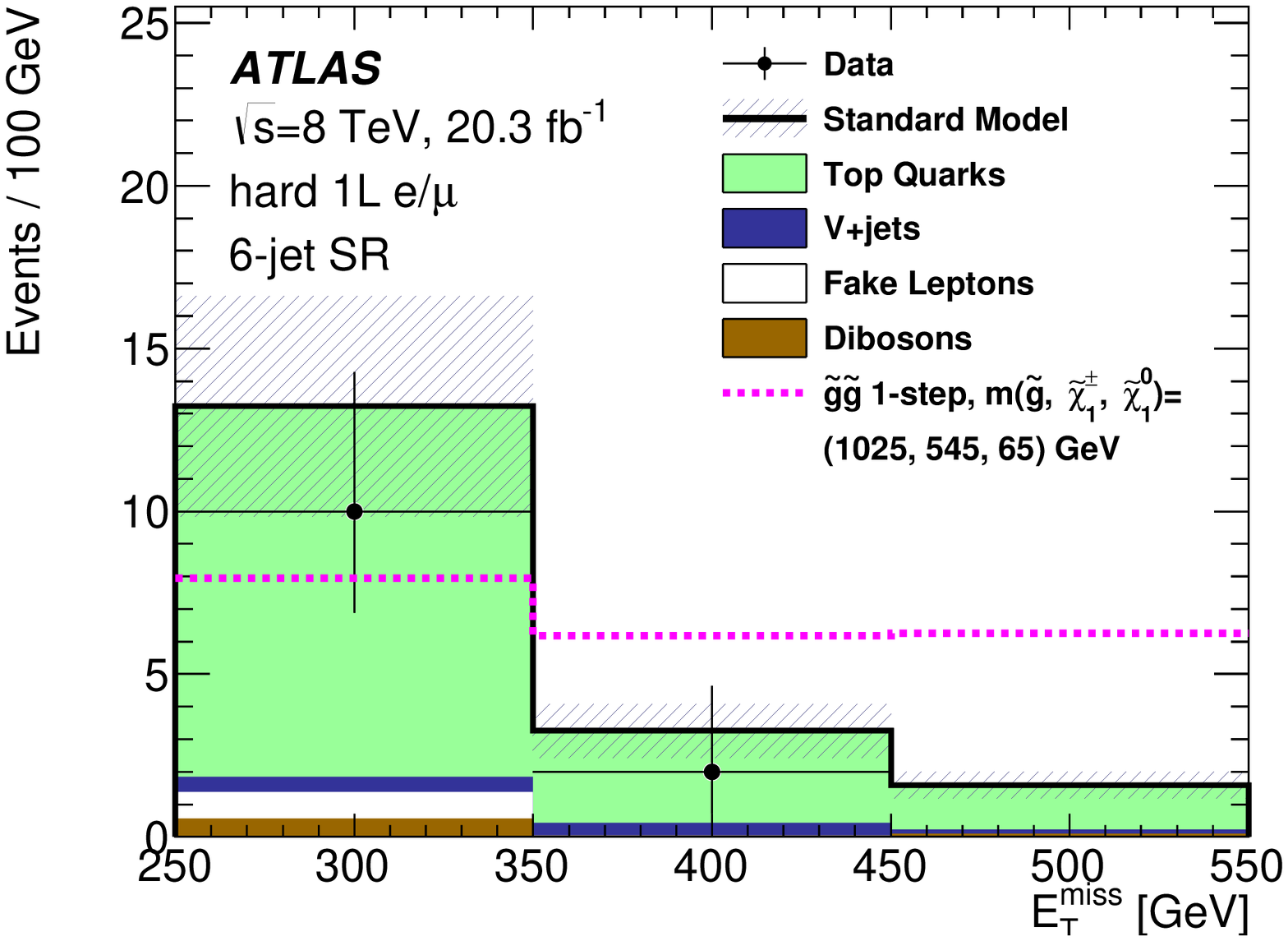}

\caption{Distribution of the inclusive effective mass $m_{\mathrm{eff}}^{\mathrm{incl}}$~in the hard single-lepton 3-jet (upper left) and 5-jet (upper right) binned signal regions 
and distribution of the missing transverse momentum \met~ in the hard single-lepton 6-jet binned signal region (bottom). 
The Standard Model expectation is derived from the fit. The uncertainty band on the Standard Model expectation shown here
combines the statistical uncertainty on the simulated event samples and the systematic uncertainties.
The last bin includes the overflow. The ``Top Quarks'' include \ttbar, single top, $t+Z$, \ttbar+$W$ and \ttbar+$WW$, while ``V+jets'' 
includes $W$+jets, $Z$+jets and Drell--Yan contributions.
For illustration, the expected signal
distributions are shown for gluino pair production 
 with $m_{\tilde{g}}=1025 \GeV, m_{\tilde{\chi}^{\pm}_{1}}=545 \GeV$ and $m_{\tilde{\chi}^0_1}=65 \GeV$.}
\label{fig:SRexclAfterFit}
\end{figure}

  \begin{figure}[hbt]
   \centering
   \includegraphics[width=0.49\textwidth]{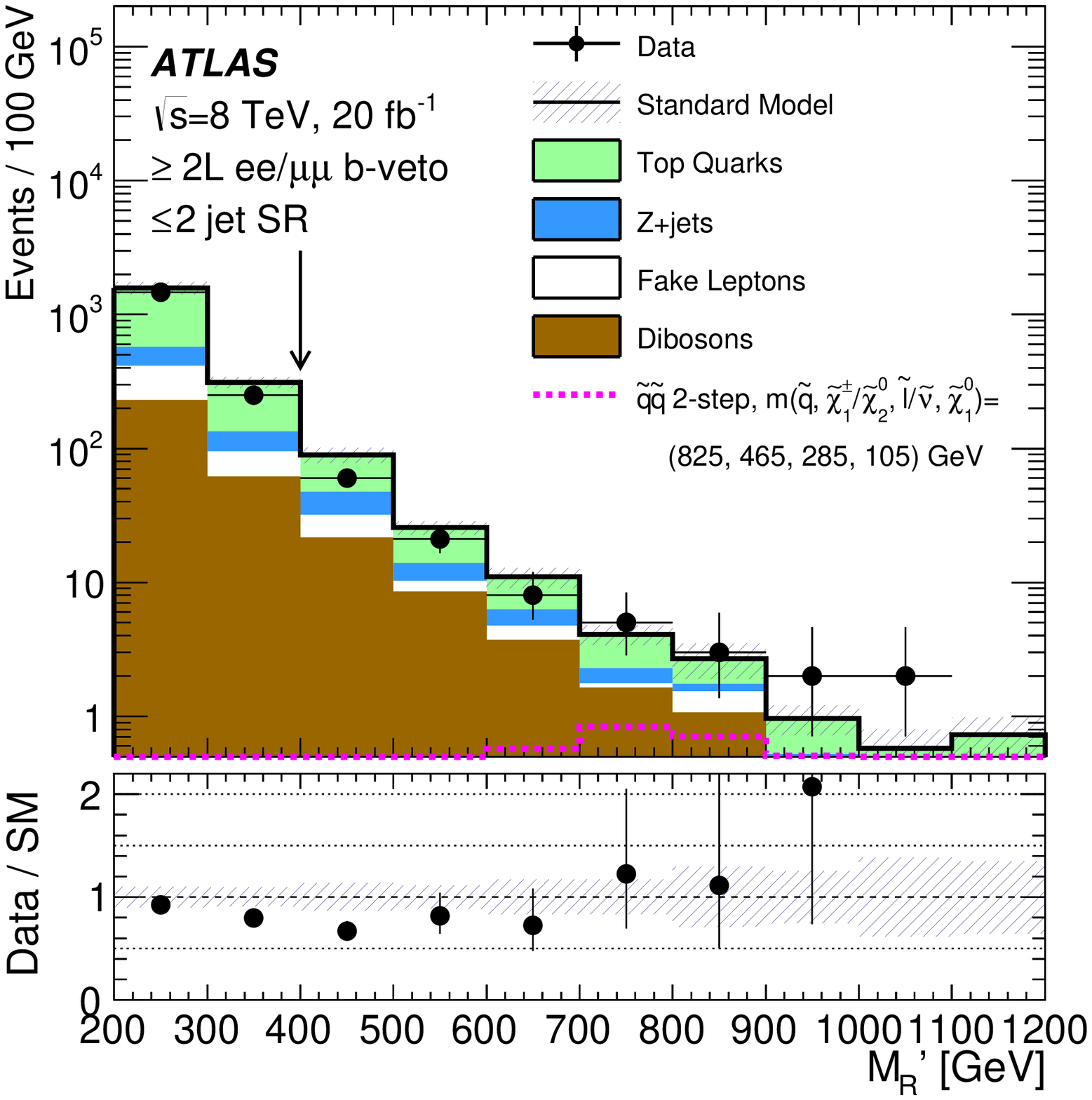}  
   \includegraphics[width=0.49\textwidth]{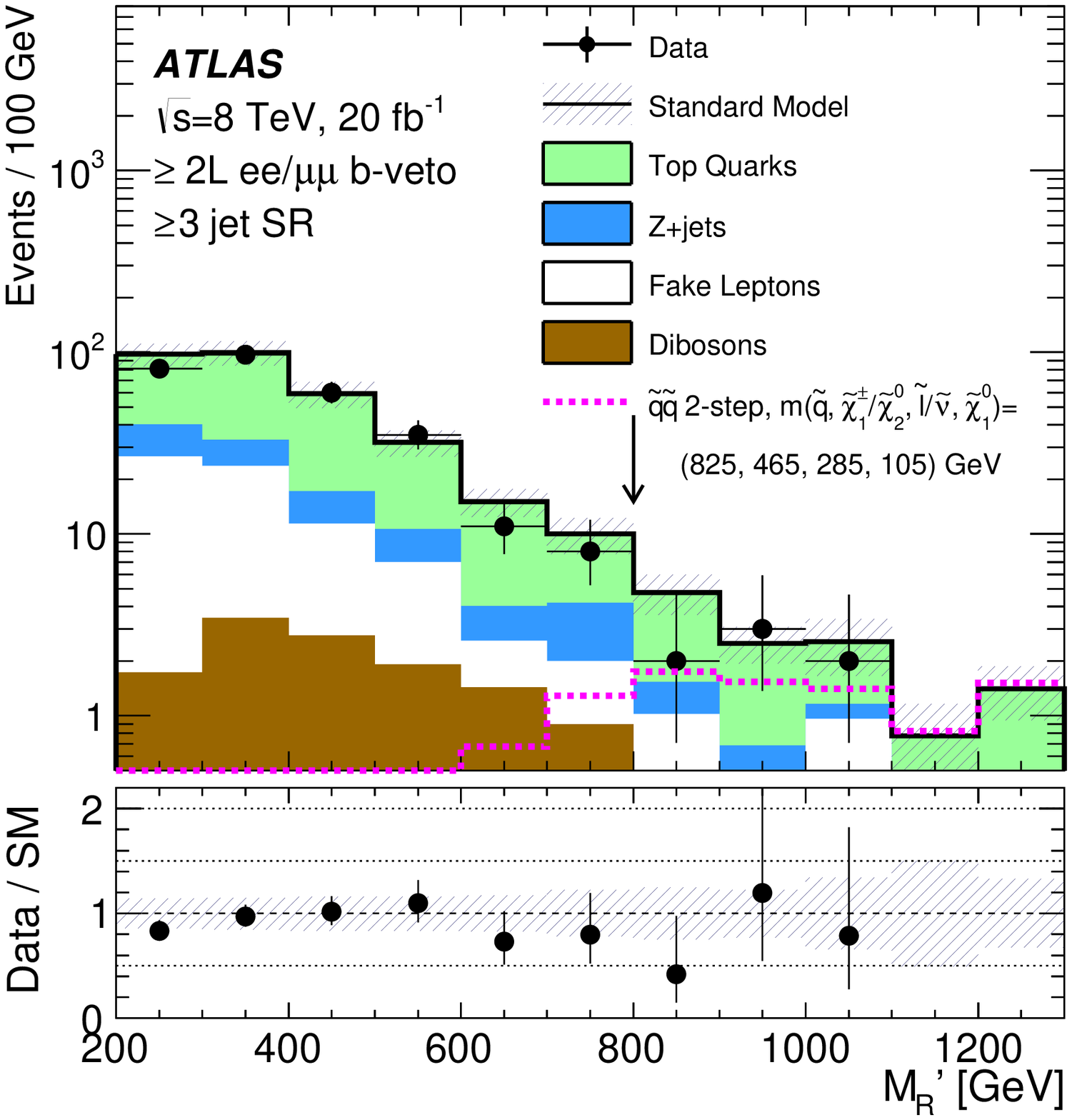}
   \includegraphics[width=0.49\textwidth]{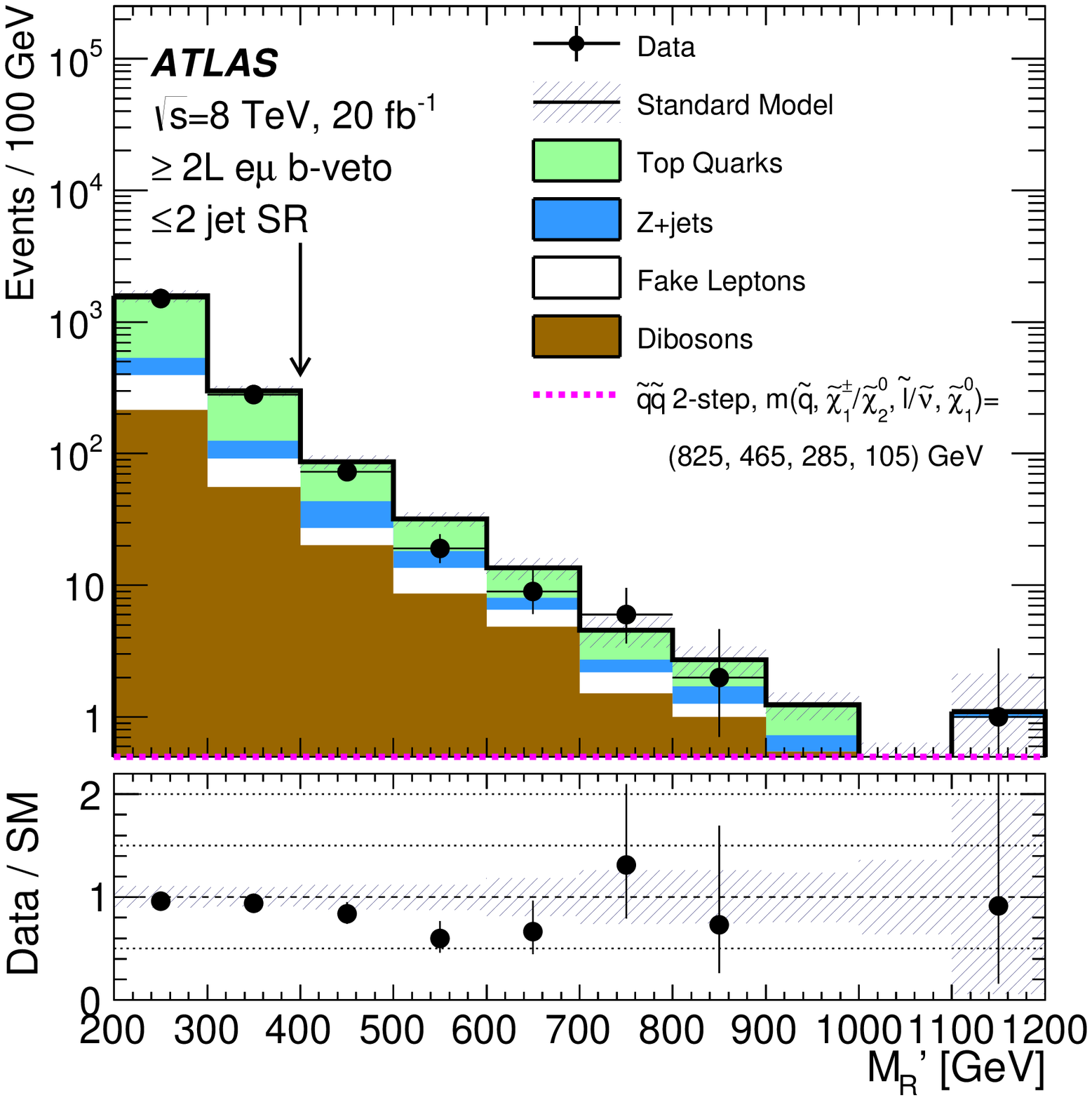}
   \includegraphics[width=0.49\textwidth]{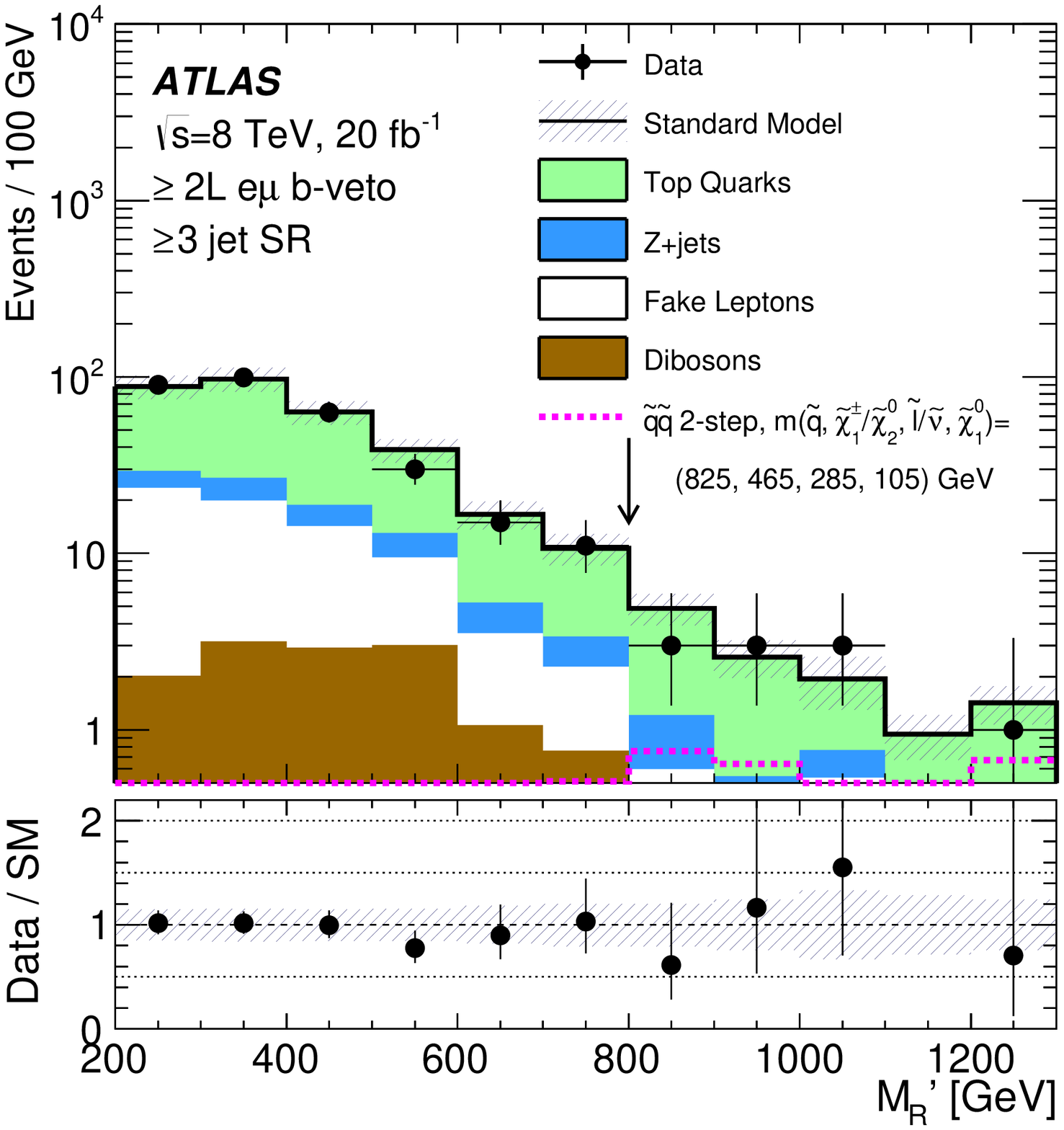}  
   \caption{Distribution of the $R$-frame mass $M_R'$ in the inclusive hard dilepton low-multiplicity (left) and 3-jet (right) signal regions for the same-flavour (top) and opposite-flavour (bottom) pairs.
The arrow indicates the requirement defining the signal region. The Standard Model expectation is derived from the fit. The uncertainty band on the Standard Model expectation shown here
combines the statistical uncertainty on the simulated event samples and the systematic uncertainties. The last bin includes the overflow.
The ``Top Quarks'' include \ttbar, single top, $t+Z$, \ttbar+$W$ and \ttbar+$WW$, while ``V+jets'' 
includes $W$+jets, $Z$+jets and Drell--Yan contributions. For illustration, the expected signal
distributions are shown for squark pair production with $m_{\tilde{q}}=825 \GeV, m_{\tilde{\chi}^{\pm}_{1}/\tilde{\chi}^{0}_{2}}=465 \GeV, m_{\tilde{\ell}/\tilde{\nu}}=285 \GeV$ and $m_{\tilde{\chi}^0_1}=105 \GeV$.
}
   \label{fig:InclSR_MR}
 \end{figure}

\begin{table}[hbt]
\begin{center}
\setlength{\tabcolsep}{0.0pc}
{\scriptsize
\begin{tabular*}{\textwidth}{@{\extracolsep{\fill}}lrrrrrr}
\noalign{\smallskip}\hline\noalign{\smallskip}
& \multicolumn{5}{c}{Soft single-lepton} & Soft dimuon\\
& 3-jet incl. & \multicolumn{2}{c}{3-jet} & \multicolumn{2}{c}{5-jet}        &       \\[-0.05cm]
&             & \multicolumn{1}{c}{Single-bin} & \multicolumn{1}{c}{Binned} & \multicolumn{1}{c}{Single-bin} & \multicolumn{1}{c}{Binned} &  \\[-0.05cm]

\noalign{\smallskip}\hline\noalign{\smallskip}
Observed events          & $34$  & 7 & 8 & $11$ & $19$ & $6$\\
\noalign{\smallskip}\hline\noalign{\smallskip}
Fitted background events         & $37.0 \pm 6.3$ & $7.5 \pm 1.4$ & $7.9 \pm 1.6$ & $15.9 \pm 3.7$ & $27.7 \pm 6.7$ & $6.0 \pm 2.6$           \\
\noalign{\smallskip}\hline\noalign{\smallskip}
        \ttbar         & $11.2 \pm 4.7$ & $2.0 \pm 0.9$ & $2.4 \pm 1.1$ & $8.5 \pm 3.4$  & $16.0 \pm 6.4$ & $1.8 \pm 0.8$\\
       Other top quarks &  $1.4 \pm 0.9$ & $0.96 \pm 0.31$ & $1.0 \pm 0.3$ & $1.1 \pm 0.4$ & $1.6 \pm 0.6$ & $0.24 \pm 0.14$ \\
       V+jets         & $15.2 \pm 2.8$ & $2.9 \pm 0.5$ & $2.9 \pm 0.5$ & $2.6 \pm 0.6$ & $5.0 \pm 1.2$  & $0.28 \pm 0.19$\\
       Diboson         & $5.1 \pm 1.2$ & $0.89 \pm 0.36$ & $0.91 \pm 0.36$ & $0.83 \pm 0.50$ & $1.3 \pm 0.7$ & $1.4 \pm 0.5$\\
       Fake leptons     & $4.2 \pm 1.9$ & $0.75_{-0.75}^{+0.78}$ & $0.64_{-0.64}^{+0.76}$ & $2.9 \pm 1.2$ & $3.9 \pm 1.7$  & $2.3_{-2.3}^{+2.4}$ \\
 \noalign{\smallskip}\hline\noalign{\smallskip}
Expected background events before the fit             & $43.1$  & $8.4$ & 8.8 & 18.3 & 32.3 & 6.8          \\
\noalign{\smallskip}\hline\noalign{\smallskip}
        \ttbar      & 14.9 & 2.4 & 2.9 & $10.2$ & 19.0 & 2.6 \\
        Other top quarks  & 1.4 & 0.96 & 1.0 & 1.1 & 1.6 & 0.24 \\
        V+jets         & 17.6 & 3.4 & 3.4 & 3.4 & 6.5 & 0.28 \\
        Diboson         & 5.1 & 0.89 & 0.91 & 0.83 & 1.3 & 1.4 \\
        Fake leptons & 4.2 & 0.75 & 0.64 & 2.9 & 3.9 & 2.3 \\
\noalign{\smallskip}\hline\noalign{\smallskip}
\end{tabular*}
}
\end{center}
\caption{ 
Background fit results (top) for the soft single-lepton and soft dimuon signal regions, for an integrated luminosity of $20.1$~\ifb. The V+jets events contain $W$+jets and $Z$+jets events.
The background expectations before the fit (bottom) are given for comparison 
(see section~\ref{bkgestimate} for a detailed description of how each background source is estimated). 
The uncertainties shown here combine the statistical uncertainty on the simulated event samples with the systematic uncertainties.}
\label{table11nominal.results.systematics.in.logL.fit.table.results.yields}
\end{table}

\begin{table}
\begin{center}
\setlength{\tabcolsep}{0.0pc}
{\scriptsize
\begin{tabular*}{\textwidth}{@{\extracolsep{\fill}}lrrrrrr}
\noalign{\smallskip}\hline\noalign{\smallskip}
           & \multicolumn{6}{c}{Hard single-lepton} \\
           & \multicolumn{2}{c}{3-jet} & \multicolumn{2}{c}{5-jet} & \multicolumn{2}{c}{6-jet}             \\
	   & \multicolumn{1}{c}{Single-bin} & \multicolumn{1}{c}{Binned} & \multicolumn{1}{c}{Single-bin} & \multicolumn{1}{c}{Binned} & \multicolumn{1}{c}{Single-bin} & \multicolumn{1}{c}{Binned} \\[-0.05cm] 
\noalign{\smallskip}\hline\noalign{\smallskip}
Observed events          & $9$              & $75$              & $5$              & $16$              & $2$              & $12$                    \\
\noalign{\smallskip}\hline\noalign{\smallskip}
Fitted background events         & $8.5\pm 1.4$          & $82.5 \pm 7.2$          & $7.3 \pm 1.7$          & $17.7 \pm 4.0$          & $4.9 \pm 1.1$          & $18.1 \pm 4.3$              \\
\noalign{\smallskip}\hline\noalign{\smallskip}
        \ttbar      & $2.2 \pm 0.5$          & $35.0 \pm 6.2$          & $4.8 \pm 1.6$          & $12.3 \pm 4.1$          & $3.7 \pm 1.3$          & $13.9 \pm 4.7$              \\
        Other top quarks         & $0.79 \pm 0.35$          & $7.6 \pm 3.0$          & $0.71 \pm 0.18$          & $2.1 \pm 0.5$          & $0.54 \pm 0.18$          & $1.7 \pm 0.5$              \\
        V+jets          & $2.5 \pm 0.4$          & $24.4 \pm 3.6$          & $0.80 \pm 0.28$          & $1.8 \pm 0.6$          & $0.5 \pm 0.4$          & $0.99 \pm 0.80$              \\
        Diboson         & $2.9 \pm 1.0$          & $14.3 \pm 4.3$          & $0.96 \pm 0.69$          & $1.5 \pm 1.0$          & $0.14 \pm 0.07$          & $0.70 \pm 0.36$              \\
        Fake leptons         & $0.09_{-0.09}^{+0.15}$          & $1.2_{-1.2}^{+1.3}$          & $0.00 _{-0.00}^{+0.01}$          & $0.00_{-0.00}^{+0.09}$          & $0.00 \pm 0.00$          & $0.82_{-0.82}^{+0.87}$              \\
 \noalign{\smallskip}\hline\noalign{\smallskip}
Expected background events before the fit              & $10.1$          & $104.4$          & $9.5$          & $23.2$          & $6.2$          & $22.3$              \\
\noalign{\smallskip}\hline\noalign{\smallskip}
        \ttbar       & $3.1$          & $49.3$          & $6.5$          & $16.5$          & $4.5$          & $17.3$              \\
        Other top quarks         & $0.79$          & $7.6$          & $0.7$          & $2.1$          & $0.54$          & $1.7$              \\
        V+jets          & $3.3$          & $32.0$          & $1.3$          & $3.1$          & $1.0$          & $1.9$              \\
        Diboson         & $2.9$          & $14.3$          & $0.96$          & $1.5$          & $0.14$          & $0.70$              \\
        Fake leptons         & $0.09$          & $1.2$          & $0.00$          & $0.00$          & $0.00$          & $0.82$              \\
\noalign{\smallskip}\hline\noalign{\smallskip}
\end{tabular*}
}
\end{center}
\caption{ 
Background fit results (top) for the hard single-lepton signal regions, for an integrated luminosity of $20.3$~\ifb. The V+jets events contain 
$W$+jets and $Z$+jets events.
The background expectations before the fit (bottom) are given for comparison %
(see section~\ref{bkgestimate} for a detailed description of how each background source is estimated). 
The uncertainties shown here combine the statistical uncertainty on the simulated event samples with the systematic uncertainties.
}
\label{table.results.yields.fit.h1L_SR3JEM_h1L_SR5JEM_h1L_SR6JEM}
\end{table}

\begin{table}
\begin{center}
\setlength{\tabcolsep}{0.0pc}
{\scriptsize 
\begin{tabular*}{\textwidth}{@{\extracolsep{\fill}}lrrrrrr}
\noalign{\smallskip}\hline\noalign{\smallskip}
         & \multicolumn{6}{c}{Hard dilepton inclusive} \\
         & \multicolumn{4}{c}{Low-multiplicity} & \multicolumn{2}{c}{3-jet} \\
         & \multicolumn{2}{c}{$ee/\mu\mu$}           & \multicolumn{2}{c}{$e\mu$}         & \multicolumn{1}{c}{$ee/\mu\mu$}        & \multicolumn{1}{c}{$e\mu$}            \\
         & \multicolumn{1}{c}{Single-bin} & \multicolumn{1}{c}{Binned} & \multicolumn{1}{c}{Single-bin} & \multicolumn{1}{c}{Binned} & & \\[-0.05cm]
\noalign{\smallskip}\hline\noalign{\smallskip}
Observed events            & $20$ & $101$              & $18$ & $110$              & $7$              & $10$                    \\
\noalign{\smallskip}\hline\noalign{\smallskip}
Fitted background events   & $20.0 \pm 3.3$ & $136 \pm 15$  & $24 \pm 4$  & $141 \pm 16$          & $11.8 \pm 2.6$          & $11.5 \pm 2.0$              \\
\noalign{\smallskip}\hline\noalign{\smallskip}
        $t\bar{t}$         & $6.7 \pm 1.3$  & $53 \pm 9$    & $7.7 \pm 1.5$         & $54 \pm 9$          & $5.9 \pm 1.5$          & $7.0 \pm 1.6$       \\
        Other top quarks   & $1.7 \pm 0.7$ & $10.8 \pm 1.8$    & $2.1 \pm 0.7$       & $11.1 \pm 2.1$          & $1.7 \pm 0.3$          & $1.4 \pm 0.3$          \\
        Diboson            & $7.4 \pm 2.4$  & $38 \pm 5$    & $8.5 \pm 2.6$        & $37 \pm 5$          & $1.3 \pm 0.2$          & $1.0 \pm 0.3$         \\
        $Z+$jets           & $2.6 \pm 0.4$ & $21 \pm 4$  & $2.8 \pm 1.4$        & $24 \pm 4$          & $1.3 \pm 0.3$          & $1.4 \pm 0.4$            \\
        Fake leptons       & $1.6 \pm 1.4$ & $13 \pm 7$  & $3.2 \pm 2.1$        & $15 \pm 9$          & $1.5_{-1.5}^{+2.0}$     & $0.6_{-0.6}^{+0.9}$      \\
 \noalign{\smallskip}\hline\noalign{\smallskip}
Expected background events before the fit    & $20.1$ & $144$  & $25$          & $150$          & $14.6$          & $14.2$              \\
\noalign{\smallskip}\hline\noalign{\smallskip}
        $t\bar{t}$         & $6.2$ & $50$   & $6.8$        & $56$          & $8.4$          & $9.4$              \\
        Other top quarks   & $1.5$ & $10.1$   & $2.1$        & $10.3$          & $1.8$          & $1.4$              \\
        Diboson            & $7.4$ & $40$   & $9.0$           & $38$          & $1.3$          & $1.1$              \\
        $Z+$jets           & $3.3$ & $30$   & $4.1$        & $31$          & $1.6$          & $1.6$              \\
        Fake leptons       & $1.6$ & $13$   & $3.2$        & $15$          & $1.5$          & $0.6$              \\
\noalign{\smallskip}\hline\noalign{\smallskip}
\end{tabular*}
}
\end{center}
\caption{Background fit results (top) for the inclusive hard dilepton signal regions, for an integrated luminosity of $20.3$~fb$^{-1}$.
        The background expectations before the fit (bottom) are given for comparison 
	(see section~\ref{bkgestimate} for a detailed description of how each background source is estimated).
  The uncertainties shown combine the statistical and systematic uncertainties. The 3-jet binned signal region uses the
same event selection as the 3-jet single-bin signal region (see table \ref{tab:hard2lSR}); therefore it has the same event yields.}
\label{table.results.yields.2l}
\end{table}

The observed and expected upper limits at 95\% confidence level (CL) on the number of events from new phenomena beyond the SM (BSM) for each signal region ($S_{\rm obs}^{95}$ and $S_{\rm exp}^{95}$) are derived using the $CL_S$ prescription \cite{Read:2002hq} 
and neglecting any
possible signal contamination in the control regions; an uncertainty on $S_{\rm exp}^{95}$ is also computed from the $\pm 1\sigma$
uncertainty on the expectation.
The observed upper limits, normalised by the integrated luminosity of the data sample, can be interpreted as upper
limits on the visible BSM cross-section, ($\langle\epsilon\sigma\rangle_{\rm obs}^{95}$), where the visible cross section is defined as the product of acceptance, 
selection efficiency and production cross-section. The results obtained using asymptotic formulae \cite{Cowan:2010js} are given in
table \ref{tab:upperlimitALL} for the single-bin signal regions.\footnote{The method was cross-checked with pseudo experiments for various configurations and the results were found 
to be consistent.} For each signal region, the results are also shown 
split by lepton flavour to allow for comparison 
with new models which could favour a specific lepton flavour.
In this table the value of $CL_B$,  the confidence level observed for the background-only hypothesis, and the one-sided discovery 
p-value, $p(s=0)$, are also given; the latter represents, for each 
signal region, the probability that the event yield obtained in a single hypothetical background-only experiment (signal $s=0$) is greater than that observed in this dataset. $CL_B$ and $p(s=0)$ are
obtained using different test statistic definitions: the former has a conditional likelihood
calculated with a signal strength $\mu$ set to $1$, while the latter uses
$\mu=0$.
For an observed number of events lower than expected, the discovery p-value is truncated at 0.5.

\begin{table}
\begin{center}
\scriptsize
\setlength{\tabcolsep}{0.0pc}
\begin{tabular*}{\textwidth}{@{\extracolsep{\fill}}lccccc}
\noalign{\smallskip}\hline\noalign{\smallskip}
{\bf Signal region}                        & $\langle\epsilon{\rm \sigma}\rangle_{\rm obs}^{95}$[fb]  &  $S_{\rm obs}^{95}$  & $S_{\rm exp}^{95}$ &
$CL_{B}$ & $p(s=0)$  \\
\noalign{\smallskip}\hline\noalign{\smallskip}
\multicolumn{6}{l}{Soft single-lepton channel}\\
\noalign{\smallskip}\hline\noalign{\smallskip}
3-jet inclusive & $0.77$ &  $15.6$  & ${17.3}^{+6.9}_{-4.9}$ & $0.37$ &  $0.50$  \\
3-jet inclusive (electron) & $0.55$ &  $11.1$  & ${10.8}^{+4.8}_{-3.0}$ & $0.52$ &  $0.48$  \\
3-jet inclusive (muon) & $0.52$ &  $10.5$  & ${12.4}^{+5.2}_{-3.4}$ & $0.31$ &  $0.50$  \\
3-jet      & $0.35$ &  $7.0$  & ${7.5}^{+3.3}_{-2.2}$ & $0.44$ &  $0.50$  \\
3-jet (electron)  & $0.26$ &  $5.2$  & ${5.7}^{+2.6}_{-1.8}$ & $0.40$ &  $0.50$  \\
3-jet (muon)  & $0.29$ &  $5.9$  & ${5.8}^{+2.6}_{-1.8}$ & $0.52$ &  $0.48$  \\
5-jet   & $0.39$ &  $7.9$  & ${10.4}^{+4.4}_{-2.8}$ & $0.20$ &  $0.50$  \\
5-jet (electron)     & $0.38$ &  $7.7$  & ${7.6}^{+3.4}_{-2.2}$ & $0.52$ &  $0.48$  \\
5-jet (muon)  & $0.21$ &  $4.3$  & ${7.1}^{+3.2}_{-2.0}$ & $0.10$ &  $0.50$  \\
\noalign{\smallskip}\hline\noalign{\smallskip}
Soft dimuon channel &$0.39$ &  $7.7$  & ${7.8}^{+3.5}_{-2.4}$ & $0.50$ & $0.50$ \\
\noalign{\smallskip}\hline\noalign{\smallskip}
\multicolumn{6}{l}{Single-bin hard single-lepton channel}\\
\noalign{\smallskip}\hline\noalign{\smallskip}
3-jet  & $0.40$ &  $8.2$  & ${7.8}^{+3.3}_{-2.2}$ & $0.59$ &  $0.41$  \\
3-jet (electron) & $0.29$ &  $5.9$  & ${6.2}^{+2.1}_{-2.0}$ & $0.39$ &  $0.50$  \\
3-jet (muon)  & $0.35$ &  $7.0$  & ${5.7}^{+2.1}_{-1.5}$ & $0.72$ &  $0.27$  \\
5-jet & $0.28$ &  $5.7$  & ${7.2}^{+3.1}_{-1.9}$ & $0.26$ &  $0.50$  \\
5jet (electron) & $0.29$ &  $6.0$  & ${5.8}^{+2.2}_{-1.2}$ & $0.55$ &  $0.50$  \\
5-jet (muon) & $0.20$ &  $4.0$  & ${5.4}^{+1.6}_{-1.4}$ & $0.10$ &  $0.49$  \\
6-jet  & $0.22$ &  $4.5$  & ${6.0}^{+2.2}_{-1.6}$ & $0.10$ &  $0.50$  \\
6-jet (electron)  & $0.24$ &  $4.8$  & ${5.3}^{+1.6}_{-1.3}$ & $0.32$ &  $0.50$  \\
6-jet (muon) & $0.16$ &  $3.2$  & ${4.9}^{+1.4}_{-1.3}$ & $0.08$ &  $0.50$  \\
\noalign{\smallskip}\hline\noalign{\smallskip}
\multicolumn{6}{l}{Single-bin inclusive hard dilepton channel}\\
\noalign{\smallskip}\hline\noalign{\smallskip}
Low-multiplicity $ee$ and $\mu\mu$ combined &	0.44	& 11.6	& $11.5_{-3.4}^{+5.0}$	& 0.50	& 0.49	\\
Low-multiplicity $\mu\mu$ &	0.46	& 9.3	& $8.4_{-2.5}^{+3.9}$	& 0.61	& 0.38	\\
Low-multiplicity $ee$ &	0.35	& 7.2	& $8.1_{-2.4}^{+3.8}$	& 0.37	& 0.50	\\
Low-multiplicity $e\mu$ &	0.46	& 8.8	& $12.4_{-3.6}^{+5.4}$	& 0.16	& 0.50	\\
3-jet $ee$ and $\mu\mu$ combined & 0.29 & 5.8 & 8.4$^{+3.9}_{-2.6}$ & 0.15 & 0.50 \\
3-jet $ee$  & 0.24 & 4.9 & $6.3^{+3.2}_{-2.0}$ & 0.26 & 0.50 \\
3-jet $\mu\mu$ & 0.22 & 4.4 & $6.0^{+3.1}_{-1.9}$ & 0.19 & 0.50 \\
3-jet $e\mu$ & 0.40 & 8.2 & 8.8$^{+4.0}_{-2.6}$ & 0.41 & 0.50 \\
\noalign{\smallskip}\hline\noalign{\smallskip}
\end{tabular*}
\end{center}
\caption[Breakdown of upper limits.]{From left to right: 95\% confidence level (CL) upper limits on the visible cross section
($\langle\epsilon\sigma\rangle_{\rm obs}^{95}$) and on the number of signal events ($S_{\rm obs}^{95}$ ); 
$S_{\rm exp}^{95}$ shows the 95\% CL upper limit on the number of signal events, given the expected number (and $\pm 1\sigma$
excursions on the expectation) of background events; two-sided $CL_B$ value, i.e.\ the confidence level observed for
the background-only hypothesis, and the discovery p-value ($p(s = 0)$). For an observed number of events lower than
expected, the discovery p-value is truncated at 0.5. More details about the $CL_S$ prescription can be found in ref. \cite{Read:2002hq}.
\label{tab:upperlimitALL}}
\end{table}

\clearpage
\subsection{Exclusion limits on specific models}

Given the absence of a significant excess above the SM background expectation, exclusion limits are also placed 
on the various specific models of physics beyond the SM described in section
 \ref{simulsection}. In this case, the fit is modified in the following way:
\begin{enumerate}
  \item there is an extra free parameter for a possible non-SM signal strength which is constrained to be non-negative;
  \item the number of events observed in the signal region is now also considered as an input to the fit;
  \item the expected contamination of the control regions by the signal is included in the fit.
\end{enumerate}
For all regions except the soft dimuon channel, binned signal regions are used. The likelihood (see section \ref{sec:bkgfit})
takes into account the model shape information in the signal regions as a further discriminant by including bin-by-bin expectations. 

The systematic uncertainties on the signal expectations 
originating from detector effects and the theoretical uncertainties on the signal acceptance are included in the fit. 
The impact of the theoretical uncertainties on the signal cross section is shown on the limit plots obtained. 
Numbers quoted in the text are evaluated from the observed exclusion
limit based on the nominal signal cross section minus its $1\sigma$ theoretical uncertainty. 

As mentioned in section~\ref{sec:trigger}, the soft-lepton channel has a slightly lower integrated luminosity
than the hard-lepton channel. For consistency in all the following limit contour figures, the
integrated luminosity is abbreviated to 20~\ifb.

\subsubsection{Limits on phenomenological models}
The limits in the ($m_0$,$m_{1/2}$) mSUGRA/CMSSM plane obtained from the hard single-lepton signal regions are shown in figure \ref{fig:sugralimit}.
The observed limit is driven above the expected limit
mainly by the shape fit to the \met~distribution in the 6-jet binned signal region (see figure~\ref{fig:SRexclAfterFit}) as no data event is found in the high-\met~bin. 
At large values of $m_0$, this analysis is able to exclude a gluino mass of up to 1.2 \TeV.
~A 1.0 \TeV~limit on the gluino mass is obtained from the hard single-lepton channel
in the bRPV model at high $m_0$, while the limit is at 1.13 \TeV~in the nGM model, as shown in 
figures~\ref{fig:sugralimit} and \ref{fig:NUHMGlimit}, respectively.
The observed exclusion limit obtained by the hard single-lepton channel for the NUHMG model is also shown in figure~\ref{fig:NUHMGlimit}. 

The limit obtained for the mUED scenario is shown in figure \ref{fig:muedlimits}                                                         
in the ($1/R_{\mathrm{c}}$,$\Lambda R_{\mathrm{c}}$) plane.
The soft dimuon channel complements the hard dilepton case, as it covers the low $\Lambda R_{\mathrm{c}}$ values.
Because the signal regions in the two channels overlap, they cannot be combined in a straightforward manner.
Instead, the limit, for each point in the plane, is obtained by using the better expected limit from either the soft dimuon or the hard dilepton channel.
The limit obtained reaches up to a compactification radius of $1/R_{\mathrm{c}}=950$ \GeV~for a cut-off scale times radius ($\Lambda R_{\mathrm{c}}$) of approximately 30.

\clearpage
\begin{figure}[!ht]
\centering
\includegraphics[width=0.88\textwidth]{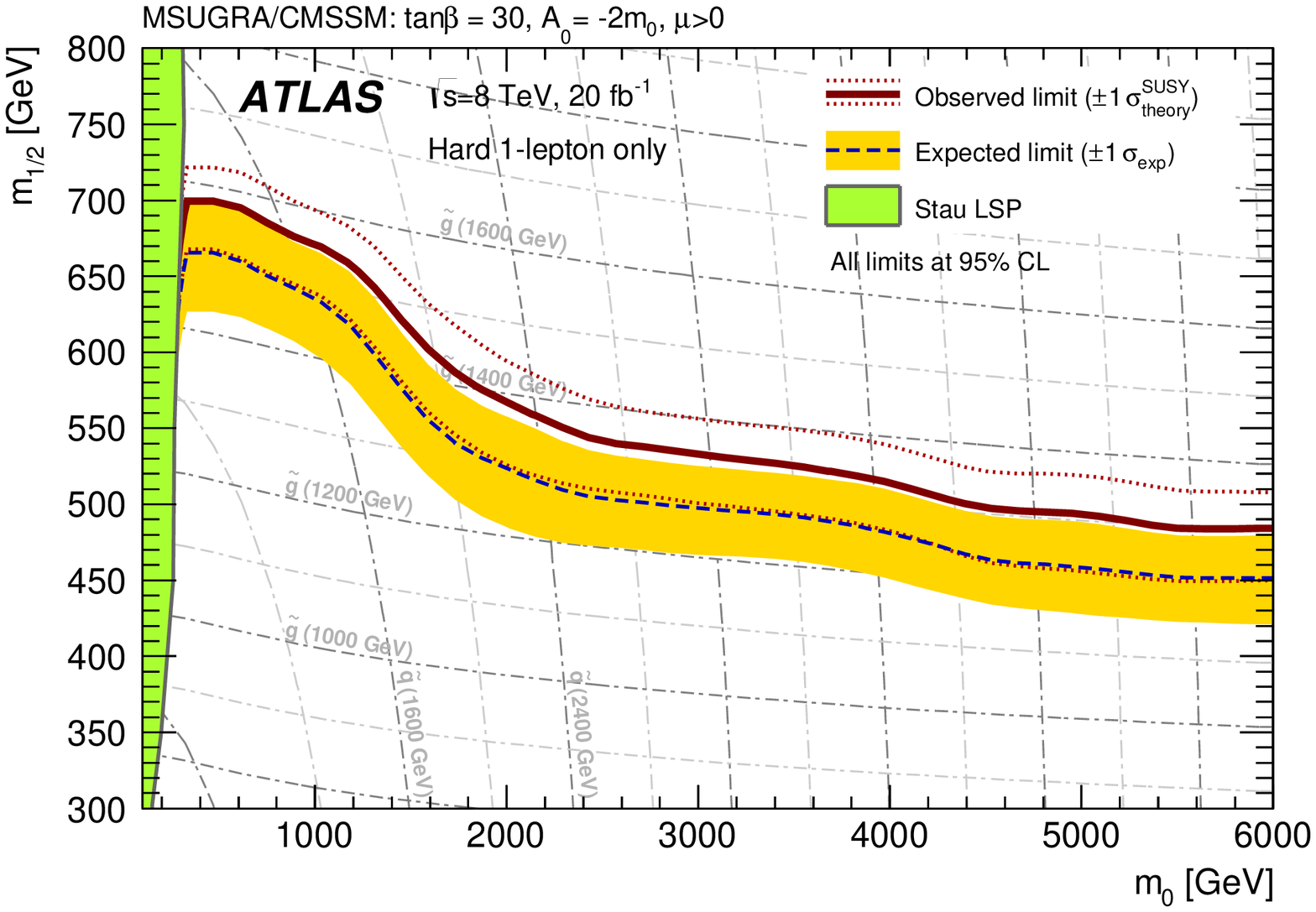}
\includegraphics[width=0.88\textwidth]{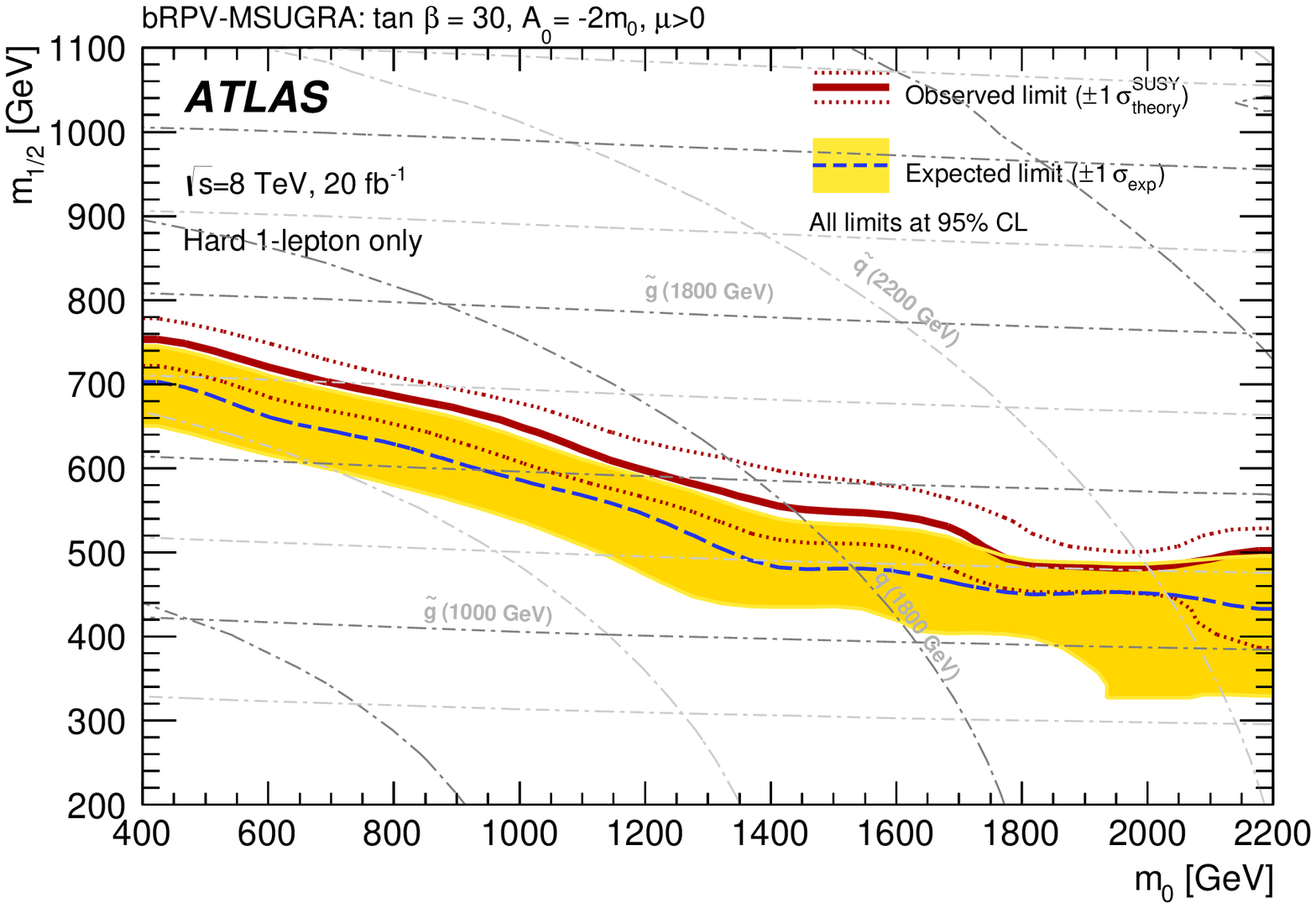}

\caption{
	95\% CL exclusion limit from the hard single-lepton channel in the ($m_0$,$m_{1/2}$) plane for the mSUGRA/CMSSM model with
$\tan\beta = 30$, $A_{0} = -2m_0$ and $\mu > 0$ 
(top) and in the same plane for the bRPV model (bottom). The green area represents the region of the parameter space for which the stau is the LSP.  
The dark blue dashed line shows the expected limits at 95\% CL, with the light (yellow) bands indicating the $\pm1\sigma$
variation on the median expected limit due to the experimental and background-only theory uncertainties. The observed nominal limit is shown by a solid
dark red line, with the dark red dotted lines indicating the $\pm1\sigma$ variation on this limit due to the theoretical scale and PDF uncertainties on 
the signal cross section. Lines of constant squark and gluino masses are also shown.} 
\label{fig:sugralimit}
\end{figure}

\begin{figure}
\begin{center}
\includegraphics[width=0.88\textwidth]{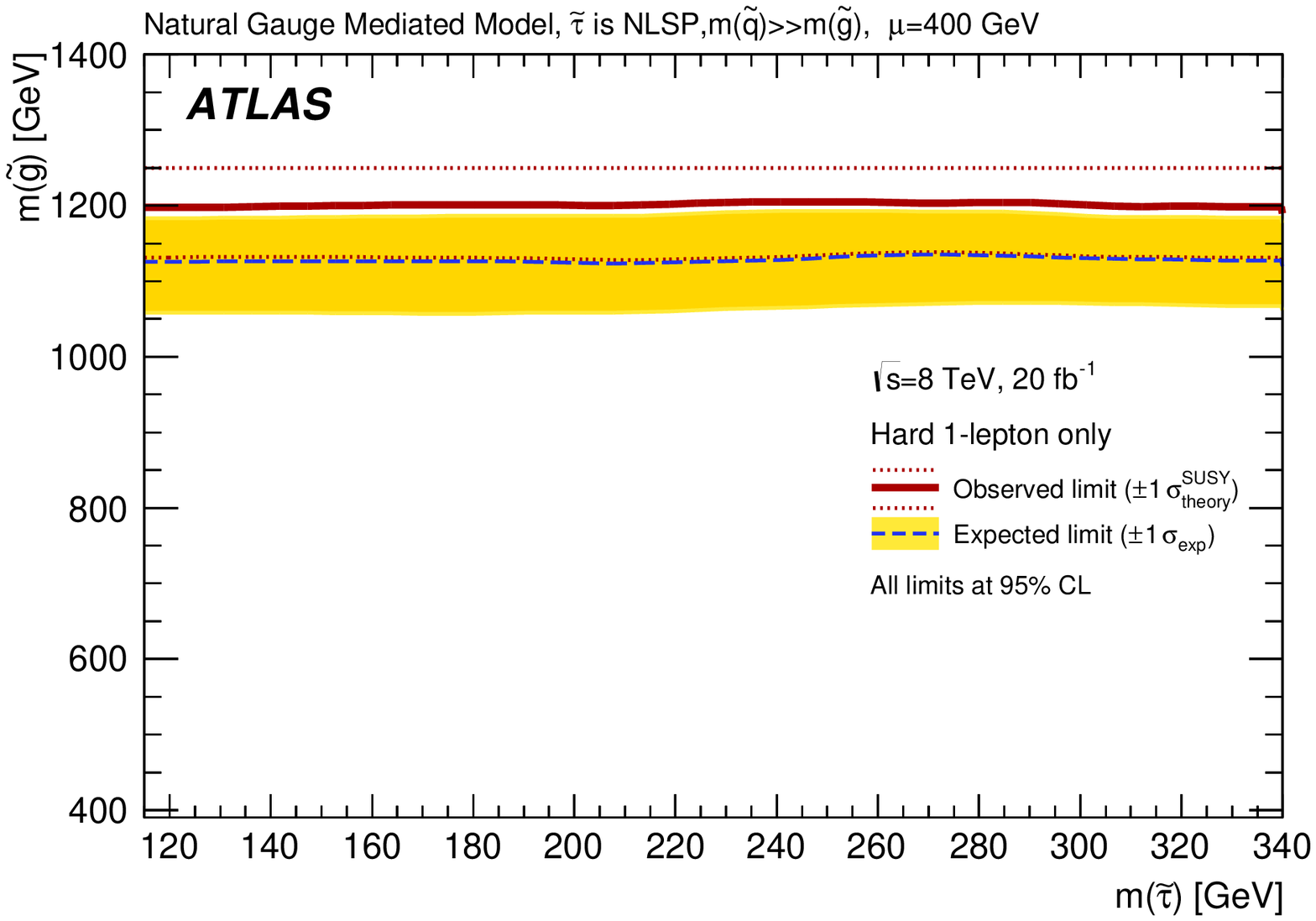}
\includegraphics[width=0.88\textwidth]{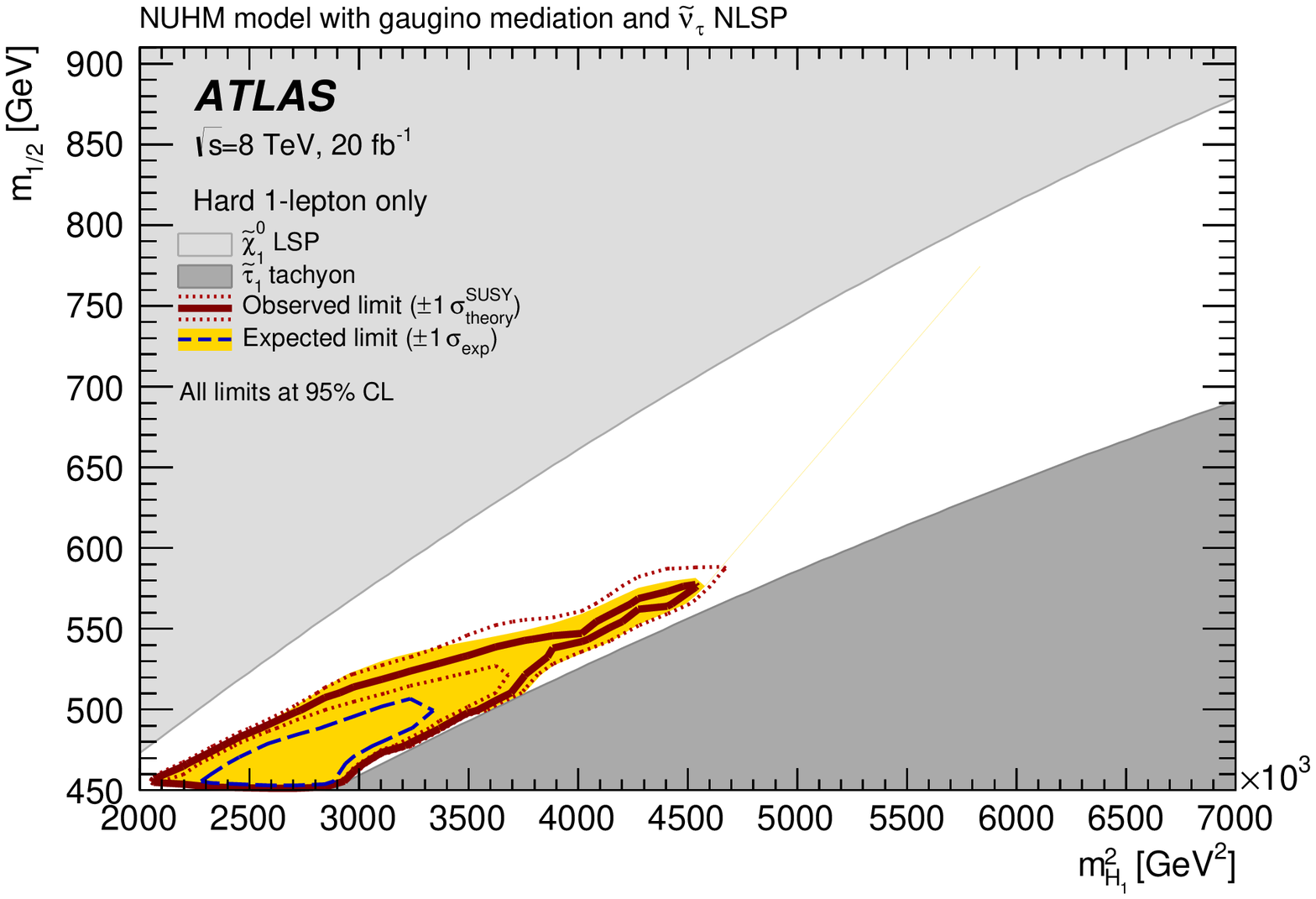} 
\caption{
	95\% CL exclusion limit from the hard single-lepton channel in the ($m_{\tilde{\tau}}$,$m_{\tilde{g}}$) plane for the nGM model (top) and in 
	the ($m_{H_1}^2$,$m_{1/2}$) plane for the NUHMG model (bottom).
The dark blue dashed line shows the expected limits at 95\% CL, with the light (yellow) bands indicating the $\pm1\sigma$
variation on the median expected limit due to the experimental and background-only theory uncertainties. 
The observed nominal limit is shown by a solid
dark red line, with the dark red dotted lines indicating the $\pm1\sigma$ variation on this limit due to the theoretical scale and PDF uncertainties on 
the signal cross section.
In the NUHMG model, the shaded areas are excluded either because the LSP is the $\tilde{\chi}^0_1$ or because the $\tilde{\tau}_1$ is a tachyon.  
}
\label{fig:NUHMGlimit}
\end{center}
\end{figure}

\begin{figure}[!ht]
\centering
  \includegraphics[width=0.89\textwidth]{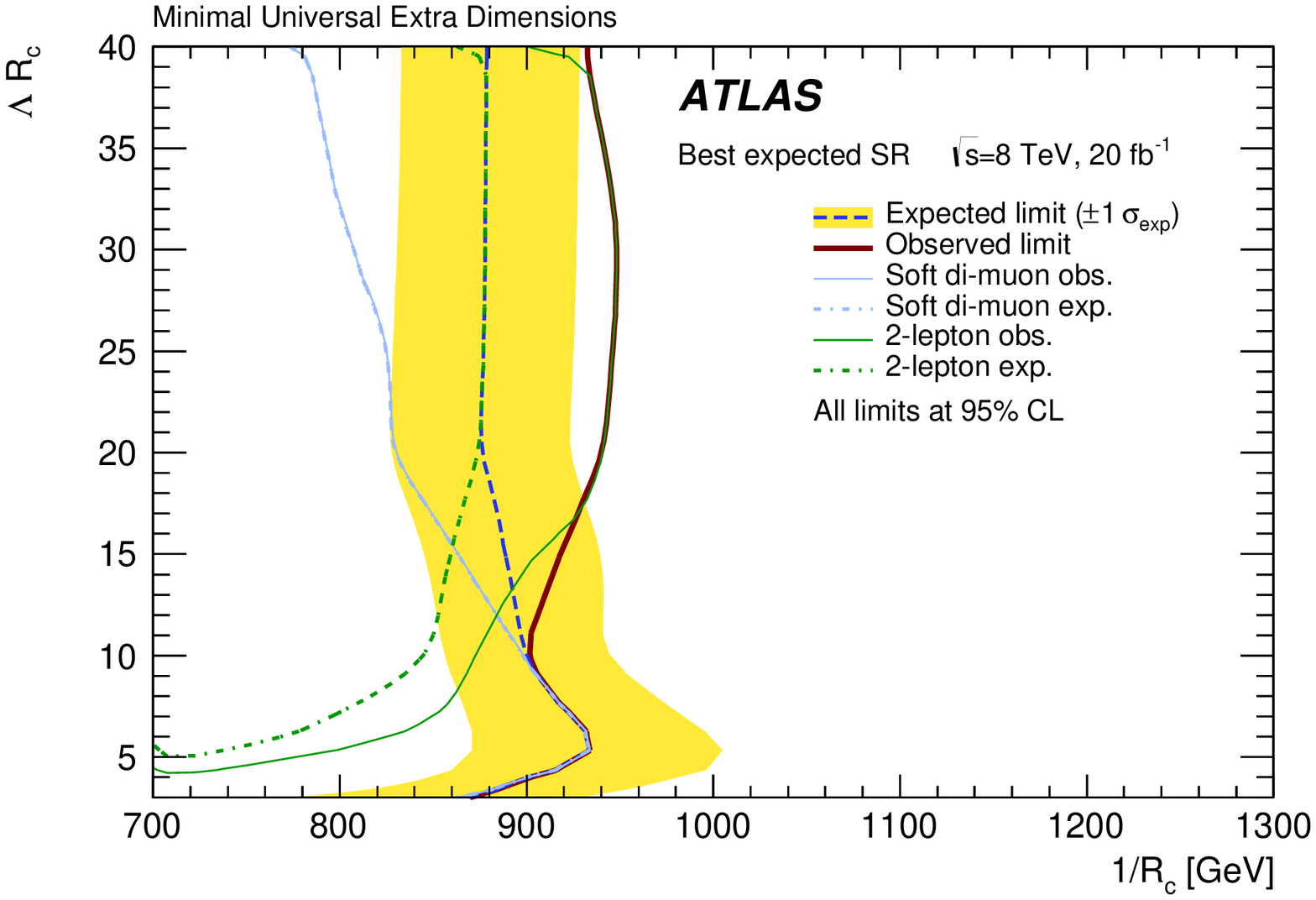}
\caption{
95\% CL exclusion limit from the combination of the soft dimuon and hard dilepton channels for the mUED model,
presented in the ($1/R_{\mathrm{c}}$,$\Lambda R_{\mathrm{c}}$) plane.
The dark blue dashed line shows the expected limits at 95\% CL, with the light (yellow) bands indicating the $\pm1\sigma$
variation on the median expected limit due to the experimental and background-only theory uncertainties. 
The observed nominal limit is shown by a solid
dark red line.
The blue and green full (dashed) lines show the observed (expected) 
exclusion obtained by the soft dimuon and hard dilepton analyses, respectively. For each point in the plane, the limit is obtained by using the better expected limit from either the soft dimuon or the hard dilepton channel. 
}
\label{fig:muedlimits}
\end{figure}

\subsubsection{Limits on simplified models}
The limits obtained from the hard single-lepton channel in the gluino-mediated top squark production
models, favoured by naturalness arguments, are shown in figure~\ref{fig:Gtt}. In the model where the mass gap between the $\stop_1$ and the
LSP is fixed at 20 \GeV, the observed exclusion reaches a gluino mass of 1.2 \TeV. In this model, the hard single-lepton analysis is complementary to the 0-lepton analysis
presented in ref.~\cite{0lepton}, in that the expected limit for the single-lepton analysis is able to cover higher top squark masses at intermediate gluino masses 
(e.g. 80 \GeV~higher at $m_{\gluino}=900$ \GeV).
In the model where the gluino exclusively decays to a pair of top quarks and the lightest neutralino, the observed exclusion reaches
a gluino mass of 1.28 \TeV.
~As in the mSUGRA/CMSSM case, the observed limit is more stringent than the expected limit mainly because of the 6-jet binned signal region.

As the soft and hard single-lepton channels are orthogonal in their signal and control region definitions, a 
full statistical combination of all soft and hard single-lepton signal regions is possible; the limits obtained using this combination are shown in
figures \ref{fig:1steplimitsNE60} for the gluino and the squark pair production simplified models. These limits are shown
in the ($m_{\tilde{g}(\tilde{q})}$,$m_{\tilde{\chi}^0_1}$) mass plane for the case in which  
$x=(m_{\tilde{\chi}^\pm_1}-m_{\tilde{\chi}^0_1})/(m_{\tilde{g}(\tilde{q})}-m_{\tilde{\chi}^0_1})=1/2$.
The soft single-lepton analysis is particularly powerful along the diagonal, where the masses of the gluino/squark
and the lightest neutralino are almost degenerate. The combination improves the exclusion where the sensitivity of the soft and hard single-lepton signal regions 
are similar; for example, at $m_{\gluino}\approx800$ \GeV, near the combined observed exclusion limit, the observed $CL_S$ values for the individual channels are 0.27 and 0.19 for the 
soft and hard single-lepton signal regions, respectively, while the combination gives a $CL_S$ value of 0.08. The combined limit on the gluino mass reaches up to 1.2 \TeV. This result complements the
 statistically independent 0-lepton analysis, as the lepton channel offers a better sensitivity to more compressed scenarios at intermediate
gluino masses while the 0-lepton analysis is able to reach slightly higher gluino masses.
The squark limits are considerably weaker than those for the gluino due to the lower production cross section and only reach up to 750 \GeV.\footnote{At 
high squark masses, the combined limit is not exactly the same as the hard single-lepton limit; this is due to the combination having slightly different $CL_S$ values.} The very low squark masses are already excluded by previous analyses \cite{paper7tev}.

Limits are shown in figure \ref{fig:1steplimitsEE60} for the gluino and squark simplified
models in which the LSP mass is set at 60 \GeV~and
the value of $x$ is varied, using again the hard single-lepton channel in combination
with the soft single-lepton channel.
In these models, a gluino mass up to 1.22 \TeV~and a squark mass
up to 800 \GeV~are excluded. Again, the single-lepton and the 0-lepton channels cover different parameter space, the latter being able to exclude larger masses at low $x$ (where it excludes a gluino mass
up to 1.13 \TeV), while the analysis presented here excludes higher squark masses at moderate to high $x$ values.

For the two-step gluino/squark simplified models with sleptons, the limits are obtained using the hard dilepton
signal regions either individually (for squark production) or in combination with the hard single-lepton signal regions (for gluino
production). These limits are shown in figure \ref{fig:2stepsleptonlimit}. A gluino mass of up to 1.32 \TeV~
and a squark mass up to 840 \GeV~can be excluded in these models. This result is complementary to the exclusion limit obtained in the 2-same-charge/3-lepton analysis
described in ref.~\cite{2lepss}. The analysis presented here is able to probe higher gluino/squark mass values (the expected limit is higher by approximately 
100 \GeV~and 50 \GeV~for the gluino and squark cases, respectively), while the analysis presented in ref.~\cite{2lepss}  is able to cover higher neutralino masses at intermediate gluino/squark mass values (e.g. the expected limits are 160~\GeV~higher at $m_{\gluino}=1000$~\GeV~and 140~\GeV~higher at $m_{\squark}=650$~\GeV). 

Using the hard single-lepton channel, a gluino mass up to 1.14 \TeV~can be excluded for the two-step
simplified model without sleptons, as can be seen in figure \ref{fig:2stepwwzzlimit}. Also here, 
the hard single-lepton channel complements the exclusion limits obtained 
in the 2-same-charge/3-lepton analysis described in ref.~\cite{2lepss} and the 0-lepton multijet analysis described in ref.~\cite{0lepmultijet}.
In particular, this analysis reaches higher neutralino masses at high gluino mass values with respect to the 2-same-charge/3-lepton analysis (e.g. the expected limit is 
80~\GeV~higher at $m_{\gluino}=1050~\GeV$) and 
has an expected limit on the neutralino mass similar to that 
from the orthogonal multijet analysis for low-mass gluinos.

\begin{figure}
 \begin{center}
    \includegraphics[width=0.89\textwidth]{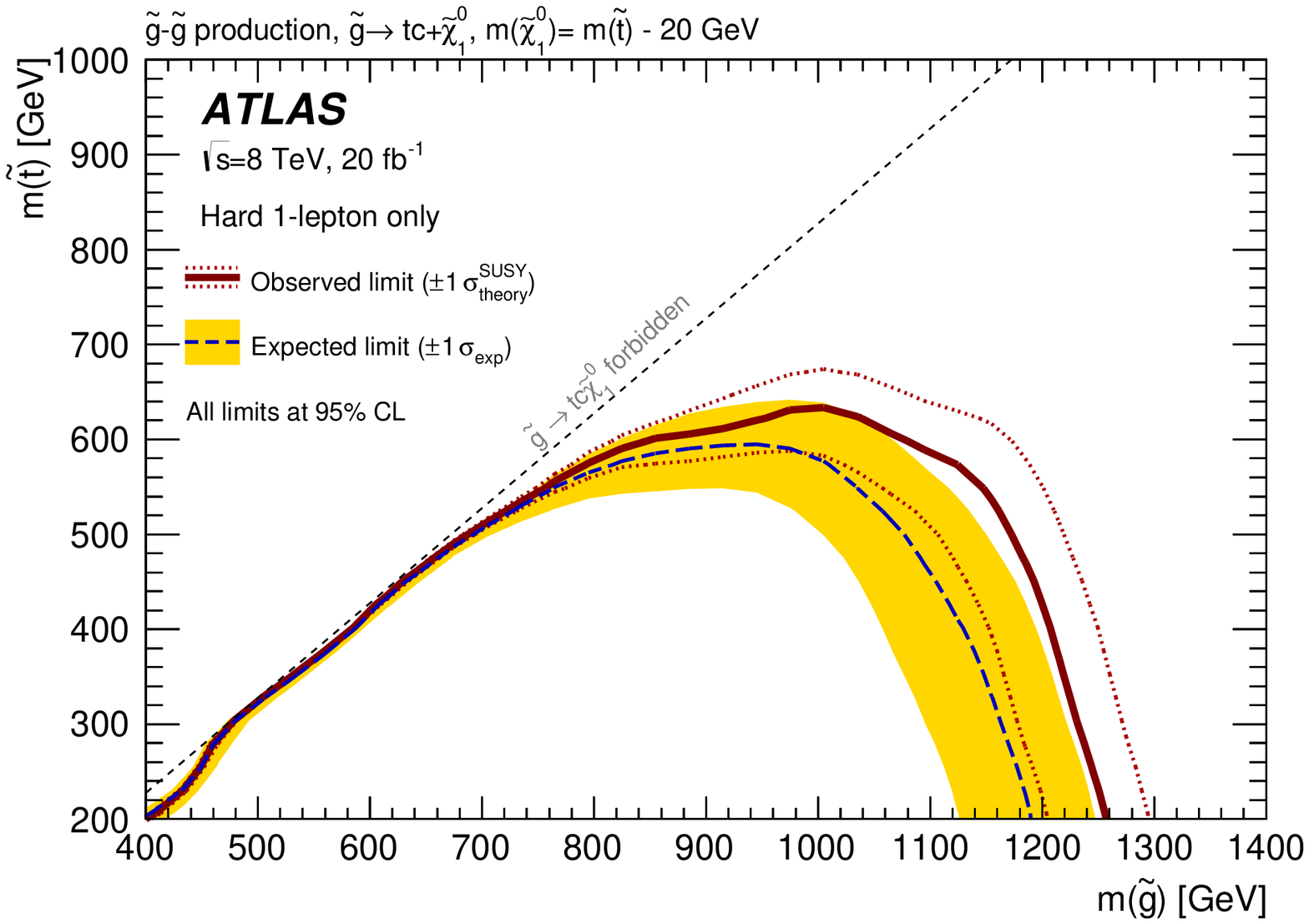}
   \includegraphics[width=0.89\textwidth]{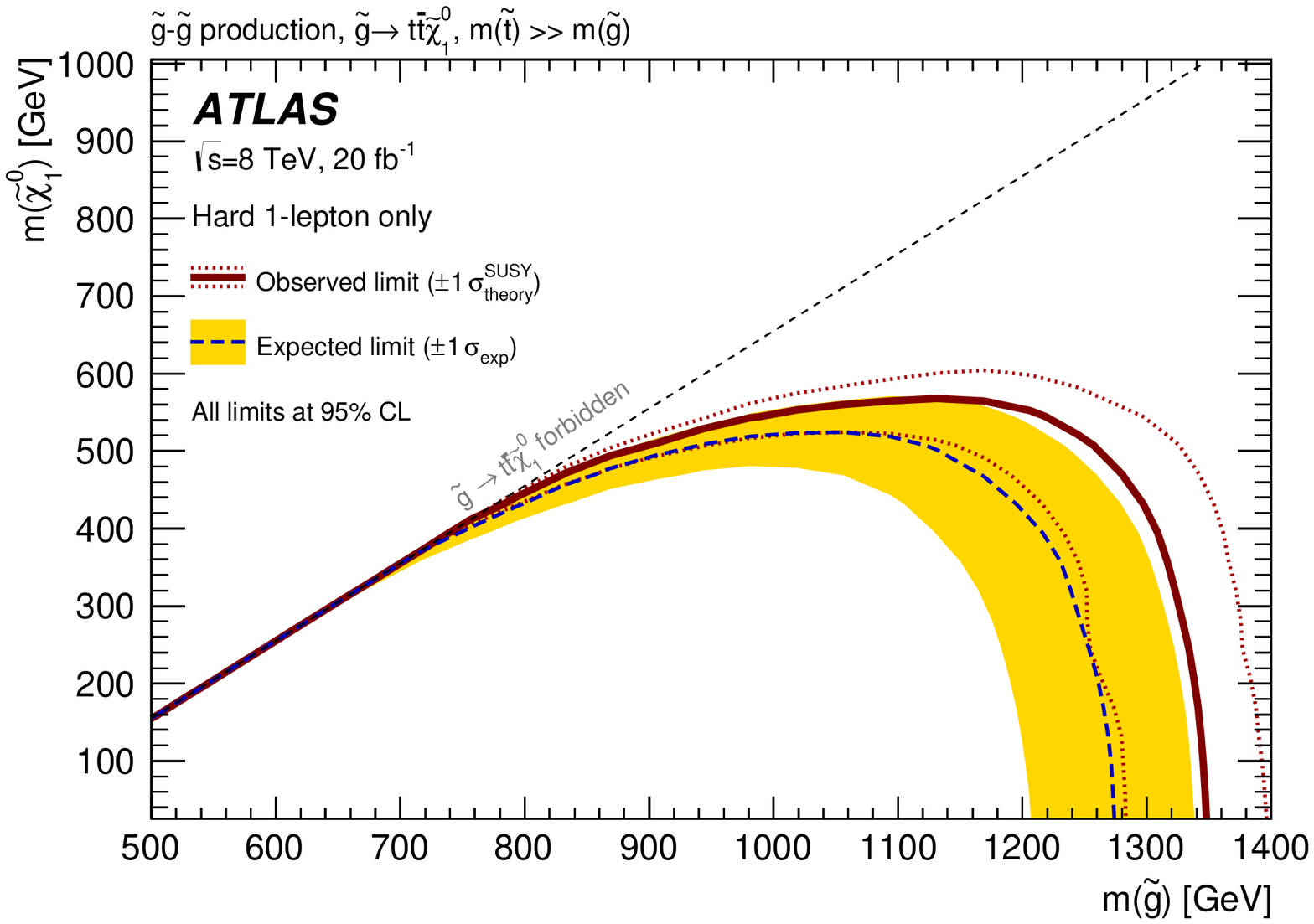}  
   \caption{95\% CL exclusion limit from the hard single-lepton channel in the ($m_{\tilde{g}}$,$m_{\tilde{t}(\tilde{\chi}_{1}^{0})}$) plane for
the simplified model with gluino-mediated top squark production, where the top squark is assumed to decay exclusively via $\tilde{t}\to c\tilde{\chi}_{1}^{0}$ (top) 
or where the gluinos are assumed to decay exclusively through a virtual top squark, $\gluino \to t\bar t\ninoone$ (bottom) . 
The dark blue dashed line shows the expected limits at 95\% CL, with the light (yellow) bands indicating the $\pm1\sigma$
variation on the median expected limit due to the experimental and background-only theory uncertainties. 
The observed nominal limit is shown by a solid
dark red line, with the dark red dotted lines indicating the $\pm1\sigma$ variation on this limit due to the theoretical scale and PDF uncertainties on 
the signal cross section.
}
  \label{fig:Gtt}
 \end{center}
\end{figure}

\begin{figure}[!ht]
\centering
    \includegraphics[width=0.89\textwidth]{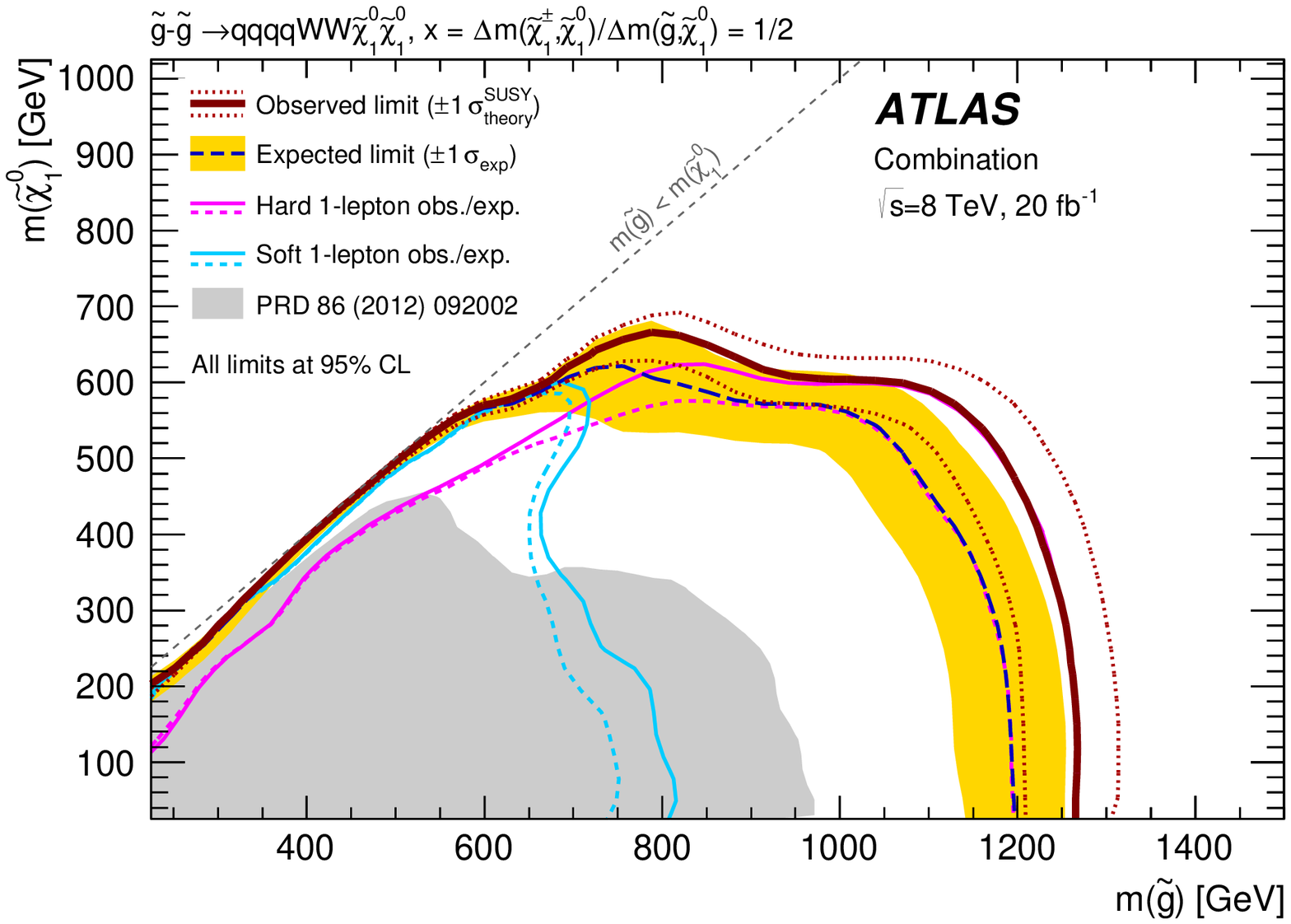}
    \includegraphics[width=0.89\textwidth]{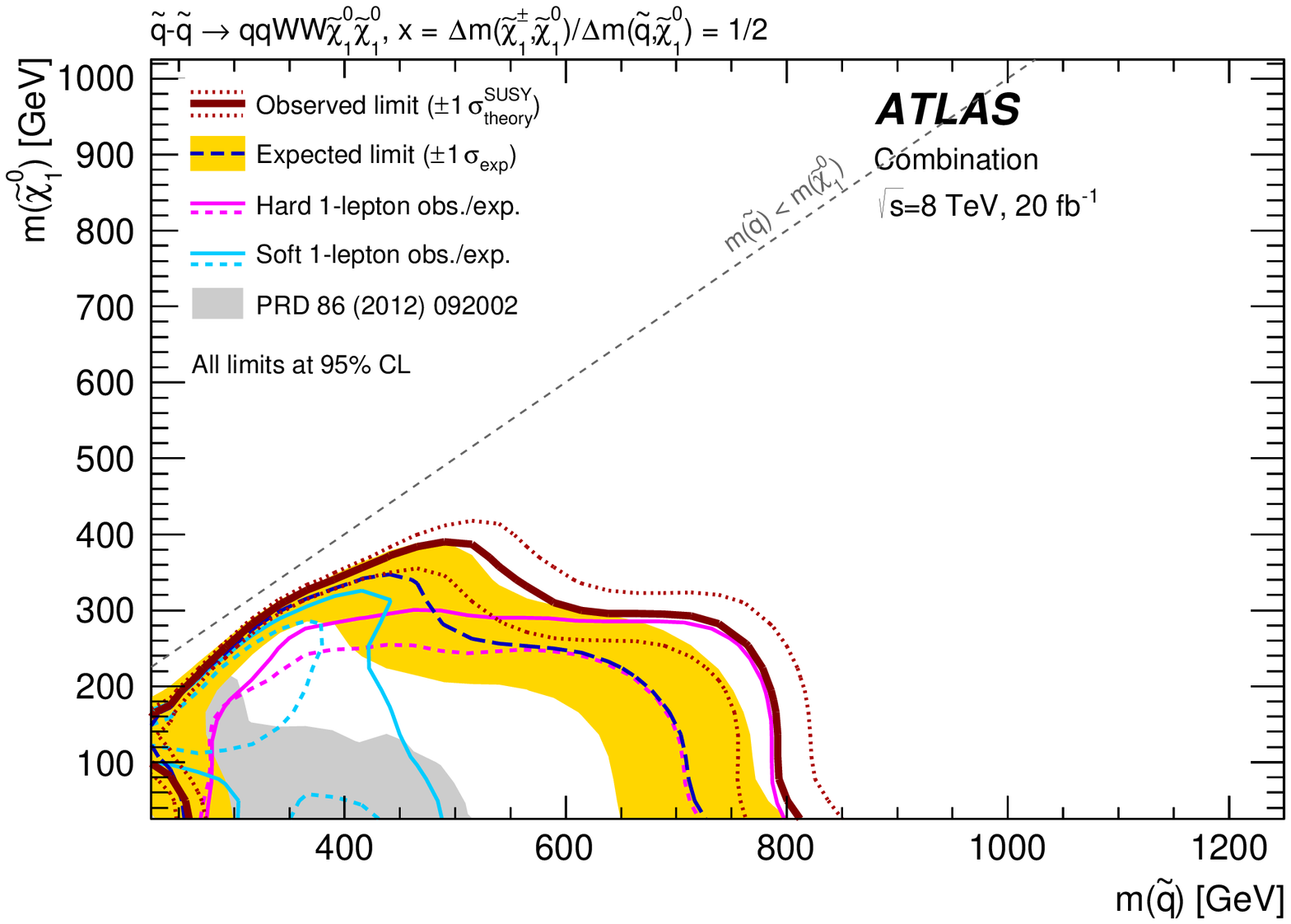}    

\caption{ 95\% CL exclusion limit for the gluino simplified model (top) and the first- and second-generation squark simplified model (bottom) from the combination of the soft and hard single-lepton analyses. 
The limits are presented in the 
($m_{\tilde{g}(\tilde{q})}$,$m_{\tilde{\chi}^0_1}$) mass plane for the case in which the 
chargino mass is fixed at $x=(m_{\tilde{\chi}^\pm_1}-m_{\tilde{\chi}^0_1})/(m_{\tilde{g}(\tilde{q})}-m_{\tilde{\chi}^0_1})=1/2$.  
The dark blue dashed line shows the expected limits at 95\% CL, with the light (yellow) bands indicating 
the $\pm1\sigma$ variation on the median expected limit due to the experimental and background-only theory uncertainties. 
The observed nominal limit is shown by a solid
dark red line, with the dark red dotted lines indicating the $\pm1\sigma$ variation on this limit due to the theoretical scale and PDF uncertainties on 
the signal cross section.
The observed limit set by the previous ATLAS analysis \cite{paper7tev} using 7~\TeV~data 
is shown as a gray area.  The light blue and purple full (dashed) lines show the observed (expected) 
exclusion obtained by the soft and hard single-lepton analyses, respectively. }
\label{fig:1steplimitsNE60}
\end{figure}

\begin{figure}[!ht]
\centering
    \includegraphics[width=0.89\textwidth]{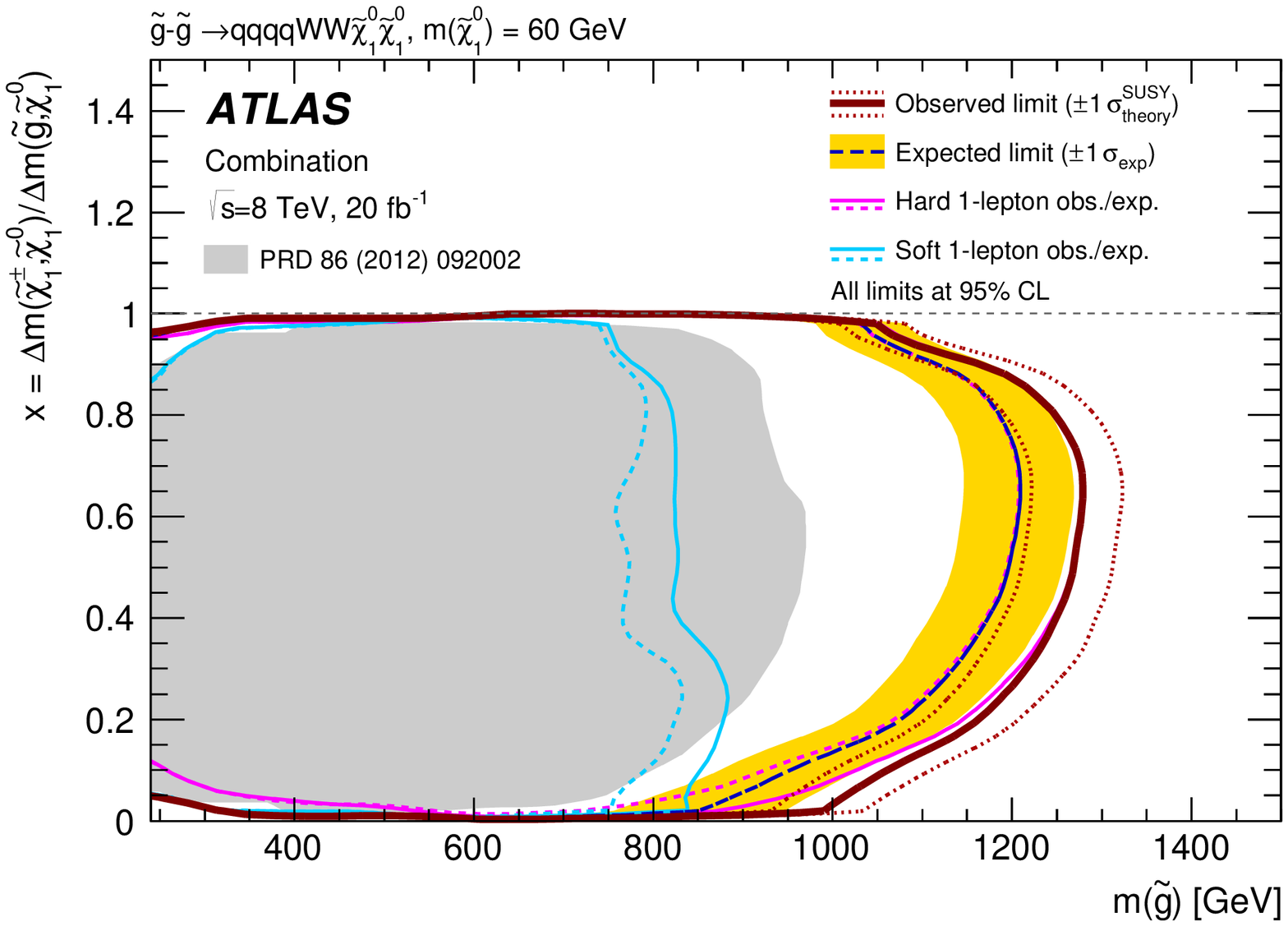}
    \includegraphics[width=0.89\textwidth]{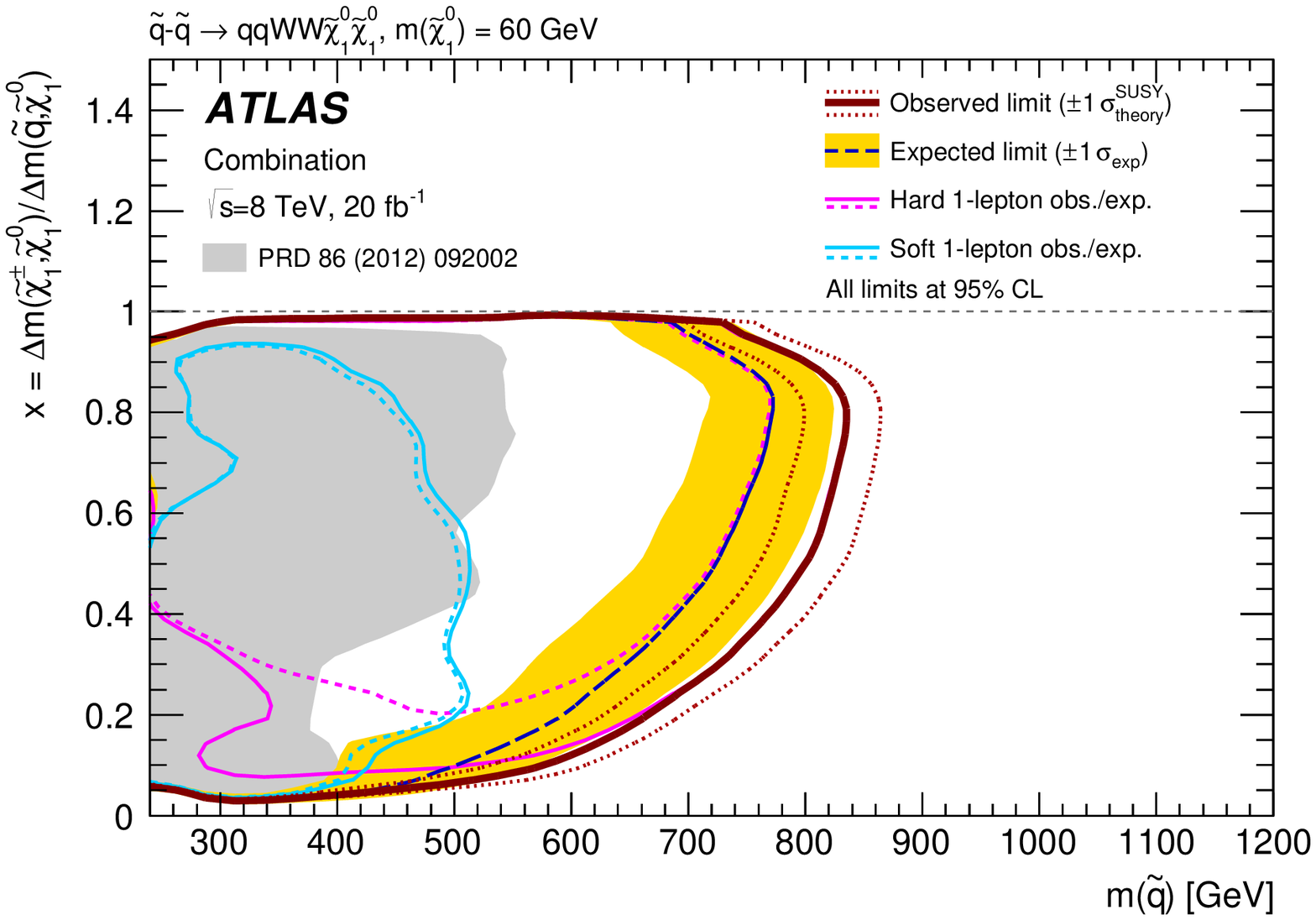}
\caption{
95\% CL exclusion limit for the gluino simplified model (top) and the
first- and second-generation squark simplified model (bottom) from the combination of the soft and hard single-lepton analyses. 
The limits are
presented in the ($m_{\tilde{g}(\tilde{q})}$,$x$) mass plane,
where ${x}=(m_{\tilde{\chi}^\pm_1}-m_{\tilde{\chi}^0_1})/(m_{\tilde{g}(\tilde{q})}-m_{\tilde{\chi}^0_1})$,
for the case in which the chargino mass is varied and the LSP mass is set at 60 \GeV.
The dark blue dashed line shows the expected limits at 95\% CL, with the light (yellow) bands indicating the $\pm1\sigma$
variation on the median expected limit due to the experimental and background-only theory uncertainties. 
The observed nominal limit is shown by a solid
dark red line, with the dark red dotted lines indicating the $\pm1\sigma$ variation on this limit due to the theoretical scale and PDF uncertainties on 
the signal cross section.
The observed limit set by the previous ATLAS analysis \cite{paper7tev} using 7~\TeV~data
is shown as a gray area. The light blue and purple full (dashed) lines show the observed (expected) 
exclusion obtained by the soft and hard single-lepton analyses, respectively. 
} \label{fig:1steplimitsEE60}
\end{figure}

\begin{figure}[!ht]
\centering
\includegraphics[width=0.89\textwidth]{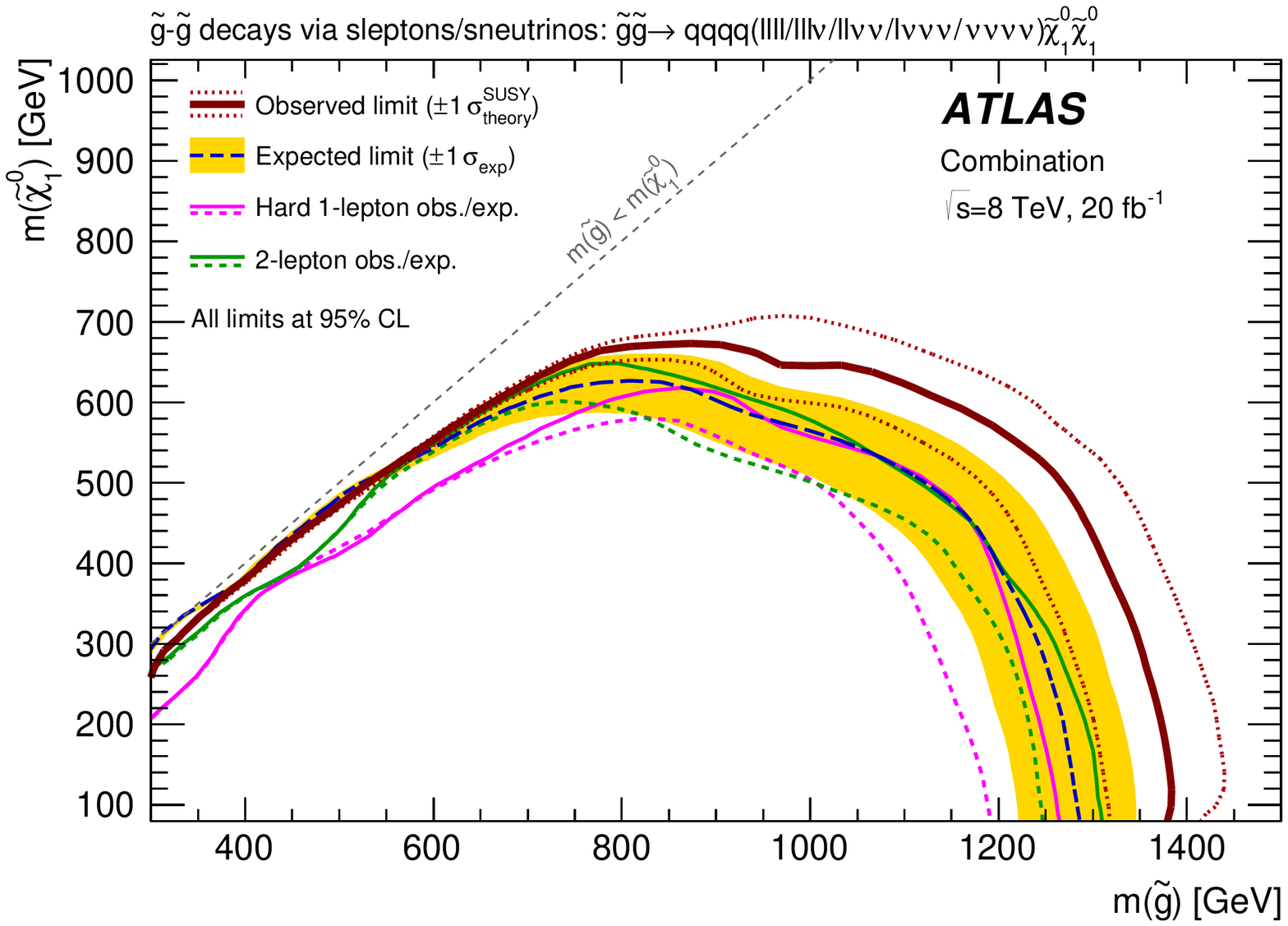}
  \includegraphics[width=0.89\textwidth]{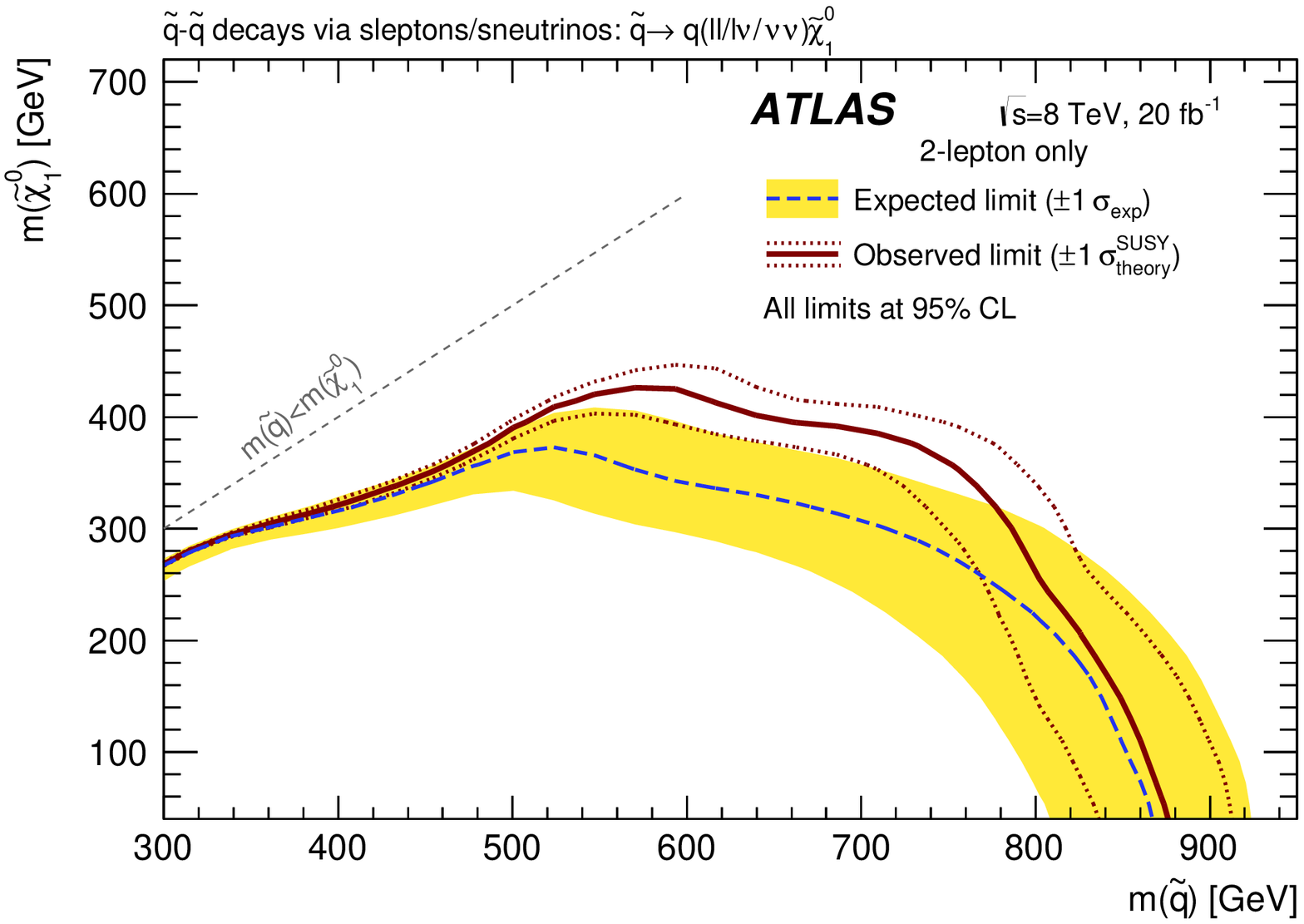}
\caption{
95\% CL exclusion limit for the two-step gluino simplified model with sleptons from the combination of the hard dilepton and single-lepton channels (top) and the two-step
first- and second-generation squark simplified model with sleptons from the hard dilepton channel (bottom). The limits are
presented in the ($m_{\tilde{g}(\tilde{q})}$,$m_{\tilde{\chi}^0_1}$) mass plane.
The dark blue dashed line shows the expected limits at 95\% CL, with the light (yellow) bands indicating the $\pm1\sigma$
variation on the median expected limit due to the experimental and background-only theory uncertainties. 
The observed nominal limit is shown by a solid
dark red line, with the dark red dotted lines indicating the $\pm1\sigma$ variation on this limit due to the theoretical scale and PDF uncertainties on 
the signal cross section.
The green and purple full (dashed) lines show the observed (expected) 
exclusion obtained by the hard dilepton and single-lepton analyses, respectively. 
} \label{fig:2stepsleptonlimit}
\end{figure}

\begin{figure}[!ht]
\centering
   \includegraphics[width=0.89\textwidth]{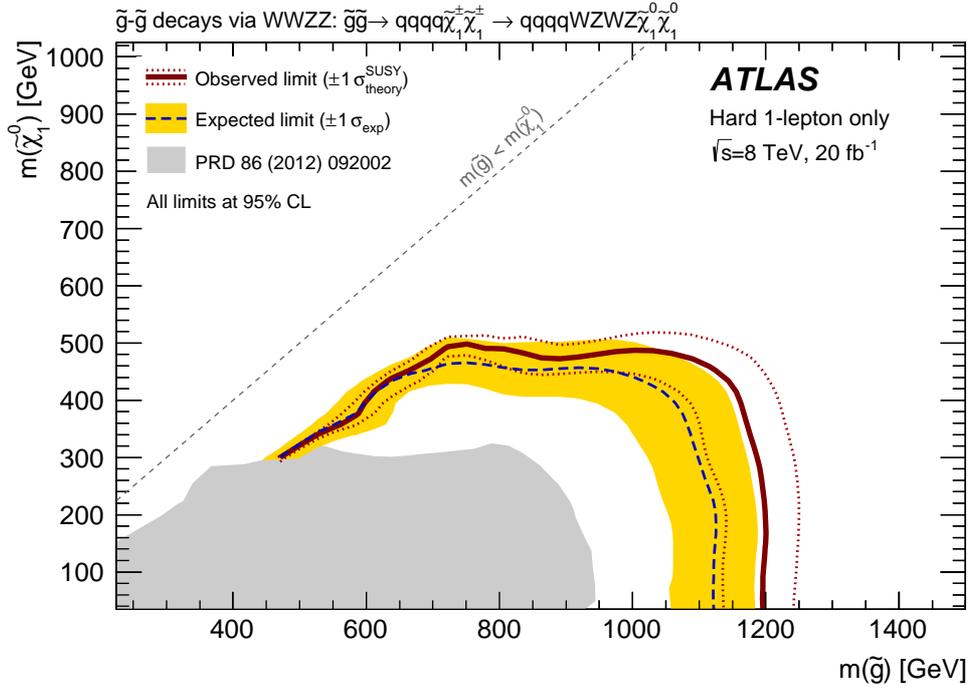}
\caption{
95\% CL exclusion limit from the hard single-lepton channel for the two-step gluino simplified model without sleptons
presented in the ($m_{\tilde{g}}$,$m_{\tilde{\chi}^0_1}$) mass plane. 
The dark blue dashed line shows the expected limits at 95\% CL, with the light (yellow) bands indicating the $\pm1\sigma$
variation on the median expected limit due to the experimental and background-only theory uncertainties. 
The observed nominal limit is shown by a solid
dark red line, with the dark red dotted lines indicating the $\pm1\sigma$ variation on this limit due to the theoretical scale and PDF uncertainties on 
the signal cross section.
The limit is not extrapolated to lower gluino/neutralino masses where no grid point was generated.
} \label{fig:2stepwwzzlimit}
\end{figure}

\clearpage
\section{Conclusion}\label{conclusion}

A search with the ATLAS detector at the LHC for SUSY in final states 
containing at least one isolated lepton (electron or muon), jets and large missing transverse momentum 
is presented. This analysis uses 20 \ifb~of proton--proton collision data collected at a centre-of-mass energy of 8~\TeV.
Several single-lepton and dilepton signal regions are used to cover a broad parameter space for a variety of models. 
The hard-lepton channel is complemented by a soft-lepton channel in order to increase the
 sensitivity to supersymmetric spectra with small mass splitting. Some signal regions are also subdivided in order
to enhance the sensitivity to gluino or squark production, by requiring higher or lower jet multiplicity, or by
placing requirements on the charge or flavour of the leptons in the dilepton channel.

Observations are in agreement with SM expectations in each signal region and limits are set 
on the visible cross section in models of new physics within the kinematic requirements of the searches. 
Exclusion limits are also placed on a large number of supersymmetric models, including a bRPV model, and on one model of mUED, 
for which a compactification radius of $1/R_{\mathrm{c}}=950$~\GeV~is excluded for a cut-off scale times radius ($\Lambda R_{\mathrm{c}}$) of approximately 30.
These limits are either new or extend the region of parameter space excluded by previous searches with the ATLAS detector.
Depending on the model of supersymmetry considered, the limits are able to exclude gluino masses up to 1.32~\TeV~and squark masses up to 840~\GeV.


\acknowledgments
We thank CERN for the very successful operation of the LHC, as well as the
support staff from our institutions without whom ATLAS could not be
operated efficiently.

We acknowledge the support of ANPCyT, Argentina; YerPhI, Armenia; ARC,
Australia; BMWFW and FWF, Austria; ANAS, Azerbaijan; SSTC, Belarus; CNPq and FAPESP,
Brazil; NSERC, NRC and CFI, Canada; CERN; CONICYT, Chile; CAS, MOST and NSFC,
China; COLCIENCIAS, Colombia; MSMT CR, MPO CR and VSC CR, Czech Republic;
DNRF, DNSRC and Lundbeck Foundation, Denmark; EPLANET, ERC and NSRF, European Union;
IN2P3-CNRS, CEA-DSM/IRFU, France; GNSF, Georgia; BMBF, DFG, HGF, MPG and AvH
Foundation, Germany; GSRT and NSRF, Greece; RGC, Hong Kong SAR, China; ISF, MINERVA, GIF, I-CORE and Benoziyo Center, Israel; INFN, Italy; MEXT and JSPS, Japan; CNRST, Morocco; FOM and NWO, Netherlands; BRF and RCN, Norway; MNiSW and NCN, Poland; GRICES and FCT, Portugal; MNE/IFA, Romania; MES of Russia and NRC KI, Russian Federation; JINR; MSTD,
Serbia; MSSR, Slovakia; ARRS and MIZ\v{S}, Slovenia; DST/NRF, South Africa;
MINECO, Spain; SRC and Wallenberg Foundation, Sweden; SER, SNSF and Cantons of
Bern and Geneva, Switzerland; NSC, Taiwan; TAEK, Turkey; STFC, the Royal
Society and Leverhulme Trust, United Kingdom; DOE and NSF, United States of
America.

The crucial computing support from all WLCG partners is acknowledged
gratefully, in particular from CERN and the ATLAS Tier-1 facilities at
TRIUMF (Canada), NDGF (Denmark, Norway, Sweden), CC-IN2P3 (France),
KIT/GridKA (Germany), INFN-CNAF (Italy), NL-T1 (Netherlands), PIC (Spain),
ASGC (Taiwan), RAL (UK) and BNL (USA) and in the Tier-2 facilities
worldwide.

\clearpage
\addcontentsline{toc}{section}{Bibliography}
\bibliographystyle{JHEP}
\bibliography{paper_LepMETJets_2013}

\clearpage
\begin{flushleft}
{\Large The ATLAS Collaboration}

\bigskip

G.~Aad$^{\rm 85}$,
B.~Abbott$^{\rm 113}$,
J.~Abdallah$^{\rm 152}$,
S.~Abdel~Khalek$^{\rm 117}$,
O.~Abdinov$^{\rm 11}$,
R.~Aben$^{\rm 107}$,
B.~Abi$^{\rm 114}$,
M.~Abolins$^{\rm 90}$,
O.S.~AbouZeid$^{\rm 159}$,
H.~Abramowicz$^{\rm 154}$
% ...
\bigskip
\\
% ...
$^{11}$ Institute of Physics, Azerbaijan Academy of Sciences, Baku, Azerbaijan\\
% ...
$^{85}$ CPPM, Aix-Marseille Universit{\'e} and CNRS/IN2P3, Marseille, France\\
% ...
$^{90}$ Department of Physics and Astronomy, Michigan State University, East Lansing MI, United States of America\\
% ...
$^{107}$ Nikhef National Institute for Subatomic Physics and University of Amsterdam, Amsterdam, Netherlands\\
% ...
$^{113}$ Homer L. Dodge Department of Physics and Astronomy, University of Oklahoma, Norman OK, United States of America\\
$^{114}$ Department of Physics, Oklahoma State University, Stillwater OK, United States of America\\
% ...
$^{117}$ LAL, Universit{\'e} Paris-Sud and CNRS/IN2P3, Orsay, France\\
% ...
$^{152}$ Institute of Physics, Academia Sinica, Taipei, Taiwan\\
% ...
$^{154}$ Raymond and Beverly Sackler School of Physics and Astronomy, Tel Aviv University, Tel Aviv, Israel\\
% ...
$^{159}$ Department of Physics, University of Toronto, Toronto ON, Canada\\
% ...
\end{flushleft}


\end{document}